\title{Bellman filtering and smoothing for state-space models}
\author{
        Rutger-Jan Lange\footnote{I thank Wisse Rutgers for research assistance, Serena Ng for helpful editorial guidance, and the anonymous AE and two referees for their valuable comments. Thansks are also due to Maksim Anisimov, Francisco Blasques, Leopoldo Catania, Dick van Dijk, Simon Donker van Heel, Jippe van Dunn{\'e}, Dennis Fok, Maria Grith, Andrew Harvey, Christiaan Heij, Elwin Kardux, Matthias Katzfuss, Onno Kleen, Erik Kole, Siem Jan Koopman, Rutger Lit, Rasmus Lonn, Andr{\'e} Lucas, Robin Lumsdaine, Jan Maciejowski, Andrea Naghi, Jochem Oorschot, Richard Paap, Andreas Pick, Krzysztof Postek, Rogier Quaedvlieg, Daniel Ralph, Bram van Os, Omiros Papaspiliopoulos, Marcel Scharth, Annika Schn{\"u}cker, Ekaterina Smetanina, Panos Toulis, Stephen Thiele, Nando Vermeer, Sebastiaan Vermeulen, Michel van der Wel, Martina Zaharieva, Mikhail Zhelonkin and Chen Zhou. Finally, I thank participants of the 2021 North American summer meeting of the Econometric Society and the 27th international conference on Computing in Economics and Finance for stimulating discussions. }
        \\
                Econometric Institute, Erasmus School of Economics,                Rotterdam, Netherlands
}
\date{\today}

\documentclass[11pt]{article}

\usepackage[hyperindex,breaklinks]{hyperref}
\usepackage{dsfont}

\usepackage{hyperref}
\hypersetup{
    colorlinks=true,
    linkcolor=black,
    citecolor=black,
    filecolor=black,
    urlcolor=black,
}

\usepackage[margin=0.78in]{geometry}
\usepackage{eqnarray,amsmath}
\usepackage{braket}
\usepackage{siunitx}
\usepackage{amsthm}
\usepackage{bm}
\usepackage{color}
\usepackage{pdflscape}
\usepackage{amsmath,amssymb}
\usepackage{natbib,float}
\newtheorem{proposition}{Proposition}
\newtheorem{theorem}{Theorem}
\newtheorem{corollary}{Corollary}
\newtheorem{definition}{Definition}
\newtheorem{remark}{Remark}
\newtheorem{assump}{Assumption}

\newtheorem{assumption}{Assumption}
\newtheorem{lemma}{Lemma}

\usepackage{setspace}
\usepackage{comment}
\usepackage{multirow}
\usepackage{booktabs}
\usepackage{tabularx}
\usepackage[para,online,flushleft]{threeparttable} 

\usepackage[demo]{graphicx}
\usepackage{subfig}

\newcommand{\dd}{\mathrm{d}}

\usepackage[vpos=1cm]{draftwatermark}
\SetWatermarkText{\textnormal{Forthcoming in} \textit{Journal of Econometrics}}
\SetWatermarkScale{0.4}
\SetWatermarkColor[gray]{0.5}
	\SetWatermarkAngle{0}

\begin{document}
\maketitle

\begin{abstract}
\noindent This paper presents a new filter for state-space models based on Bellman's dynamic-programming principle, allowing for nonlinearity, non-Gaussianity and degeneracy in the observation and/or state-transition equations. The resulting Bellman filter is a direct generalisation of the (iterated and extended) Kalman filter, enabling scalability to higher dimensions while remaining computationally inexpensive. It can also be extended to enable smoothing. Under suitable conditions, the Bellman-filtered states are stable over time and contractive towards a region around the true state at every time step. 
Static (hyper)parameters are estimated by maximising a filter-implied  pseudo log-likelihood decomposition. In univariate simulation studies, the Bellman filter performs on par with state-of-the-art simulation-based techniques at a fraction of the computational cost. In two empirical applications, involving up to $150$ spatial dimensions or highly degenerate/nonlinear state dynamics, the Bellman filter outperforms competing methods in both accuracy and speed. 
\end{abstract}

\lineskip=0pt

\section{Introduction}
\label{section1}

\subsection{State-space models}
State-space models allow observations to be affected by an unobserved state that changes stochastically over time. For discrete times $t=1,2,\ldots,n$, the observation $\bm{y}_t\in\mathbb{R}^l$ is drawn from a conditional distribution, $p(\bm{y}_t|\bm{\alpha}_t)$, while the latent state $\bm{\alpha}_t\in\mathbb{R}^m$ follows a first-order Markov process with a state-transition density, $p(\bm{\alpha}_{t+1}|\bm{\alpha}_{t})$, and some initial condition, $p(\bm{\alpha}_1)$, i.e.
\begin{eqnarray}
\label{DGP0}
\bm{y}_t \sim p(\bm{y}_t|\bm{\alpha}_t),\hspace{0.7cm}   \bm{\alpha}_{t+1} \sim p(\bm{\alpha}_{t+1}|\bm{\alpha}_{t}),\hspace{0.7cm}   \bm{\alpha}_1 \sim p(\bm{\alpha}_1).
\end{eqnarray} 
In a slight abuse of notation, $p(\cdot|\cdot)$ and $p(\cdot)$ denote \emph{generic} conditional and marginal densities; i.e.\ any two $p$'s need not denote the same probability density function (e.g.\ \citealp[p.\ 6]{durbin2000time}). For a given model, the functional form of all $p$'s is considered known. These densities may further depend on a static (hyper)parameter ${\bm{\psi}}$, which for notational simplicity is suppressed. They may also depend on lags of $\bm{y}_{t}$ or, more generally, any $\mathcal{F}_{t-1}$-measurable variables, where $\mathcal{F}_{t-1}$ denotes the information set at time $t-1$. This potential dependence on $\mathcal{F}_{t-1}$ is likewise suppressed for the sake of readability. Both the observation and state-transition densities may involve non-Gaussianity, nonlinearity and degeneracy. 

Observations $\bm{y}_t$ may take either continuous or discrete values in $\mathbb{R}^l$; in the case of discrete observations, $p(\bm{y}_t|\bm{\alpha}_{t})$ is interpreted as a probability rather than a density. Latent states are assumed to take continuous values in $\mathbb{R}^m$; hence, the state space can be viewed as `infinite dimensional' even as $m$ remains finite. This is in contrast with Markov-switching models  (also known as hidden Markov models; see e.g.\ \citealp[p.\ 109]{kunsch2001complex} and \citealp[p.\ 2026]{fuh2006efficient}), in which the state takes a finite number of (discrete) values.

Myriad examples of model~\eqref{DGP0} can be found in engineering, biology, geological physics, economics and mathematical finance (for a comprehensive overview, see \citealp{kunsch2001complex}, or \citealp{doucet2001sequential}). Examples in financial econometrics with continuous state spaces include models for count data (\citealp{singh1992state},  \citealp{fruhwirth2006auxiliary}), intensity  \citep{bauwens2006stochastic}, duration  \citep{bauwens2004stochastic}, volatility (\citealp{harvey1994multivariate}, \citealp{ghysels1996}, \citealp{jacquier2002bayesian}, \citealp{taylor2008modelling}) and dependence structure \citep{hafner2012dynamic}.

Model~\eqref{DGP0} presents researchers and practitioners with  three important problems: (a) filtering, (b) smoothing and (c) parameter estimation.
The filtering problem
concerns the real-time estimation of the current state~$\bm{\alpha}_t$ conditional on the real-time data~$\bm{y}_{1},\ldots,\bm{y}_t$, where the static parameter~${\bm{\psi}}$ is considered known. The smoothing problem concerns the ex-post estimation of all latent states~$\bm{\alpha}_1,\ldots,\bm{\alpha}_n$ conditional on the full sample~$\bm{y}_{1},\ldots,\bm{y}_n$, still assuming that ${\bm{\psi}}$ is known.  
The parameter-estimation problem entails determining the parameter~${\bm{\psi}}$, where both this parameter and the latent states are assumed to be unknown. 

The filtering and smoothing problems can be solved in closed form when model~\eqref{DGP0} is linear and Gaussian. \citeauthor{kalman1960new}'s (\citeyear{kalman1960new}) filter then computes the real-time expectation of the state (i.e.\ the mean) and the most likely state (i.e.\ the mode), which are identical for these models (see Table~\ref{table0}).  The Rauch, Tung and Striebel (RTS, \citeyear{rauch1965maximum}) smoother, colloquially known as the `Kalman smoother', computes ex-post state estimates by complementing the (forward) Kalman filter with a subsequent backward recursion. Parameter estimation is typically performed by numerically maximising the log-likelihood function, which is known in closed form through the standard prediction-error decomposition (e.g\ \citealp[p.\ 126]{harvey1990forecasting}). 

For the majority of state-space models, however, no exact methods are available for filtering, smoothing or likelihood computation.
Here I present an approximate filter and smoother for the general state-space model~\eqref{DGP0}, followed by an approximate parameter-estimation method. This paper thus addresses all three problems mentioned above. 

\subsection{Primary contribution: Filtering and smoothing using Bellman's equation}

This article develops an approximate filter and smoother that are generally applicable and computationally efficient even in higher dimensions. My point of departure is the view that optimisation may be computationally more attractive than integration---especially in higher dimensions. For this reason, I consider a filter and smoother based not on the mean but on the mode,  which is also known as the \emph{maximum a posteriori} (MAP) estimate (e.g.\ \citealp{koyama2010efficient},   \citealp{liu2013variational}) or the posterior mode (e.g \citealp{fahrmeir1992posterior},  \citealp{durbin1997monte}, \citealp{jungbacker2007monte}). In line with the literature, this approach relies on the assumption that the mode exists and is unique. This assumption is not overly restrictive in practice, although it is possible to formulate models for which it does not hold.\footnote{E.g.\ when the observation equation reads $y_t=\alpha_t^2+\varepsilon_t$ with $\varepsilon_t{\sim}\mathrm{N}(0,\sigma_\varepsilon^2)$.}
 
\begin{table}[t]
\caption{\label{table0} Categorisation of filtering methods.}
\hspace{-1cm}
\renewcommand{\arraystretch}{1}
\center
\begin{threeparttable}
\begin{footnotesize}
\begin{tabular}
{l@{\hspace{3mm}}l@{\hspace{3mm}}l@{\hspace{-16mm}}ll}
  \toprule
 & \bf{Discrete states} &  {\bf{Continuously varying states}}
 \\
 & &  Linear $\&$ Gaussian & Nonlinear and/or non-Gaussian
  \\
&  Exact filters  &  Exact filters & Approximate filters
 \\
   \cmidrule(r{5pt}){1-1}  \cmidrule(r){2-2} \cmidrule(r){3-3}\cmidrule{4-4}
\bf{Mean} & \cite{baum1966statistical} 
 &\cite{kalman1960new}
 & Iterated extended KF (e.g.\ \citealp{anderson2012optimal})
\\
&
\cite{hamilton1989new}
& & Unscented KF \citep{julier1997new}
\\
& & & 
\citeauthor{masreliez1975approximate} (\citeyear{masreliez1975approximate}) filter
\\
&  && Numerical integration filter (\citealp{kitagawa1987non})
\\
&&& Discretisation filter \citep{farmer2021discretization}
 \\
   \cmidrule(r{5pt}){1-1}  \cmidrule(r){2-2} \cmidrule(r){3-3}\cmidrule{4-4}
 \bf{Mode} & \cite{viterbi1967error}
   & \cite{kalman1960new} & Bellman filter (BF, this article)
\\
 & & & Special cases of BF: \citeauthor{fahrmeir1992posterior}'s (\citeyear{fahrmeir1992posterior}) mode estimator
 \\
 & & & and \citeauthor{koyama2010approximate}'s (\citeyear{koyama2010approximate}) Laplace  Gaussian filter \\
 \bottomrule
\end{tabular}
\begin{tablenotes}
\item Note: The table should be considered indicative rather than exhaustive, and, for brevity, excludes simulation-based approaches. KF = Kalman filter. BF = Bellman filter. 
\end{tablenotes}
\end{footnotesize}
\end{threeparttable}
\end{table}

Computing the mode in real time using plain-vanilla optimisation methods is, however, computationally cumbersome. A naive approach would be to re-estimate, at each time step $t$, all previous states of dimension $m$, requiring us to continually solve $m \times t$ dimensional optimisation problems. Computing times per time step then scale as $O(m^3 t^3)$, implying a cumulative computing effort, up to time $t$, of $O(m^3 t^4)$. This escalating complexity over time may explain why the mode estimator has to date received scant attention as a potential filtering method.
 
My proposed solution to this drawback is to apply \citeauthor{bellman1957dynamic}'s (\citeyear{bellman1957dynamic}) dynamic-programming principle, which yields a forward recursion in function space. The solution to this recursion at any time step is referred to as the \emph{value function}, which maps the state space $\mathbb{R}^m$ to values in $\mathbb{R}$ and summarises the researcher's knowledge of the state at time $t$. First, the argmax of the value function represents the most likely state at time $t$ conditional on $\bm{y}_1,\ldots,\bm{y}_t$; hence, it acts as our  filtered state estimate. Second, the negative Hessian matrix evaluated at the peak is indicative of the precision of this state estimate: a `sharper' peak corresponds to a more precise state estimate. Recursively solving Bellman's equation thus yields a feasible filtering method, producing at each time step both a filtered state and an associated measure of uncertainty.

Importantly for the present purpose, computing the argmax of the value function entails maximisation over a \emph{single} state of dimension $m$ for each time step. The required computing cost per time step remains constant at $O(m^3)$. The resulting cumulative computational complexity over $t$ time steps then amounts to $O(m^3 t)$, which is identical to that of the (information form of the) Kalman filter. On the one hand, the computational complexity of $O(t)$ means the Bellman filter can be classed as a filter in the strict sense of the term. On the other, the complexity of $O(m^3)$ offers  full scalability to higher dimensional state spaces; e.g.\ up to $150$ dimensions in the application in section~\ref{section8}. 

The price we pay for this reduced computational complexity is that Bellman's recursion generally lacks an analytic solution; hence, we must resort to approximation, which can be viewed as a form of approximate dynamic programming (e.g.\ \citealp{bertsekas2012approximate}). One possibility is to discretise the (continuous) state space $\mathbb{R}^m$, forcing the state to take a finite number of (discrete) values.  Bellman's equation can then be solved exactly, yielding \citeauthor{viterbi1967error}'s (\citeyear{viterbi1967error}) algorithm (see Table~\ref{table0}), which has proven highly successful in engineering.  However, this approach quickly becomes infeasible  due to the curse of dimensionality (\citealp[p.\ 125]{kunsch2001complex}, \citealp[p.\ 29]{liu2008monte}), as it requires the computation and storage of $N^m$ values for each time step, where $N$ is the number of gridpoints in each of $m$ spatial directions (e.g.\ $N=100$ and $m=5$ is infeasible).

Instead, I take inspiration from another exact solution to Bellman's forward recursion. As it turns out, Bellman's recursion allows an exact solution if the entire model~\eqref{DGP0} is linear and Gaussian, yielding \citeauthor{kalman1960new}'s (\citeyear{kalman1960new}) filter.  The solution to Bellman's equation is then
a \emph{function}, rather than a finite-dimensional object as in Viterbi's case. This value function 
has a particularly simple form: it is multivariate quadratic at every time step, with a unique argmax that corresponds to  \citeauthor{kalman1960new}'s filtered state.  Moreover, its negative Hessian matrix equals the inverse of the usual Kalman-filtered covariance matrix. Hence, the Kalman filter represents an \emph{exact function-space solution} to Bellman's equation. This was long recognised in the engineering literature
 (e.g.\ \citealp[ch.\ 12]{whittle1996optimal}; \citealp{whittle2004}) before finding its way into the econometrics literature (\citealp[ch.\ 8]{hansen2013recursive}). Perhaps less widely known is the fact that the RTS (\citeyear{rauch1965maximum}) smoother similarly corresponds to an exact---also multivariate quadratic---solution to a combination of Bellman's forward and backward recursions (see section~\ref{sec:backward recursion}).
 
The basic premise of this article is that Bellman's forward and backward recursions remain valid in the context of the general state-space model~\eqref{DGP0}. Motivated by the exact solutions leading to the Kalman filter and RTS smoother, I deviate from the literature in exploring \emph{function-space approximations} of value functions rather than discretising. Computing at every time step some parametric approximation of the value function yields a new class of (Bellman) filters and smoothers.  Within the class of function-space approximations, I employ arguably the simplest non-trivial option: a multivariate quadratic function. This quadratic approximation is exact for linear Gaussian models and---given that value functions in filtering applications are typically smooth and possess global maxima---broadly applicable. The approximation can also be viewed as a second-order Taylor expansion of a generic smooth value function. This simple approximation approach yields immediate and novel extensions of the Kalman filter and smoother. The main contribution of this article is the insight that using function-space rather than discrete approximations allows us to avoid the curse of dimensionality, leading to a new class of filters and smoothers that are computationally frugal  and turn out to be remarkably accurate.  

\subsection{Secondary contribution: Parameter estimation using likelihood approximation}

To address the parameter-estimation problem, I deviate from the literature that relies on simulation-based approaches (e.g.\ \citealp{malik2011particle}, \citealp{koopman2015numerically},
\citealp{koopman2016predicting}) by presenting a deterministic and computationally efficient---albeit approximate---method  based on the output of the Bellman filter. While no formal guarantees are offered, an extensive simulation study (section~\ref{section7}) demonstrates that the proposed estimator is no less accurate or efficient than (asymptotically exact) simulation-based methods, while requiring a fraction of the computational cost.  
Establishing the asymptotic properties of the estimator remains an open question.

Specifically, I propose to maximise an approximate version of the log-likelihood function that is immediately computable from the output of the Bellman filter.  First, the (exact) log-likelihood function is  decomposed into (a)~the `fit' of the Bellman-filtered states in view of the data, minus (b)~the realised Kullback-Leibler (KL, see \citealp{kullback1951information}) divergence between filtered and predicted state densities. While the former is known in closed form, the latter typically is not---except in the case of linear Gaussian state-space models, in which case it is multivariate quadratic. Second, I approximate this KL divergence term using a multivariate quadratic term computed from the output of the Bellman filter. The resulting pseudo log-likelihood function remains exact in the case of linear Gaussian models; more generally, it can be viewed as a second-order approximation of the log-likelihood function. It can be optimised using standard gradient-based numerical optimisers, making approximate parameter estimation for the general state-space model~\eqref{DGP0} as simple and fast as maximum-likelihood estimation of the Kalman filter.

\subsection{Limitations of existing methods}

Existing approaches to filtering, smoothing and parameter estimation can be classified as either approximation- or simulation-based, each with their own disadvantages. First, approximate filtering methods (see Table~\ref{table0}) tend to be specialised in their applications.
 The extended and unscented Kalman filters account for nonlinearity, but assume additive noise and maintain the normality assumption. Conversely, \cite{west1981robust} relaxes the normality assumption, while maintaining the linearity assumption.  \citeauthor{masreliez1975approximate}'s (\citeyear{masreliez1975approximate}) filter is robust in the case of heavy-tailed observation noise but, due to the need to approximate integrals, computationally inefficient in higher dimensions. Similarly, numerical integration (\citealp{kitagawa1987non}) and other discretisation methods (\citealp{farmer2021discretization}) are flexible in theory, but  restricted in practice by the curse of dimensionality. \citeauthor{fahrmeir1992posterior}'s
(\citeyear{fahrmeir1992posterior}) method applies to observations drawn from an exponential distribution. \cite{durbin2000time} and  \cite{koyama2010approximate} mostly rely on a linear Gaussian state equation. \cite{muller2010efficient} assume that deviations of the latent state from its equilibrium value are small. In the literature, no approximate filters seem to be available at the level of generality of model~\eqref{DGP0}. Moreover, the aforementioned approaches tend to neglect the smoothing and parameter-estimation problems.  

Second, simulation-based methods such as particle filters are widely applicable and easy to implement (for a textbook treatment, see e.g.\ \citealp{chopin2020introduction}). However, the curse of dimensionality means that particle filters may struggle with high-dimensional state spaces (\citealp{surace2019avoid}). For the same reason, the importance-sampling method by \cite{koopman2015numerically,koopman2016predicting,koopman2017intraday}
has not been applied in situations in which the state-space dimension exceeds two. Particle smoothing (as opposed to filtering) tends to be even more computationally expensive, as the computational cost scales with the number of particles squared (\citealp{kantas2015particle}). Particle filters have also been applied to the parameter-estimation problem, but this remains  challenging (\citealp{liu2001combined},  \citealp{kantas2015particle}); e.g.\ \citeauthor{malik2011particle}'s (\citeyear{malik2011particle}) method applies only when the state space is one dimensional.  

\section{ Main idea: Filtering using Bellman's principle}
\label{section2}
\label{sec:forward recursion}

The state-space model under consideration is given in equation~\eqref{DGP0}. A realised path is denoted by $(\bm{y}_{1},\ldots,\bm{y}_t)(\omega)$ for every event $\omega \in \Omega$, where $\Omega$ denotes the event space of the underlying complete probability space of interest, denoted $(\Omega, \mathcal{F}, \mathbb{P})$.  The logarithm of joint and conditional densities are written using generic notation as $\ell(\cdot,\cdot):=\log p(\cdot,\cdot)$ and $\ell(\cdot|\cdot):=\log p(\cdot|\cdot)$, respectively, for potentially different $p$'s. This section considers the filtering problem; any dependence on $\bm{\psi}$ is suppressed.

The joint log-likelihood function of the states and the data is written as $L_{1:t}(\bm{a}_1,\ldots,\bm{a}_t):\Omega \times \mathbb{R}^m \times \ldots \times \mathbb{R}^m \to \mathbb{R}$. Here, the data~$\bm{y}_{1},\ldots,\bm{y}_t$ are considered fixed and known, as indicated by the subscript, while the states~$\bm{a}_{1},\ldots,\bm{a}_t$ in Roman font are considered  variables to be evaluated along any path. The true states $\bm{\alpha}_1,\ldots,\bm{\alpha}_t$ in Greek font  remain unknown. For the state-space model~\eqref{DGP0}, the joint log likelihood of the data and the states follows from the `probability chain rule' (\citealp[p. 156]{godsill2004monte}):
\begin{eqnarray}
L_{1:t}(\bm{a}_1,\ldots,\bm{a}_t)
=   \overset{t}{\underset{i=1}{\sum}} \ell(\bm{y}_{i}| \bm{a}_i)\,+\,  \overset{t}{\underset{i=2}{\sum}}
\ell(\bm{a}_i|\bm{a}_{i-1}) \,+\,  \ell(\bm{a}_1),\qquad t\leq n.\label{loglik}
\end{eqnarray}
This joint log likelihood is, \textit{a priori}, a random function of the observations $\bm{y}_{1},\ldots,\bm{y}_{t}$, even though the data are considered known and fixed \textit{ex post}.  For clarity, I formalise the assumption that for some sufficiently large $t$, there exists a unique sequence of states, denoted $\bm{a}_{1|t},\ldots,\bm{a}_{t|t}$, that maximise equation~\eqref{loglik}.

\begin{assump}[Existence of the mode] There exists some $t_0\geq 1$, such that for all $t\geq t_0$, the mode $({\bm{a}}_{1|
t},{\bm{a}}_{2|
t},\ldots,{\bm{a}}_{t|
t})$ exists and is unique, where
\begin{eqnarray}
\label{mode estimator}
({\bm{a}}_{1|
t},{\bm{a}}_{2|
t},\ldots,{\bm{a}}_{t|
t}):=
 \underset{ ({\bm{a}}_{1},{\bm{a}}_{2},\ldots,{\bm{a}}_{t}) \in \mathbb{R}^{m\times t} }{\;\arg\;\max\;} L_{1:t}(\bm{a}_1,\ldots,\bm{a}_t). 
\end{eqnarray}
\end{assump}
This assumption is labelled ``E'' for existence, because it is required to underpin the main idea; later, Assumption~1-3 (in section~\ref{section5}) are used to derive the theoretical properties of the filter.

As equation~\eqref{mode estimator} illustrates, elements of the mode at time $t$ are denoted by ${\bm{a}}_{i|t}$ for $i\leq t$, where $i$ denotes the state that is estimated, $t$ the information set used. The entire solution is a collection of~$t$ vectors, each of length $m$. Iterative solution methods for solving~\eqref{mode estimator} were proposed in~\citet{durbin2000time} and \citet{so2003posterior}. When the mode~\eqref{mode estimator} is computed for \emph{each} time step $t\geq t_0$, we can extract a sequence of real-time state estimates $\{\bm{a}_{t|t}\}_{t\geq t_0}$, where each estimate $\bm{a}_{t|t}$ is extracted from a \emph{different} mode~\eqref{mode estimator}.
 
As time progresses, however, the computation of filtered states $\{\bm{a}_{t|t}\}_{t}$ becomes ever more complicated---note that optimisation problem~\eqref{mode estimator} involves $m \times t$ optimisation variables at each time $t$. Indeed, solving problem~\eqref{mode estimator} may become practically infeasible for large $t$.
This raises the question whether it is possible to proceed in real time without solving an optimisation problem of  ever-increasing complexity. As shown next, this can be achieved using Bellman's dynamic-programming principle. To this end, I define the \emph{value function} by maximising the joint log-likelihood function~\eqref{loglik}
with respect to all states apart from the most recent state~$\bm{a}_t\in \mathbb{R}^m$; such functions are also known as `profile' log-likelihood functions~\citep{murphy2000profile} in statistics and `stress' functions in engineering \citep[p.\ 769]{whittle1981risk}.

\begin{definition}[Value function] \label{def value function}  Let Assumption~E hold.  For $t\geq t_0$, the value function $V_{t}: \Omega \times \mathbb{R}^{m} \to \mathbb{R}$ is 
\begin{align}
\label{value function}
V_{t}(\bm{a}_{t}) 
:= \max_{({\bm{a}}_{1},{\bm{a}}_{2},\ldots,{\bm{a}}_{t-1})\in \mathbb{R}^{m\times (t-1)}} L_{1:t}(\bm{a}_1,\ldots,\bm{a}_t),\qquad \bm{a}_t\in \mathbb{R}^m.
\end{align}
\end{definition}

The value function $V_{t}(\cdot)$ encodes our knowledge of the state at time $t$, as indicated by the subscript, and  depends on past and current data $\bm{y}_{1},\ldots,\bm{y}_t$, which are considered fixed, as well as on its argument $\bm{a}_{t}$, which is a continuous variable in $\mathbb{R}^m$. 
Naturally, ${\bm{a}}_{t|t}=\arg \max_{\bm{a}_t}V_t(\bm{a}_t)$, such that the last element of the mode~\eqref{mode estimator} can be recovered from the value function. Usefully, the value function~\eqref{value function} satisfies a forward recursive equation, known as Bellman's equation, which can be used for the purpose of filtering.

\begin{proposition}[Filtering using Bellman's equation] \label{prop1} Let Assumption~E hold.  The value function~\eqref{value function} satisfies Bellman's forward recursion:
\begin{eqnarray}
\label{Bellman}
V_{t}(\bm{a}_{t}) \;=\; \ell(\bm{y}_{t}|\bm{a}_{t})\;+ \underset{\bm{a}_{t-1} \in \mathbb{R}^m}{\max}\; \Big\{ \ell(\bm{a}_{t}|\bm{a}_{t-1}) \; +\; V_{t-1}(\bm{a}_{t-1}) \Big\},\qquad
 \bm{a}_t\in \mathbb{R}^m,
\end{eqnarray}
for all $t_0<t\leq n$. Further, 
\begin{eqnarray}
\label{filtered estimate}
{\bm{a}}_{t|t} \;\;:=\;\; \underset{ \bm{a}_{t} \in \mathbb{R}^m}{\arg\max} \;\; V_{t}(\bm{a}_{t}),\quad t_0 \leq t \leq n.
\end{eqnarray}
\end{proposition}
Bellman's equation~\eqref{Bellman} is a forward recursion that relates the value function $V_{t}(\bm{a}_{t})$ to the (previous) value function $V_{t-1}(\bm{a}_{t-1})$ by adding one term reflecting the state transition, $\ell(\bm{a}_{t}|\bm{a}_{t-1})$; one term reflecting the observation density, $\ell(\bm{y}_{t}|\bm{a}_{t})$; and a subsequent maximisation over a single state variable, $\bm{a}_{t-1}\in\mathbb{R}^m$. The value function $V_{t}(\bm{a}_{t})$ still depends on the data $\bm{y}_{1},\ldots,\bm{y}_{t-1}$, but only indirectly, i.e.\ through the previous value function $V_{t-1}(\bm{a}_{t-1})$. Apart from assuming the existence of the mode, no (additional) assumptions are imposed on the log densities $\ell(\bm{y}_{t}|\bm{a}_{t})$ and $\ell(\bm{a}_{t}|\bm{a}_{t-1})$; the proof in Supplement~\ref{A:proof} uses only standard dynamic-programming arguments. As such, Bellman's equation~\eqref{Bellman} is of quite general applicability. As the researcher receives the data $\bm{y}_1$ through $\bm{y}_t$, she can iteratively compute a sequence of value functions~\eqref{Bellman}, which imply a sequence of filtered state estimates via the respective maximisers~\eqref{filtered estimate}. 

\begin{remark} \label{corol0} For Markov-switching models, in which the latent state takes a finite number of (discrete) values, Bellman's equation~\eqref{Bellman} can be solved exactly for all time steps, yielding \citeauthor{viterbi1967error}'s (\citeyear{viterbi1967error}) algorithm. Exact solubility of~\eqref{Bellman} tends to be lost when the states take continuous values.  \end{remark}

When latent states take values in a continuum, as in the present article, the solution to Bellman's equation~\eqref{Bellman} is a \emph{function} rather than a (finite-dimensional) vector as in Viterbi's algorithm. While the value function cannot generally be found exactly, there is an exception to this rule,  as highlighted next. 

\begin{corollary}[Kalman filter as a special case] \label{corol1} Take a linear 
Gaussian state-space model with observation equation $\bm{y}_t =  \bm{d} +\bm{Z}\,\bm{\alpha}_t + \bm{\varepsilon}_t$, where $\bm{\varepsilon}_t\sim\text{i.i.d. } \mathrm{N}(\bm{0},\bm{H})$, and state-transition equation $\bm{\alpha}_t = \bm{c}+\bm{T}\,\bm{\alpha}_{t-1} + \bm{\eta}_{t}$, where $\bm{\eta}_t \sim \text{i.i.d.}\, \mathrm{N}(\bm{0},\bm{Q})$ with a positive semidefinite covariance matrix $\bm{Q}$, such that \citeauthor{kalman1960new}'s (\citeyear{kalman1960new}) filter applies. Assume the Kalman-filtered covariance matrices, denoted $\{\bm{P}_{t|t}\}$, are positive definite. Then (a) the value function is exactly multivariate quadratic at every time step,
(b) the Bellman-filtered states are identical to the Kalman-filtered states, and (c) the negative Hessian matrix of the value function equals $\bm{P}_{t|t}^{-1}$ at every time step.
\end{corollary}

The proof of Corolary~\ref{corol1} is contained in section~\ref{section4}, where I treat the case of a linear Gaussian state equation but a general observation density. As is well known in engineering (e.g.\ \citealp[ch.\ 12]{whittle1996optimal}), the exact solubility of  Bellman's equation in the case of linear Gaussian models is attributable to the quadratic nature of all terms appearing on its right-hand side. The left-hand side turns out to be quadratic as well, preserving exact solubility over time.  

A key contribution  of this article is the insight that Bellman's equation continues to hold for state-space models that are not necessarily linear and Gaussian, even if analytic solubility is lost. In this case, I deviate from the literature in considering function-space approximations in solving Bellman's recursion~\eqref{Bellman}.  I consider a particularly simple approximation---the multivariate quadratic  function---which happens to be exact for linear Gaussian state-space models. A different class of Bellman filters, not explored here, would be obtained by using non-parametric approximations.

\section{Bellman filter for general state-space models}
\label{section3}

\subsection{Non-degenerate case}
\label{section31}

This section develops the Bellman filter for the general state-space model~\eqref{DGP0} by approximating the value function, at every time step, by a multivariate quadratic function. I assume here that the observation and state-transition densities are non-degenerate; an extension to the degenerate case is set out below.  

The Bellman-filtered state~\eqref{filtered estimate} requires a maximisation with respect to the current state, $\bm{a}_t$, while Bellman's equation~\eqref{Bellman} additionally contains a maximisation with respect to the lagged state, $\bm{a}_{t-1}$. Merging both steps generates a joint optimisation problem in both state variables:
\begin{equation}
\label{bivariate optimisiation}
\left[\begin{array}{l} \bm{a}_{t|t} \\ \bm{a}_{t-1|t} \end{array} \right] \;\; := \; \underset{\quad\scriptsize{
 \left[\begin{array}{l}\bm{a}_t\\ \bm{a}_{t-1} \end{array}\right]\,\in\, \mathbb{R}^{2m}}}{\arg \max} \Big\{ \,\ell(\bm{y}_{t}|\bm{a}_{t})\;+\; \ell(\bm{a}_{t}|\bm{a}_{t-1}) \; +\; V_{t-1}(\bm{a}_{t-1}) \Big\}.
\end{equation}
The left-hand side features the filtered state, $\bm{a}_{t|t}$, as well as the revised estimate of the previous state, denoted $\bm{a}_{t-1|t}$. The computation of the latter, while not our main focus, is inherent to Bellman's equation and cannot be avoided. The right-hand side features two log densities denoted $\ell(\cdot|\cdot):=\log p(\cdot|\cdot)$, which are given in closed form by the state-space model~\eqref{DGP0}.

While the lagged value function $V_{t-1}(\cdot)$ on the right-hand side of optimisation~\eqref{bivariate optimisiation} 
is typically unavailable in closed form, the shape around its peak turns out to be most relevant in the determination of the filtered state $\bm{a}_{t|t}$. I thus propose to approximate $V_{t-1}(\bm{a}_{t-1})$ by a multivariate quadratic function that is parametrised by its argmax, denoted $\bm{a}_{t-1|t-1}\in \mathbb{R}^m$, and the negative Hessian matrix, denoted $\bm{I}_{t-1|t-1}\in \mathbb{R}^{m\times m}$, which is assumed positive definite and can be interpreted as an information (or `precision') matrix. The approximation thus reads
\begin{eqnarray}
\label{normalapproximation00}
V_{t-1}(\bm{a}_{t-1})\; = \;  - \frac{1}{2} (\bm{a}_{t-1} - \bm{a}_{t-1|t-1} )' \,\bm{I}_{t-1|t-1}\,(\bm{a}_{t-1} - \bm{a}_{t-1|t-1} )\,+\,\text{constants},\qquad \bm{a}_{t-1}\in \mathbb{R}^m,
\end{eqnarray}
which for simplicity is written with equality. Constants can be ignored in the context of optimisation~\eqref{bivariate optimisiation}. Substituting the quadratic approximation~\eqref{normalapproximation00} into maximisation~\eqref{bivariate optimisiation} yields a viable function-space algorithm. For linear Gaussian state-space models, approximation~\eqref{normalapproximation00} is exact and the bivariate optimisation~\eqref{bivariate optimisiation} can be performed analytically, leading to (the information form of) the Kalman filter.

While optimisation~\eqref{bivariate optimisiation} does not generally allow closed-form solutions, it is typically straightforward to write out analytically the steps of e.g.\ Newton's method (\citealp{nocedal2006numerical}):
\begin{equation}
\scriptstyle
\left[
\begin{array}{c}
\bm{a}_{t}\\
\bm{a}_{t-1}
\end{array}
\right] \leftarrow \left[
\begin{array}{c}
\bm{a}_{t}\\
\bm{a}_{t-1}
\end{array}
\right]+\left[\begin{array}{cc}\bm{J}^{11}_{t} -\frac{\dd^2\ell(\bm{y}_{t}|\bm{a}_{t})}{\dd\bm{a}_t\dd\bm{a}_t' } & \bm{J}^{12}_{t}\\ \bm{J}^{21}_{t} & \bm{I}_{t-1|t-1} + \bm{J}^{22}_{t} \end{array} \right]^{-1}\left[\begin{array}{l@{\hspace{0.5mm}}l} \bm{J}^{1}_t &+\;\frac{\dd\ell(\bm{y}_{t}|\bm{a}_{t})}{\dd\bm{a}_t}   \\  \bm{J}^{2}_{t} &-\;\bm{I}_{t|t-1} (\bm{a}_{t-1}-\bm{a}_{t-1|t-1})\end{array} \right],
\label{Newton general}
\end{equation}
where, for notational simplicity, I use the assignment symbol; this allows the iterates (which appear on both the left- and right-hand sides) to be denoted by $\bm{a}_t$ and $\bm{a}_{t-1}$. In Newton's step~\eqref{Newton general}, derivatives related to the state-transition density are
\begin{equation}
\label{J}
\left[\begin{array}{cc}\bm{J}^{1}_{t} \\ \bm{J}^{2}_{t}  \end{array} \right] := \left[\begin{array}{cc} \frac{\dd\ell(\bm{a}_{t}|\bm{a}_{t-1})}{\dd\bm{a}_t } \\  \frac{\dd\ell(\bm{a}_{t}|\bm{a}_{t-1})}{\dd\bm{a}_{t-1}} \end{array} \right],\qquad 
\left[\begin{array}{cc}\bm{J}^{11}_{t} & \bm{J}^{12}_{t} \\ \bm{J}^{21}_{t} & \bm{J}^{22}_{t} \end{array} \right] := -\left[\begin{array}{cc} \frac{\dd^2\ell(\bm{a}_{t}|\bm{a}_{t-1})}{\dd\bm{a}_t \dd\bm{a}_{t}'} &  \frac{\dd^2\ell(\bm{a}_{t}|\bm{a}_{t-1})}{\dd\bm{a}_t \dd\bm{a}_{t-1}'} \\  \frac{\dd^2\ell(\bm{a}_{t}|\bm{a}_{t-1})}{\dd\bm{a}_{t-1} \dd\bm{a}_{t}'} & \frac{\dd^2\ell(\bm{a}_{t}|\bm{a}_{t-1})}{\dd\bm{a}_{t-1} \dd\bm{a}_{t-1}'} \end{array} \right].
\end{equation}
Fisher's optimisation method is obtained by replacing $\dd^2 \ell(\bm{y}_t|\bm{a}_t)/(\dd\bm{a}_t\dd\bm{a}_t')$ in equation~\eqref{Newton general} with its expectation conditional on $\bm{a}_t$.  When the observation and state-transition densities in model~\eqref{DGP0} are given, it is straightforward (if tedious) to compute all required derivatives. As $\bm{I}_{t-1|t-1}$ is assumed to be invertible, analytic block-matrix inversion can be used for each Newton step~\eqref{Newton general}, reducing the size of matrices to be numerically inverted from $2m\times 2m$ to $m\times m$ (see Supplement~\ref{app:block matrix} for details). The resulting algorithm is shown under step 4 in Table~\ref{table2}. Alternatively, black-box numerical optimisers may be used to solve~\eqref{bivariate optimisiation}, obviating the need for manual computations; this will  save researcher time but potentially increase the required computer time.
 The optimisation can be started using $(\bm{a}_t,\bm{a}_{t-1}) \leftarrow (\bm{a}_{t|t-1},\bm{a}_{t-1|t-1})$, where $\bm{a}_{t|t-1}:=\arg \max_{\bm{a}} \ell(\bm{a}|\bm{a}_{t-1|t-1})$, as indicated under steps 2 and 3 in Table~\ref{table2}. This prediction $\bm{a}_{t|t-1}$ can often be computed in closed form. 
 
To facilitate the proposed recursive method, the left-hand side of Bellman's equation~\eqref{Bellman} must also be approximated by a multivariate quadratic function. To this end, I compute the negative Hessian matrix (with respect to $\bm{a}_t$) of the value function, i.e.\ $V_t(\bm{a}_t)=\ell(\bm{y}_t|\bm{a}_t)+\max_{\bm{a}_{t-1}}\{ \ell(\bm{a}_t|\bm{a}_{t-1})+V_{t-1}(\bm{a}_{t-1})\}$. The negative Hessian may be then be evaluated at the peak. Employing the second-order envelope theorem (Supplement~\ref{app:envelope}) yields
\begin{align}
\bm{I}_{t|t}\,&:=\, \bm{J}^{11}_{t} -\bm{J}^{12}_{t}(\bm{I}_{t-1|t-1}+\bm{J}^{22}_{t}
)^{-1}\bm{J}^{21}_{t} -\frac{\dd^2 \ell(\bm{y}_t|\bm{a}_t)}{\dd\bm{a}_t\dd\bm{a}_t'} \Big|_{\bm{a}_t=\bm{a}_{t|t},\bm{a}_{t-1}=\bm{a}_{t-1|t}}
\label{general information update}
\end{align}
as shown in Table~\ref{table2} under step 6.  Fisher's version is obtained by taking a conditional expectation of the last term. For linear Gaussian state-space models, Newton and Fisher versions of update~\eqref{general information update} are identical and equal to the information update of the Kalman filter (Supplement~\ref{app: Kalman info}). Update~\eqref{general information update} can also be viewed as a `realised' version of the recursion for the inverse of Cram{\' e}r-Rao lower bounds \citep[eq.\ 21]{tichavsky1998posterior}---the difference being that equation~\eqref{general information update} has no expectations. The predicted information $\bm{I}_{t|t-1}$, given in step 2 of Table~\ref{table2}, is similar in form and used for static-parameter estimation purposes in section~\ref{section6}. 

\begin{table}[t!]
\caption{\label{table2} Bellman filter for model~\eqref{DGP0}.}
\center
\renewcommand{\arraystretch}{1}
\begin{footnotesize}
\begin{threeparttable}
\begin{tabularx}{\textwidth}{l@{\hspace{2mm}}l@{\hspace{2mm}}X}
  \toprule
  \bf{Step} & \bf{Method} & \bf{Computation}
 \\
\midrule  
1.\ Initialise & & Set $\bm{a}_{0|0}$ equal to the unconditional mean of the latent state (or treat it as a static parameter to be estimated) and set $\bm{I}_{0|0}$ equal to some sufficiently large multiple of the identity matrix. Set $t=1$.
\\
2.\ Predict &&$\bm{a}_{t|t-1}=\arg \max_{\bm{a}_t \in \mathbb{R}^m} \ell(\bm{a}_t|\bm{a}_{t-1|t-1})$
\\
&& $\bm{I}_{t|t-1} =\bm{J}^{11}_{t} -\bm{J}^{12}_{t}(\bm{I}_{t-1|t-1}+\bm{J}^{22}_{t}
)^{-1}\bm{J}^{21}_{t} \big|_{\bm{a}_t=\bm{a}_{t|t-1},\bm{a}_{t-1}=\bm{a}_{t-1|t-1}}$
\\
3.\ Start & & Set $\bm{a}_{t}\leftarrow \bm{a}_{t|t-1}$ and $\bm{a}_{t-1}\leftarrow \bm{a}_{t-1|t-1}$.
\\
4.\ Optimise &Newton & $\displaystyle \bm{S}_t \leftarrow \bm{J}^{11}_{t} -\bm{J}^{12}_{t}(\bm{I}_{t-1|t-1}+\bm{J}^{22}_{t}
)^{-1}\bm{J}^{21}_{t} -\frac{\dd^2 \ell(\bm{y}_t|\bm{a}_t)}{\dd\bm{a}_t\dd\bm{a}_t'} ,\qquad \displaystyle \bm{D}_t \leftarrow  \bm{I}_{t-1|t-1} + \bm{J}_t^{22}, $ 
\\
&  & $\displaystyle \bm{G}^{1}_t \leftarrow   \bm{J}_t^{1}+ \frac{\dd\ell(\bm{y}_{t}|\bm{a}_t)}{\dd\bm{a}_t} ,\qquad \displaystyle \bm{G}^{2}_t \leftarrow  \bm{J}_t^2-\bm{I}_{t-1|t-1} (\bm{a}_{t-1}-\bm{a}_{t-1|t-1}), $ 
\\
&& $\displaystyle \bm{a}_{t} 
\leftarrow 
\bm{a}_{t}+\bm{S}_t^{-1} \bm{G}_t^{1} -\bm{S}_t^{-1} \bm{J}_{t}^{12} \bm{D}_t^{-1}\bm{G}_t^{2},
$  
\\
&& $\displaystyle \bm{a}_{t-1} 
\leftarrow 
\bm{a}_{t-1} -\bm{D}_t^{-1} \bm{J}_t^{21} \bm{S}_{t}^{-1} \bm{G}_t^{1}+ ( \bm{D}_t^{-1}+ \bm{D}_t^{-1} \bm{J}_{t}^{21} \bm{S}_{t}^{-1} \bm{J}_{t}^{12} \bm{D}_{t}^{-1} ) \bm{G}_t^2$.
\\
&Fisher & Like Newton's method, but with  $\bm{S}_t$ adjusted to include 
 $\mathbb{E}[\dd^2 \ell(\bm{y}_t|\bm{a}_t)/(\dd\bm{a}_t\dd\bm{a}_t')|\bm{a}_t]$.
\\
5.\ Stop && Stop if some convergence criterion is satisfied or after a predetermined number of iterations. 
\\
6.\ Update 
& &  $\bm{a}_{t|t}=\bm{a}_t$ and $\bm{a}_{t-1|t}=\bm{a}_{t-1}$.
\\ &Newton & $\displaystyle \bm{I}_{t|t} =\bm{J}^{11}_{t} -\bm{J}^{12}_{t}(\bm{I}_{t-1|t-1}+\bm{J}^{22}_{t}
)^{-1}\bm{J}^{21}_{t} -\frac{\dd^2 \ell(\bm{y}_t|\bm{a}_t)}{\dd\bm{a}_t\dd\bm{a}_t'}\Big|{}_{\bm{a}_t=\bm{a}_{t|t},\bm{a}_{t-1}=\bm{a}_{t-1|t}}$
\\
&Fisher&  $\displaystyle \bm{I}_{t|t} =\bm{J}^{11}_{t} -\bm{J}^{12}_{t}(\bm{I}_{t-1|t-1}+\bm{J}^{22}_{t}
)^{-1}\bm{J}^{21}_{t} -\mathbb{E}\left[\frac{\dd^2 \ell(\bm{y}_t|\bm{a}_t)}{\dd\bm{a}_t\dd\bm{a}_t'}\Big| \bm{a}_t\right]\Big|{}_{\bm{a}_t=\bm{a}_{t|t},\bm{a}_{t-1}=\bm{a}_{t-1|t}}$
\\
6.\ Proceed && Set $t=t+1$ and return to step 2.
\\
\bottomrule
\end{tabularx}
\begin{tablenotes}
\item \footnotesize{\emph{Note:} The log-likelihood functions $\ell(\bm{y}_t|\bm{a}_t)$ and $\ell(\bm{a}_t|\bm{a}_{t-1})$ are known in closed form and can be read off from the data-generating process~\eqref{DGP0}. Various derivatives of $\ell(\bm{a}_t|\bm{a}_{t-1})$ are defined in equation~\eqref{J}. Two (intentionally vanilla) optimisation methods are listed under steps 4 and 6. Users may also implement more sophisticated and/or black-box optimisation methods based on maximisation~\eqref{bivariate optimisiation}.
}
\end{tablenotes}
\end{threeparttable}
\end{footnotesize}
\end{table}

The resulting Bellman filter in Table~\ref{table2} has a computational complexity of $O(m^3 t)$, which is attributable to the need to invert $m\times m$ matrices at every time step. This complexity matches that of (the information form of) the Kalman filter, thus offering scalability to at least moderately high dimensions $m$. I am unaware of other approximate filters offering the same breadth of applicability and computational efficiency.\footnote{In related work, \citet[p.\ 173]{koyama2010approximate} report a computational complexity of $O(m^2 t)$, purportedly as $O(m^2)$ is the `complexity of matrix manipulations'. This result comes with two important caveats. First, it relies on having a linear and Gaussian state equation; otherwise, their prediction step requires the (numerical) evaluation of an integral in $m$ dimensions. Second, it overlooks the fact that the (dense) matrix inversion required by Newton's method typically requires $O(m^3)$ computational effort; not even the best linear solvers achieve $O(m^2)$.}

\subsection{Extension to the degenerate case}
\label{section32}

When some elements of $\bm{a}_{t-1|t-1}$ are known to be pinpoint accurate, the corresponding diagonal values of the precision matrix $\bm{I}_{t-1|t-1}$ in equation~\eqref{normalapproximation00} are unbounded. Such infinite diagonal values make optimisation~\eqref{bivariate optimisiation} easier rather than harder, as some elements of $\bm{a}_{t-1}$ are constrained and need not be numerically optimised; rather, they can be fixed by hand. When the relevant restriction is implemented, the unbounded contributions in the quadratic term~\eqref{normalapproximation00} can be dropped. Similarly, when the state-transition density $\ell(\bm{a}_t|\bm{a}_{t-1})$ is degenerate, some elements of the current state are deterministic functions of the previous state. When these restrictions are implemented, the degenerate part of the transition density can be dropped. Indeed, this procedure will be used for the model in section~\ref{section9}, which involves degenerate state dynamics. Finally, when the observation density $\ell(\bm{y}_t|\bm{a}_t)$ is degenerate, as when some elements of $\bm{a}_t$ are fully revealed by the observation~$\bm{y}_t$, optimisation~\eqref{bivariate optimisiation} requires that some elements of $\bm{a}_t$ take a specific functional form of $\bm{y}_t$. From an optimisation perspective, therefore, degeneracies correspond to equality constraints that can typically be implemented by hand, reducing the dimension of the  numerical optimisation problem to be solved. This capacity to deal with (partially) deterministic state dynamics forms an advantage over e.g.\ particle-filtering methods, which may struggle in such situations.

\section{Bellman filter for models with linear Gaussian state dynamics}
\label{section4}

This section applies the general idea developed in the previous section to models in which the state-transition equation remains linear and Gaussian. The advantage of this special case is that the  `inner' optimisation in Bellman's equation~\eqref{Bellman}, i.e.\ with respect to the lagged state $\bm{a}_{t-1}$, can now be performed in closed form. The `outer' optimisation with respect to the current state $\bm{a}_{t}$ remains numerical. Models with linear Gaussian state equations are written as in \cite{koopman2015numerically,koopman2016predicting}:
\begin{eqnarray}
\label{DGP0.3}
\bm{y}_t \sim p(\bm{y}_t\,|\, \bm{\alpha}_t),\quad  \bm{\alpha}_{t+1} = \bm{c}+\bm{T}\,\bm{\alpha}_{t} + \bm{\eta}_{t+1},\quad \bm{\eta}_t \sim \text{i.i.d.}\, \mathrm{N}(\bm{0},\bm{Q}),\quad \bm{\alpha}_1\sim p(\bm{\alpha_1}),
\end{eqnarray} 
where $t=1,\ldots,n$, and the state-transition equation contains the system vector $\bm{c}\in \mathbb{R}^m$ and system matrix $\bm{T}\in \mathbb{R}^{m \times m}$.  The state innovation $\bm{\eta}_t$ is controlled by a positive semidefinite covariance matrix $\bm{Q}\in \mathbb{R}^{m \times m}$, which presents no loss of generality compared to authors who write the innovation as $\bm{R}\bm{\eta}_t$ for some matrix $\bm{R}$.\footnote{Indeed, my $\bm{Q}$ could throughout be replaced by $\bm{R}\bm{Q}\bm{R}'$; for a similar comment, see \citet[p.\ 43]{durbin2000time}.} The observation density $p(\bm{y}_t\,|\, \bm{\alpha}_t)$ may still be non-Gaussian and involve nonlinearity.

\subsection{Inner maximisation}

Taking Bellman's equation~\eqref{Bellman}, substituting the quadratic approximation~\eqref{normalapproximation00} and the (similarly quadratic) logarithmic state-transition density from model~\eqref{DGP0.3} yields
\allowdisplaybreaks
\begin{eqnarray}
\label{approximate Bellman} 
V_{t}(\bm{a}_{t})&=& \ell(\bm{y}_{t}|\bm{a}_{t}) + \underset{\bm{a}_{t-1}\in \mathbb{R}^m}{\mbox{max}} \Big\{-\frac{1}{2}(\bm{a}_{t}-\bm{c}-\bm{T}\bm{a}_{t-1})'\;\bm{Q}^{-1}\;(\bm{a}_{t}-\bm{c}-\bm{T}\bm{a}_{t-1})
\\
&&\hspace{1cm}
- \frac{1}{2}(\bm{a}_{t-1}-\bm{a}_{t-1|t-1})'\;\bm{I}_{t-1|t-1}\;(\bm{a}_{t-1}-\bm{a}_{t-1|t-1})\Big\}+\text{constants},\quad \bm{a}_t\in \mathbb{R}^m.\notag
\end{eqnarray}
While $\bm{Q}^{-1}$ is assumed to exist in writing equation~\eqref{approximate Bellman}, the results derived below will remain valid when $\bm{Q}$ is only positive semidefinite; this follows from standard limiting arguments (e.g.\ \citealp[p.\ 78]{chopin2020introduction}). Here I focus on the maximisation over the lagged state variable $\bm{a}_{t-1}$.

As the variable $\bm{a}_{t-1}$ appears at most quadratically on the right-hand side of equation~\eqref{approximate Bellman}, its maximisation can be performed in closed form. Importantly for the development below, the solution, denoted $\bm{a}_{t-1}^\ast\in \mathbb{R}^m$, depends linearly on the variable $\bm{a}_t\in \mathbb{R}^m$, which is involved in the outer maximisation. Hence $\bm{a}_{t-1}^\ast$ is a vector function $\bm{a}_{t-1}^\ast: \mathbb{R}^m \to \mathbb{R}^m$, whose expression following from the standard first-order condition can be usefully expressed (after some algebra, see Supplement~\ref{app:inner maximisation}) as
\begin{eqnarray}
\bm{a}_{t-1}^\ast &=& \bm{a}_{t-1|t-1} +  \bm{I}_{t-1|t-1}^{-1}\, \bm{T}'
\, \bm{I}_{t|t-1} \,  \big(\bm{a}_t-\bm{a}_{t|t-1}\big),
\label{alphastarmain}
\end{eqnarray}
which employs the definitions of the predicted state $\bm{a}_{t|t-1}$ and the predicted precision matrix $\bm{I}_{t|t-1}$ given under step 2 in Table~\ref{table3}. Expression~\eqref{alphastarmain} can be  recognised the one-period version of RTS (\citeyear{rauch1965maximum}) smoother, providing the best estimate of $\bm{a}_{t-1}$ conditional on the best estimate of next state, $\bm{a}_t$, which at this point remains to be found; i.e.\ the optimal $\bm{a}_{t-1}^\ast$ is a function of the (still to be optimised) state  variable $\bm{a}_t$.

Regarding the predicted precision matrix $\bm{I}_{t|t-1}$, the first expression in step 2 of Table~\ref{table3}
relies on the positive definiteness of the matrix $\bm{Q}$. The second expression, which holds by the Woodbury matrix identity, remains valid even when~$\bm{Q}$ becomes singular; a similar argument is made in \citet[p.\ 78]{chopin2020introduction}. Hence the algorithm in Table~\ref{table3} remains valid when $\bm{Q}$ is singular. While the derivation here is different, the resulting prediction step 2 in Table~\ref{table3} is in fact identical to that of the (information form of the) Kalman filter  (e.g.\ \citealp[p.\ 106]{harvey1990forecasting}).  Hence, while the usual derivation of the Kalman filter is based on taking expectations, the optimisation approach presented here yields the same result.

\begin{table}[t!]
\caption{\label{table3} Bellman filter and smoother for model~\eqref{DGP0.3}.}
\center
\renewcommand{\arraystretch}{1}
\begin{footnotesize}
\begin{threeparttable}
\begin{tabularx}{\textwidth}{l@{\hspace{2mm}}l@{\hspace{2mm}}X}
  \toprule
  \bf{Step} & \bf{Method} & \bf{Computation}
 \\
\midrule  
1.\ Initialise & Unconditional &Set $\bm{a}_{0|0}=(\mathds{1}_{m\times m}-\bm{T})^{-1}\bm{c}$ and $\text{vec}(\bm{I}^{-1}_{0|0})=(\mathds{1}_{m^2 \times m^2}-\bm{T}\otimes \bm{T})^{-1}\text{vec}(\bm{Q})$. Set $t=1$.
\\
& Estimation & Treat $\bm{a}_{0|0}$ as a static parameter to be estimated and set  $\bm{I}_{0|0}$ equal to a large multiple of the identity matrix. Set $t=1$.
\\
& Diffuse & Possible if $\arg \max_{\bm{a}} \ell(\bm{y}_1|\bm{a})$ exists. Set $\bm{I}_{0|0}$ equal to a small multiple of the identity matrix. Set $t=1$.
\\
2.\ Predict && $
\bm{a}_{t|t-1} =\bm{c}+ \bm{T}\,\bm{a}_{t-1|t-1}$.
\\
&& $\bm{I}_{t|t-1} = \bm{Q}^{-1}-\bm{Q}^{-1}\bm{T}\big(\bm{I}_{t-1|t-1}+\bm{T}'\bm{Q}^{-1}\bm{T}\big)^{-1}\,\bm{T}'\bm{Q}^{-1}= (\bm{T}\bm{I}_{t-1|t-1}^{-1}\bm{T}' +\bm{Q})^{-1}$.  \\
3.\ Start & & Set $\bm{a}_{t}\leftarrow \bm{a}_{t|t-1}$. \\
&& Alternatively, set $\bm{a}_{t} \leftarrow \arg \max_{\bm{a}} \ell(\bm{y}_t|\bm{a})$ if this quantity exists. 
\\
4.\ Optimise
& Newton&  $\displaystyle\bm{a}_t 
\leftarrow
 \bm{a}_{t}+ \Big[ \bm{I}_{t|t-1}-\frac{\dd^2\ell(\bm{y}_{t}|\bm{a}_{t})}{\dd\bm{a}_{t}\,\dd\bm{a}_{t}'} \Big]^{-1}\,\Big[\frac{\dd\ell(\bm{y}_{t}|\bm{a}_{t})}{\dd\bm{a}_{t}} - \bm{I}_{t|t-1}\big( \bm{a}_{t}-\bm{a}_{t|t-1}\big)\Big] $.
\\
& Fisher& Like Newton step, but replace $\dd^2\ell(\bm{y}_{t}|\bm{a}_{t})/(\dd\bm{a}_{t}\,\dd\bm{a}_{t}')$ by $\mathbb{E}[\dd^2\ell(\bm{y}_{t}|\bm{a}_{t})(\dd\bm{a}_{t}\,\dd\bm{a}_{t}') |  \bm{a}_{t} ]$.
\\
& BHHH & Like Newton step, but replace $\dd^2\ell(\bm{y}_{t}|\bm{a}_{t})/(\dd\bm{a}_{t}\,\dd\bm{a}_{t}')$ by $-\dd\ell(\bm{y}_{t}|\bm{a}_{t})/\dd\bm{a}_{t} \times \dd\ell(\bm{y}_{t}|\bm{a}_{t})/\dd\bm{a}_{t}'$ .
\\
5.\ Stop && Stop at if some convergence criterion is satisfied or after a predetermined number of iterations.
\\
6.\ Update && $\bm{a}_{t|t} = \bm{a}_t$.
\\
& Newton &  $\displaystyle \bm{I}_{t|t} = \bm{I}_{t|t-1} -\left. \frac{\dd^2\ell(\bm{y}_{t}|\bm{a}_t)}{\dd\bm{a}_t\,\dd\bm{a}_t'}\right|_{\bm{a}_t=\bm{a}_{t|t}}$ if the realised information is positive semidefinite
\\
&Fisher& Like Newton update, but replace $\dd^2\ell(\bm{y}_{t}|\bm{a}_{t})/(\dd\bm{a}_{t}\,\dd\bm{a}_{t}')$ by $\mathbb{E}[\dd^2\ell(\bm{y}_{t}|\bm{a}_{t})(\dd\bm{a}_{t}\,\dd\bm{a}_{t}') |  \bm{a}_{t} ]$.
\\
& BHHH & Like Newton update, but replace $\dd^2\ell(\bm{y}_{t}|\bm{a}_{t})/(\dd\bm{a}_{t}\,\dd\bm{a}_{t}')$ by $-\dd\ell(\bm{y}_{t}|\bm{a}_{t})/\dd\bm{a}_{t} \times \dd\ell(\bm{y}_{t}|\bm{a}_{t})/\dd\bm{a}_{t}'$.
\\
7.\ Proceed && Set $t=t+1$ and return to step 2.
\\
\midrule  
8.\ { Smooth} && Run the Bellman filter and store $\bm{a}_{t|t}$, $\bm{P}_{t|t}=\bm{I}^{-1}_{t|t}$ and $\bm{P}_{t|t-1}=\bm{I}^{-1}_{t|t-1}$ for all $1\leq t \leq n$.
\\
 &&  Start with $t=n-1$ and iterate the following recursions backwards until $t=1$ is reached:
\\
&&  $\bm{a}_{t|n} =\bm{a}_{t|t} +\bm{P}_{t|t} \bm{T}' \bm{I}_{t+1|t} (\bm{a}_{t+1|n}-\bm{c}-\bm{T}\bm{a}_{t|t})$,  and
\\
&& $\bm{P}_{t|n}=\bm{P}_{t|t}-\bm{P}_{t|t}\bm{T}' \bm{I}_{t+1|t} ( \bm{P}_{t+1|t} - \bm{P}_{t+1|n} ) \bm{I}_{t+1|t} \bm{T} \bm{P}_{t|t}$.
\\
\bottomrule
\end{tabularx}
\begin{tablenotes}
\item \footnotesize{\emph{Note:} BHHH = Berndt-Hall-Hall-Hausman. The log-likelihood function $\ell(\bm{y}_t|\bm{\alpha}_t)$ is known in closed form and can be read off from the data-generating process~\eqref{DGP0.3}. The corresponding score and the realised and expected information quantities are written as $\dd\ell(\bm{y}_t|\bm{a})/\dd\bm{a}$,  $-\dd^2\ell(\bm{y}_t|\bm{a})/(\dd\bm{a}\dd\bm{a}')$ and $\mathbb{E}[-\dd^2\ell(\bm{y}_t|\bm{a})/(\dd\bm{a}\dd\bm{a}')|\bm{a}]$, respectively, which are viewed as functions of $\bm{a}$, to be evaluated at some state estimate. Steps 4 and 6 list three (intentionally vanilla) optimisation methods, which may but need not be identical for both steps. Users may also implement more sophisticated optimisation methods based on the argmax~\eqref{updatingrule v2}.  The expressions in the (optional) smoother step 8 are derived in section~\ref{sec:backward recursion}. 
}
\end{tablenotes}
\end{threeparttable}
\end{footnotesize}
\end{table}

\subsection{Outer maximisation}

Substituting the vector function $\bm{a}^\ast_{t-1}:\mathbb{R}^m\to \mathbb{R}^m$ of equation~\eqref{alphastarmain} back into Bellman's equation~\eqref{approximate Bellman}, we obtain (after some algebra, see Supplement~\ref{app:B}) the value function with a single argument, $\bm{a}_t$, as follows:
\begin{eqnarray}
\label{approximate Bellman v2}
V_{t}(\bm{a}_t)\; =\;  \ell(\bm{y}_{t}|\bm{a}_t) -\frac{1}{2}(\bm{a}_t-\bm{a}_{t|t-1})'\,\bm{I}_{t|t-1}\,(\bm{a}_t-\bm{a}_{t|t-1}) +\text{constants},\qquad \bm{a}_t\in \mathbb{R}^m,
\end{eqnarray}
where predicted quantities $\bm{a}_{t|t-1}\in \mathbb{R}^m$ and $\bm{I}_{t|t-1}\in \mathbb{R}^{m\times m}$ were derived above (see step 2 of Table~\ref{table3}). The (approximate) value function~\eqref{approximate Bellman v2} involves two terms: (a) the log-likelihood contribution of $\bm{y}_t$ evaluated at the state variable $\bm{a}_t$ and (b) a quadratic term that penalises deviations of $\bm{a}_t$ from $\bm{a}_{t|t-1}$.  The filtered state at time $t$ maximises the sum of both terms, i.e.\
\begin{eqnarray}
\bm{a}_{t|t}\;:=\;
\underset{\bm{a}_t\in \mathbb{R}^m}{\mbox{argmax}}\, V_{t}(\bm{a}_t) = \underset{\bm{a}_t\in \mathbb{R}^m}{\mbox{argmax}}
\left\{
\ell(\bm{y}_{t}|\bm{a}_t) -\frac{1}{2}(\bm{a}_t-\bm{a}_{t|t-1})'\,\bm{I}_{t|t-1}\,(\bm{a}_t-\bm{a}_{t|t-1}) \right\}.\label{updatingrule v2}
\end{eqnarray}
The optimisation can be performed in closed form when the observation density is Gaussian with mean $\bm{d}+\bm{Z}\bm{a}_t$, as in Corollary~\ref{corol0}, in which case $\ell(\bm{y}_{t}|\bm{a}_t)$ is multivariate quadratic in $\bm{a}_t$; this yields the standard Kalman filter (see Supplement~\ref{comparison} for details). 
In general, the potentially complicated functional form of $\ell(\bm{y}_t|\bm{a}_t)$ implies that optimisation~\eqref{updatingrule v2} cannot be performed in closed form. Some plain-vanilla applications of optimisation methods are included in Table~\ref{table3} under step~4. The presence of the score in this optimisation step is distinctive for the Bellman filter and guarantees its robustness if the observation density is heavy tailed.  As before, the computational complexity of the resulting filter is $O(m^3 t)$. 

A unique argmax~\eqref{updatingrule v2} is guaranteed when the precision matrix $\bm{I}_{t|t-1}$ is positive definite and the log-likelihood function $\ell(\bm{y}_t|\bm{a}_t)$ is concave in the state variable $\bm{a}_t\in \mathbb{R}^m$. When the smallest eigenvalue of the precision matrix $\bm{I}_{t|t-1}$ is sufficiently large, a unique argmax is still guaranteed to exist even when $\ell(\bm{y}_t|\bm{a}_t)$ fails to be concave in $\bm{a}_t$. 
 In the non-concave case, it is possible that $\bm{I}_{t|t-1}$ is insufficiently `large' to pin down the update. This may be solved by adding to $\bm{I}_{t|t-1}$ some positive multiple of the identity matrix or skip the optimisation altogether; in the simulation study in section~\ref{section7}, this situation never arose.

Before proceeding to the next time step, the value function~\eqref{approximate Bellman v2} must be approximated by a multivariate quadratic function. Because constants are irrelevant and  the argmax has already been found, what remains is to determine the negative matrix of second derivatives evaluated at the peak, denoted $\bm{I}_{t|t}$, as indicated in Table~\ref{table3} under step~6. Intuitively, one expects  $\smash{\bm{I}_{t|t}\geq \bm{I}_{t|t-1}}$, where the weak inequality means that the left-hand side minus the right-hand side is positive semidefinite. The intuition derives from the fact that missing observations can be dealt with as in the Kalman filter by setting $\bm{a}_{t|t}=\bm{a}_{t|t-1}$ and $\smash{\bm{I}_{t|t}= \bm{I}_{t|t-1}}$. Any (existing) observation should be weakly more informative than a nonexistent one, implying $\smash{\bm{I}_{t|t}\geq \bm{I}_{t|t-1}}$. The lower bound may be reached in the limit for extreme observations (i.e.\ outliers), which are uninformative. While Newton's updating method under step~6 has the advantage of explicitly utilising the observation $\bm{y}_t$, enabling it to recognise that some observations carry little information, the inequality $\smash{\bm{I}_{t|t}\geq \bm{I}_{t|t-1}}$ is not guaranteed unless the realised information quantity is positive semidefinite. For Fisher's updating method under step~6, the situation is reversed, failing to utilise the realisation $\bm{y}_t$ while ensuring $\smash{\bm{I}_{t|t}\geq \bm{I}_{t|t-1}}$. For some models it is possible to formulate a hybrid version, e.g.\ by taking a weighted average of Newton's and Fisher's updating methods, that achieves the best of both worlds (I use this hybrid method for some models in section~\ref{section7}).

\subsection{Special cases of Bellman filter with linear Gaussian states}

Special cases of the algorithm in Table~\ref{table3}  include the Kalman filter (Supplement~\ref{comparison}), the iterated extended Kalman filter (Supplement~\ref{app:E}), \citeauthor{fahrmeir1992posterior}'s (\citeyear{fahrmeir1992posterior}) approximate mode estimator (Supplement~\ref{app fahrmeir}), \citeauthor{koyama2010approximate}'s (\citeyear{koyama2010approximate}) Laplace Gaussian filter (Supplement~\ref{app LGF}), and \citeauthor{toulis2017asymptotic}'s (\citeyear{toulis2017asymptotic}) implicit stochastic gradient method for the estimation of states that are constant over time (Supplement~\ref{app ISG}). The key difference with implicit stochastic gradient methods is that the Bellman filter, like the Kalman filter, generally remains perpetually responsive and does not converge to a `true' parameter value.

\section{Theory: Contractivity,  error bounds and stability }
\label{section5}

This section investigates the theoretical properties of the Bellman filter derived in the previous section, i.e.\ under the assumption of linear and Gaussian state dynamics. Under appropriate conditions, this section will show that (a) at a fixed time step, the Bellman filtering step is contractive in quadratic mean to a small region around the true state,  (b) over time, the mean squared filtering error remains uniformly bounded (i.e.\ approximation errors cannot accumulate), and (c) the effect of the initialisation of the filter vanishes asymptotically and exponentially fast, an important property known as invertibility (\cite{straumann2006quasi} or stability (\citealp[Th.\ 4]{koyama2010approximate}).

\subsection{Contractivity at a fixed time step}

Here the time step~$t\geq 1$ is considered fixed. Similarly, in the Bellman-filter update~\eqref{updatingrule v2}, predictions $\bm{a}_{t|t-1}\in \mathbb{R}^m$ and $\bm{I}_{t|t-1}\in \mathbb{R}^{m\times m}$ are fixed. Update~\eqref{updatingrule v2} can generally be viewed as a stochastic version of \citeauthor{rockafellar1976monotone}'s (\citeyear{rockafellar1976monotone}) proximal point algorithm, which similarly combines a target function to be optimised, in this case $\ell(\bm{y}_t|\bm{a}_t)$, with a quadratic penalty centred at a previous iterate, in this case $\bm{a}_{t|t-1}$. Indeed, optimisation~\eqref{updatingrule v2} can be classed as  a stochastic proximal point method (e.g.\ \citealp{ryu2016stochastic}, \citealp{bianchi2016ergodic}, \citealp{patrascu2018nonasymptotic}, \citealp{asi2019stochastic}).  Its intuitive link with proximal optimisation methods suggests that update~\eqref{updatingrule v2} should remain both applicable and reasonably accurate outside the classic Kalman-filtering context.  Theorem~\ref{thrm1} below confirms this intuition.
\smallskip

\textbf{Notation:} For vectors $\bm{x}\in \mathbb{R}^m$, the Euclidean norm is denoted by $\|\bm{x}\|:=\sqrt{\bm{x}'\bm{x}}$. For a positive definite weight matrix $\bm{W}>\bm{0}$, the  weighted Euclidean vector norm is denoted $\|\bm{x}\|_{\bm{W}}:=\sqrt{\bm{x}'\bm{W}\bm{x}}$, while for a matrix $\bm{M}\in \mathbb{R}^{m \times m}$, the induced matrix norm is denoted $\|\bm{M}\|_{\bm{W}}:=\max\{ \|\bm{M}\bm{x}\|_{\bm{W}} :\|\bm{x}\|_{\bm{W}}= 1\}$ (see e.g.\ \citealp[Def.\ 2.8]{jungers2009joint}). 
The gradient and Hessian of $\ell(\bm{y}|\bm{a})$ with respect to $\bm{a}$ are written as $\nabla \ell(\bm{y}|\bm{a})$ and $\nabla^2 \ell(\bm{y}|\bm{a})$, respectively. The smallest and largest eigenvalues of a matrix $\cdot$ are denoted $\lambda_{\min}(\cdot)$ and $\lambda_{\max}(\cdot)$, respectively.
The $m\times m$ identity matrix is denoted by $\mathds{1}_{m\times m}$. 

\begin{assumption}[Concavity] 
With probability one in the random draw $\bm{y}$, the observation log density $\ell(\bm{y}|\cdot)$ maps $\mathbb{R}^m$ to $\mathbb{R}$,  and is either (a) concave, or (b) strictly concave, or (c) strongly concave with parameter $\epsilon>0$.
\end{assumption}

\begin{assumption}[Differentiability] With probability one in the random draw $\bm{y}$, the observation log density $\bm{a}\mapsto \ell(\bm{y}|\bm{a})$ is (a) once or (b) twice continuously differentiable on all of $\mathbb{R}^m$.
\end{assumption}

\begin{assumption}[Bounded information] $\mathbb{E}[\| \nabla \ell(\bm{y}_t|\bm{\alpha}_t)\|^2] \leq \sigma^2<\infty$, where $\bm{\alpha}_t$ is the true latent state that generates $\bm{y}_t\sim p(\bm{y}_t|\bm{\alpha}_t)$.
\end{assumption}

\begin{theorem}[Contractivity of the mean squared error] \label{thrm1} Fix the time step $t\geq 1$. Let $\bm{a}_{t|t-1}\in \mathbb{R}^m$ and $\bm{I}_{t|t-1}\in \mathbb{R}^{m\times m}$ be given and fixed, where the latter is symmetric and positive definite with eigenvalues satisfying $0<\lambda_{\min}(\bm{I}_{t|t-1})\leq\lambda_{\max}(\bm{I}_{t|t-1})< \infty$. Let update $\bm{a}_{t|t}$ be defined by~\eqref{updatingrule v2}. 
\begin{enumerate}
\item \textbf{Boundedness of updates:} Under Assumption 1a, with probability one, the update $\bm{a}_{t|t}$ is well defined and satisfies
\begin{equation}
\smash{\frac{1}{2}\big\|\bm{a}_{t|t}-\bm{a}_{t|t-1} \big\|_{\bm{I}_{t|t-1}}^2  \; \leq\;   \ell(\bm{y}_{t}|\bm{a}_{t|t})-\ell(\bm{y}_{t}|\bm{a}_{t|t-1})}\;.
\end{equation}

\item \textbf{Stability for a single time step:} Let Assumption 2b hold. Let $\lambda_{\min}(\bm{I}_{t|t-1})>\max\{0,\lambda_{\max}(\nabla^2 \ell(\bm{y}|\bm{a})) \}$ for all $\bm{a}\in \mathbb{R}^m$ and with probability one in $\bm{y}$. Then, with probability one, 
\begin{equation}
\label{norm}
\left\| \frac{\mathrm{d} \bm{a}_{t|t} }{\mathrm{d} \bm{a}_{t|t-1}' } \right\|_{\bm{I}_{t|t-1}} \; \leq \;
  1-\frac{\lambda_{\min}(-\nabla^2\ell (\bm{y}_t|\bm{a}_{t|t}))}{\lambda_{\max}(\bm{I}_{t|t-1})+\lambda_{\max}(-\nabla^2\ell (\bm{y}_t|\bm{a}_{t|t}))}.
\end{equation}

The right-hand side does not exceed (is strictly less than) unity under the additional Assumption~1a (1b).

\item \textbf{Contractivity of the quadratic error:} Under Assumptions 1c, 2a and 3, 
 \begin{equation}
 \label{inequality}
\mathbb{E}\left( \left\| \bm{a}_{t|t}-\bm{\alpha}_t\right\|_{\bm{I}_{t|t-1}+2\epsilon\, \mathds{1}_{m\times m}}^2 \right)
\; \leq \;\mathbb{E}
\left( \left\| \bm{a}_{t|t-1}-\bm{\alpha}_t\right\|_{\bm{I}_{t|t-1}}^2\right)
  \,+\, \frac{\sigma^2}{\lambda_{\min}(\bm{I}_{t|t-1})}.
 \end{equation}
\end{enumerate}
\end{theorem}

The proof is presented in Supplement~\ref{proof of theorem 1}. Compared with other results for approximate filters (e.g.\ \citealp{koyama2010approximate}), Theorem~\ref{thrm1} is attractive because the assumptions are (a) more easily verifiable (relating to model inputs instead of outputs) and (b) less stringent. For example, Theorem~\ref{thrm1} applies to the Kalman filter, while the theory developed in \cite{koyama2010approximate} does not.\footnote{ \cite{koyama2010approximate} require logarithmic observation densities with five uniformly bounded derivatives, ruling out the Gaussian case in which the logarithmic density is quadratic, implying unbounded first derivatives on $\mathbb{R}^m$.}

Part~1 of Theorem~\ref{thrm1} indicates that the update is well-defined, while Part~2 demonstrates that the Bellman-filtered state $\bm{a}_{t|t}$ is stable in the prediction $\bm{a}_{t|t-1}$. This stability property can be used to establish the stability of the Bellman filter (see section~\ref{subsec:stability}). Part~3 of Theorem~\ref{thrm1} says that the quadratic filtering error is contractive in expectation towards a small region around the true state. Inequality~\eqref{inequality} features a weighted norm on both sides, in which the predicted information matrix $\bm{I}_{t|t-1}$ plays a key role. The weight matrix on the left-hand side of inequality~\eqref{inequality} contains the additional term $2 \epsilon \mathds{1}_{m\times m}$ such that the diagonal is `reinforced': this drives the contraction. Intuitively, when the weight matrix is `bigger' (i.e.\ has larger eigenvalues), the vector inside the norm must be `smaller' in magnitude. Of course, an improvement is impossible when the prediction is perfect, such that the additive term $\sigma^2/\lambda_{\min}(\bm{I}_{t|t-1})$ on the right-hand side of equation~\eqref{inequality} is unavoidable. Hence updates are contractive in quadratic mean towards a `noise-dominated region' (NDR) around the true state (e.g.\ \citealp[p.\ 3]{patrascu2018nonasymptotic}).

Theorem~\ref{thrm1} also relates to \citet[p.\ 1291]{toulis2016towards}, who present the seemingly stronger result that proximal updates are `contracting almost surely' when the log-likelihood function is strongly concave; however, their result relies on a nonstandard definition of strong concavity that rules out important cases of interest, e.g.\ the Kalman filter (see Supplement~\ref{comparison toulis} for a detailed comparison).

\subsection{Error bounds over time}

While Theorem~\ref{thrm1} involved a fixed time step, it is equally important to investigate how  filtered quantities behave over extended time periods. When the latent state is stationary, even a trivial filter may asymptotically achieve a bounded mean squared error (MSE), e.g.\ by setting the filter output equal to zero for all time steps. Hence a more pertinent question is whether the filter can asymptotically achieve a bounded MSE in the case of unit-root states. As this section shows, in the long run, the Bellman filter achieves a bounded MSE  even if the true process is free to roam. 

\begin{figure}
\caption{Illustration of mean squared error (MSE) contraction due to inequality~\eqref{inequality simplified}}
\begin{threeparttable}
\hspace{2.5cm}
\includegraphics[width = 4in]{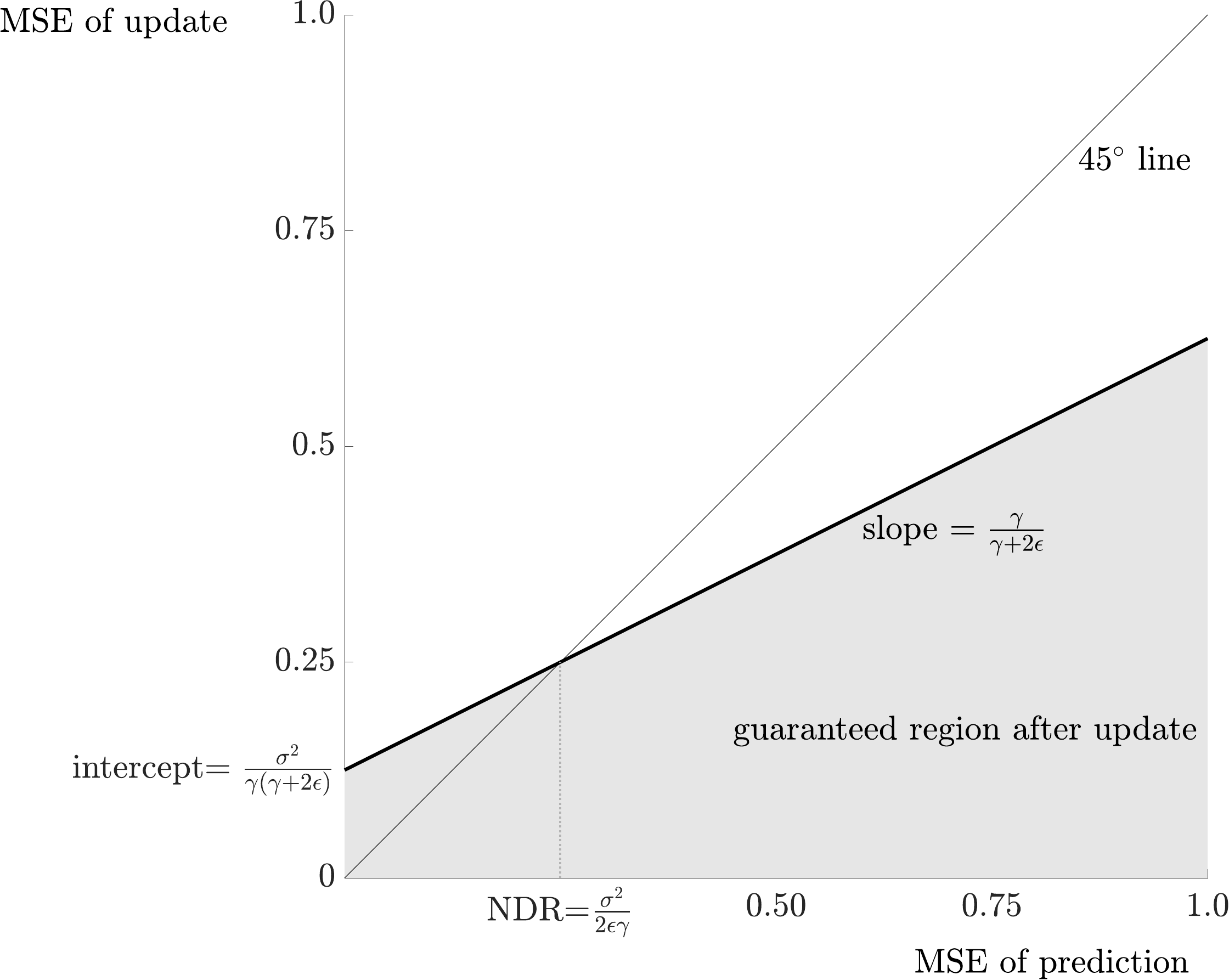}
\begin{tablenotes}
\footnotesize
\emph{Note}: NDR = noise-dominated region. The grey area corresponds to possible values of the MSE after updating, which is conditional on the MSE before updating. Purely for illustrative purposes, the parameters are $\sigma=\epsilon=1$ and $\gamma=2$. 
\end{tablenotes}
\end{threeparttable}
\label{fig1}
\end{figure}

For simplicity I focus on the case in which $\bm{I}_{t|t-1}$ is a constant multiple of the identity matrix; hence $\bm{I}_{t|t-1}=\gamma \mathds{1}_{m \times m}$, where $\gamma>0$ can be interpreted as a smoothing parameter, and $\lambda_{\min}(\bm{I}_{t|t-1})=\lambda_{\max}(\bm{I}_{t|t-1})=\gamma$.  The weighted MSE contraction~\eqref{inequality} for a fixed time step then reduces to a standard MSE contraction:
\begin{equation}
\label{inequality simplified}
\underbrace{\mathbb{E}
 \left( \left\| \bm{a}_{t|t}-\bm{\alpha}_t\right\|^2 \right)}_{\text{MSE of update}}
 \; \leq\; \underbrace{ \frac{\gamma}{\gamma+2\,\epsilon}}_{<1}\; \Bigg[ \underbrace{\mathbb{E}
 \left(
 \left\| \bm{a}_{t|t-1}-\bm{\alpha}_t\right\|^2\right)}_{\text{MSE of prediction}}  + \underbrace{\frac{\sigma^2}{\gamma^2}}_{>0} \Bigg].
\end{equation}
Inequality~\eqref{inequality simplified} features a multiplicative constant on its right-hand side that is strictly less than unity, which gives rise to the contraction. As illustrated in Figure~\ref{fig1}, the inequality says that the MSE of the update is bounded above by a linear function of the MSE of the prediction. The slope of this line is $\gamma/(\gamma+2\epsilon)<1$, while the intercept is $\sigma^2/(\gamma(\gamma+2\epsilon))>0$. The area below the line, shaded in grey, shows the contraction due to inequality~\eqref{inequality simplified}. When the prediction error is large,  the contractive property dominates and the update is expected to be beneficial: the grey area lies below the $45^\circ$ line.  When the prediction happens to be pinpoint accurate (i.e.\ the corresponding MSE is zero), the MSE of the update need not be zero, as can be seen in Figure~\ref{fig1} from the fact that the grey area  stretches above the $45^\circ$ line close to the origin. 
This is unavoidable with noisy data: when predictions are perfect, updates cannot be better. In the limit $\epsilon \to 0$, whereby the target function is concave but not strongly so, inequality~\eqref{inequality simplified} is closely related to Theorem~3.2 in \cite{asi2019stochastic}. 

MSE contraction~\eqref{inequality simplified} is used below in Proposition~\ref{corol unit root} (see Supplement~\ref{S geometric convergence} for the proof) to demonstrate that the filtering MSE remains uniformly bounded over time. Proposition~\ref{corol unit root} applies to the Kalman filter, which can similarly track unit-root states in the long run, but holds more generally for strictly concave logarithmic observation densities. 

\begin{proposition}[Uniformly bounded MSE]
\label{corol unit root} Assume $\bm{\alpha}_{t}=\bm{\alpha}_{t-1}+\bm{\eta}_{t}$ with $\bm{\eta}_{t}\sim \textnormal{i.i.d. }(\bm{0},\bm{Q})$, which need not be Gaussian, and $\sigma_\eta^2=\text{Trace}(\bm{Q})<\infty$. Set $\bm{a}_{t+1|t}=\bm{a}_{t|t}$ and take $\bm{I}_{t+1|t}=\gamma \mathds{1}_{m \times m}$ for some $\gamma>0$ and all $t \geq 1$. Let $\bm{a}_{t|t}$ be given by update~\eqref{updatingrule v2}.  Denote $\textnormal{MSE}_{t|t}:=\mathbb{E}\|\bm{a}_{t|t}-\bm{\alpha}_t\|^2$ and $\textnormal{MSE}_{t|t-1}:=\mathbb{E}\|\bm{a}_{t|t-1}-\bm{\alpha}_t\|^2$. In the setting of part 3 of Theorem~1, 
 \begin{equation}
\textnormal{MSE}_{t|t}
 \leq  \frac{\gamma}{\gamma+2\,\epsilon}\; \Big[ \textnormal{MSE}_{t|t-1}+ \frac{\sigma^2}{\gamma^2} \Big], \qquad \qquad 
 \textnormal{MSE}_{t+1|t}  =   \textnormal{MSE}_{t|t} +\sigma_\eta^2,\qquad t\geq 1.\label{MSE2}
\end{equation} 
Irrespective of the initial value $\textnormal{MSE}_{1|0}$, the long-run filtering error remains uniformly bounded:
\begin{equation}
\label{upper bound}
\underset{t\to \infty}{\textnormal{lim sup}}\;\; \textnormal{MSE}_{t|t} \;\; \leq  \;\; \frac{\sigma^2}{2\,\gamma\,\epsilon} + \frac{\gamma\, \sigma_\eta^2}{2\,\epsilon}.
\end{equation}
Minimising the bound with respect to $\gamma$ yields $\gamma=\sigma/\sigma_\eta$.
\end{proposition}

\subsection{ Stability}
\label{subsec:stability}

 As emphasised by~\citet[p.\ 63]{anderson2012optimal}, `a question of vital interest [...] is whether or not the filter is stable'. A filter can be considered stable if deviations in the initial conditions `tend to be reduced, rather than amplified, by conditioning on further observations' \citep{koyama2010approximate}.  To this end, it is sufficient that filtered paths with different initialisations---but based on identical data---converge exponentially fast over time, a concept known as `invertibility' (e.g.\ \citealp{straumann2006quasi}). This section demonstrates the stability of a time-invariant version of the Bellman filter.

Stability analyses of the Kalman filter rely on the fact that, in the  time-invariant version of the filter, the matrix $\mathrm{d}\bm{a}'_{t|t}/\mathrm{d}\bm{a}_{t-1|t-1}$ is static, as $\bm{a}_{t|t}$ is then a linear function of $\bm{a}_{t|t-1}$ with a static coefficient matrix. Stability follows when the spectral radius of this coefficient matrix is strictly exceeded by one. Unfortunately, the stability analysis here is complicated by the fact that each derivative matrix $\mathrm{d}\bm{a}'_{t|t}/\mathrm{d}\bm{a}_{t-1|t-1}$ is stochastic, depending on the observations as well as the filtered states. Moreover, an analysis based on the spectral radius is ruled out because it fails to be a norm. I follow the classic literature in investigating a time-invariant setting, which implies that the predicted information matrix $\bm{I}_{t|t-1}=\bm{I}\in \mathbb{R}^{m\times m}$ is taken to be static over time. I deviate by basing the result  not on the spectral radius but the (weighted) matrix norm $\|\cdot\|_{\bm{I}}$.

\begin{theorem}[Stability of the time-invariant Bellman filter.] \label{thrm2} Let the initialisation $\bm{a}_{0|0}\in \mathbb{R}^m$ be given. For all $t\geq 1$, (a) set $\bm{a}_{t|t-1}=\bm{c} +\bm{T}\bm{a}_{t-1|t-1}$, where $\bm{c}\in \mathbb{R}^m$ and $\bm{T}\in \mathbb{R}^{m \times m}$ are given, and (b) let update $\bm{a}_{t|t}$ be defined by maximisation~\eqref{updatingrule v2}, where $\bm{I}_{t|t-1}=\bm{I}\in \mathbb{R}^{m \times m}$ is a time-invariant (i.e.\ static) positive-definite matrix with eigenvalues in the range $(\nu_{\min},\nu_{\max})$. Assume that, with probability one, the observation log density $\ell(\bm{y}|\bm{a})$ is twice continuously differentiable, while the negative Hessian matrix $-\nabla^2 \ell(\bm{y}|\bm{a})$ has eigenvalues in the range $(\mu_{\min},\mu_{\max})$ uniformly for $\bm{a}\in \mathbb{R}^m$, where $\max\{0,-\mu_{\min}\}<\nu_{\min}$. Then, with probability one,
\begin{equation}
\left\|\frac{\dd \bm{a}_{t|t} }{\dd \bm{a}_{0|0} '} \right\|_{\bm{I}} \; \leq \; \left(1- \min\left\{\frac{\delta}{\nu_{\min}},\frac{\delta}{\nu_{\max}}\right\}
\right)^{t/2} \left(1-\frac{\mu_{\min} }{\nu_{\max}+\mu_{\max} } \right)^t ,
\label{stability}
\end{equation}
where $\delta:=\lambda_{\min}(\bm{I}-\bm{T}'\bm{I}\bm{T})\leq \nu_{\min}$.
As $t\to \infty$, exponential almost sure convergence to zero is guaranteed under the following sufficient condition:
\begin{equation}
\label{log condition}
\frac{1}{2} \log  \left(1-\min\left\{\frac{\delta}{\nu_{\min}},\frac{\delta}{\nu_{\max}}\right\}
\right) + \log \left(1-\frac{\mu_{\min} }{\nu_{\max}+\mu_{\max} } \right)\; < \; 0.
\end{equation}
\end{theorem}

The proof is presented in Supplement~\ref{app:thrm2}. Theorem~\ref{thrm2} assumes that $\bm{I}$ is positive definite while its smallest eigenvalue $\nu_{\min}>0$ is sufficiently large. For concave log densities (i.e.\ $\mu_{\min}\geq 0$), it is required only that $\nu_{\min}>0$ such that $\bm{I}$ is positive definite. For log densities that fail to be concave (i.e.\ $\mu_{\min}<0$), the stronger condition $\nu_{\min}>\max\{0,-\mu_{\min}\}$ is imposed to ensure that optimisation problem~\eqref{updatingrule v2} is well-defined and leads to unique solution $\bm{a}_{t|t}$ for all $t$. The sufficient condition~\eqref{log condition} for invertibility is automatically satisfied if the prediction and updating steps are both non-expansive (both $\delta\geq 0$ and $\mu_{\min}\geq 0$), while at least one is strictly contractive ($\delta> 0$ and/or $\mu_{\min}> 0$). For example, the observation log density could be strictly concave (i.e.\ $\mu_{\min}>0$) while $\bm{T}$ is the identity matrix (in which case $\delta=0$); hence, unit root dynamics are permitted. Moreover, inequality~\eqref{log condition} will always be satisfied if the observations point adequately to the underlying state. More specifically, if $\mu_{\min}$ and $\mu_{\max}$ approach infinity at the same rate (such that the measurement is exceedingly precise), then the second logarithm in condition~\eqref{log condition} approaches negative infinity such that the condition is satisfied. For sufficiently informative observations, therefore, even explosive state dynamics may be accommodated. 

\section{ Smoothing using Bellman's principle}
\label{sec:backward recursion}

Here the general method in section~\ref{section2} is extended to present a unified method for both filtering and smoothing using Bellman's dynamic-programming principle. Readers purely interested in filtering can skip this section without loss of continuity. While the approach below is general, I present the most explicit result in the case of a linear Gaussian state equation. This specialised setting allows me to show that the classic Rauch, Tung and Striebel (RTS, \citeyear{rauch1965maximum}) smoother expressions remain valid, albeit as approximations, for a general (i.e.\ non-Gaussian) observation density---an insight that may be useful in practice. 

Below I introduce three value functions, based on (a) past data, (b) future data and (c) all data. All three are based on the partial log-likelihood function $L_{t_1:t_2}:\Omega\times\mathbb{R}^m\times \ldots \times \mathbb{R}^m\to \mathbb{R}$ involving states and observations from time $t_1$ to $t_2$ as follows:
\begin{equation}
\label{partial sum}
L_{t_{1}:t_{2}}(\bm{a}_{t_1},\ldots, \bm{a}_{t_2}) := \sum_{i=t_1}^{t_2} \ell(\bm{y}_i|\bm{a}_i)  + \sum_{i=t_1+1}^{t_2} \ell(\bm{a}_{i}|\bm{a}_{i-1}) +\mathds{1}_{t_1=1}\; \ell(\bm{a}_1), \quad 1\leq t_1 \leq t_2 \leq n,
\end{equation}
where sums containing no terms are understood to be zero.  Equation~\eqref{partial sum} generalises equation~\eqref{loglik}, which is a special case with $t_1=1$ and $t_2=t$.  The new function $L_{t_1:t_2}(\cdots)$  depends on observations $\bm{y}_{t_1}$ through $\bm{y}_{t_2}$, which are considered fixed, and involves $t_2-t_1$ state transitions from $\bm{a}_{t_1}$ to $\bm{a}_{t_2}$.  For definiteness, I assume that $L_{t_1:t_2}(\cdot,\cdots,\cdot)$ can be maximised with respect to each input argument; this assumption is too strong but sufficient for the development below.

\begin{assumption}\label{assumption4} For all $1\leq t_1\leq t_2\leq n$, the partial log-likelihood function $L_{t_1:t_2}(\cdot,\cdots,\cdot)$ defined in equation~\eqref{partial sum} has a unique maximum with respect to each state variable $\bm{a}_t$, i.e.\ for each $ t_1 \leq t \leq t_2$. \end{assumption}

Assumption~\ref{assumption4} allows us to define three value functions $V_t(\cdot),W_t(\cdot),Z_t(\cdot):\Omega\times\mathbb{R}^m\to \mathbb{R}$ as follows:
\begin{alignat}{4}
\text{using past data:} \qquad && V_{t}(\bm{a}_t) &:=\hspace{7mm}\underset{\bm{a}_{1},\ldots,\bm{a}_{t-1}}{\max} &&\;\;L_{1:t}(\bm{a}_{1},\ldots, \bm{a}_{t}),
\label{V}
\\
\text{using future data:}\qquad && W_{t}(\bm{a}_t) &:=\hspace{7mm} \underset{\bm{a}_{t+1},\ldots,\bm{a}_{n}}{\max} &&\;\;L_{t:n}(\bm{a}_{t},\ldots, \bm{a}_{n}),
\label{W}
\\
\text{using all data:}\qquad && Z_{t}(\bm{a}_t) &:=\underset{\bm{a}_1,\ldots,\bm{a}_{t-1},\bm{a}_{t+1},\ldots,\bm{a}_{n}}{\max} &&\;\; L_{1:n}(\bm{a}_{1},\ldots, \bm{a}_{n}),
\label{Z}
\end{alignat}
where $1\leq t\leq n$.
Maximisations are written as $\max_{\bm{a}}$ instead of $\max_{\bm{a}\in \mathbb{R}^m}$; i.e.\ it is implicitly understood that each state variable takes values in the state space $\mathbb{R}^m$. The backward-looking value function $V_t(\cdot)$ is identical to that in Definition~\ref{def value function}. 
The forward-looking value function ${W}_{t}(\cdot)$ is based on current and future data and specialises to that in \citet[eq.\ 18]{mayne1966solution} for linear Gaussian state-space models. The convention that any maximisation involving no variables can be ignored gives the correct initial and terminal conditions for $t=1$ and $t=n$, respectively. 
Function $Z_{t}(\cdot)$ is based on all data and implies a smoothed state estimate via ${\bm{a}}_{t|n}:=\text{argmax}_{\bm{a}} {Z}_{t}(\bm{a})$. The usefulness of the above definitions lies in the fact that the first two value functions satisfy forward and backward recursions, respectively, while jointly implying the third:

\begin{proposition}[Bellman's forward and backward recursions.] \label{prop3}  Let Assumption~\ref{assumption4} hold.  Then
\begin{align}
\textnormal{forward recursion:} && V_{t}(\bm{a}_{t}) &= \ell(\bm{y}_{t}|\bm{a}_{t})\,+\, \underset{\bm{a}_{t-1} }{\max} \Big\{ \ell(\bm{a}_{t}|\bm{a}_{t-1}) \, +\, V_{t-1}(\bm{a}_{t-1}) \Big\}, && 1< t\leq  n,
\label{forward recursion}
\\
\textnormal{backward recursion:} && W_{t}(\bm{a}_{t}) &= \ell(\bm{y}_{t}|\bm{a}_{t})\,+\, \underset{\bm{a}_{t+1} }{\max} \Big\{ \ell(\bm{a}_{t+1}|\bm{a}_{t}) \, +\, W_{t+1}(\bm{a}_{t+1}) \Big\}, && 1\leq t< n,
\label{backward recursion}
\\
\textnormal{relation between both:} & & Z_{t}(\bm{a}_t) & =  V_{t}(\bm{a}_{t}) + \underset{\bm{a}_{t+1}}{\max} \Big\{
\ell(\bm{a}_{t+1}|\bm{a}_{t})\,+\, W_{t+1}(\bm{a}_{t+1})\Big\}, && 1\leq t<  n,
\label{relation1}
\\
& &  & =W_{t}(\bm{a}_{t})+\underset{\bm{a}_{t-1}}{\max} \Big\{\ell(\bm{a}_{t}|\bm{a}_{t-1})\,+\,V_{t-1}(\bm{a}_{t-1})\Big\} , && 1< t \leq n.
\label{relation2}
\end{align}
\end{proposition}
The proof, being a straightforward extension of that of Proposition~\ref{prop1}, is omitted. Forward recursion~\eqref{forward recursion} is identical that in Proposition~\ref{prop1}, while backward recursion~\eqref{backward recursion} can be derived using similar arguments; for linear Gaussian state-space models, the latter collapses to the backward recursion in~\citet[eq.\ 27]{mayne1966solution}. Function $Z_t(\cdot)$ can be constructed by combining the output of both recursions, where either the forward or backward recursion extends to time $t$ as in equations~\eqref{relation1} and \eqref{relation2}, respectively.  In both cases, a single-state transition log-density is added, followed by an optimisation involving a single state variable. 

Interestingly, equations~\eqref{relation1} and~\eqref{relation2} do not (explicitly) contain the observation density. Instead, they contain only two value functions (one using past data, one using future data) that are linked through a single state-transition density. When both value functions are quadratic, and the state-transition equation is linear and Gaussian, such that $\ell(\bm{a}_t|\bm{a}_{t-1})$ is also quadratic, then equations~\eqref{relation1} and~\eqref{relation2} contain only quadratic terms and should thus be analytically soluble. As illustrated below, this yields the classic RTS smoother expressions. However, the main innovation of this article is to consider quadratic value functions even when inexact. As the next proposition shows, if we are willing to accept that value functions may be reasonably approximated by quadratic functions, then the resulting expression is still 
given by the classic RTS smoother. This insight appears to be new, and considerably extends the domain of applicability of the RTS smoother, at least as an approximation. In practice, it means that the Bellman filter developed in section~\ref{section4} can be executed and its output  used in the standard RTS smoothing formulas to obtain approximate smoothed state estimates---which the simulation study in section~\ref{section7} finds to be highly accurate. 

\begin{proposition}[Bellman smoother with linear Gaussian state equation] \label{prop2}  Let Assumption~\ref{assumption4} hold.  Assume $\bm{\alpha}_t = \bm{c}+\bm{T}\,\bm{\alpha}_{t-1} + \bm{\eta}_{t}$ with $\bm{\eta}_t \sim \text{i.i.d.}\, \mathrm{N}(\bm{0},\bm{Q})$.  Suppose that both value functions on the right-hand side of equation~\eqref{relation1} are approximated as quadratic functions; in particular let $V_{t}(\cdot)$ have argmax $\bm{a}_{t|t}$ and negative Hessian $\bm{I}_{t|t}=\bm{P}_{t|t}^{-1}>\bm{0}$. Under this approximation,  $Z_t(\cdot)$ on the left-hand side of equation~\eqref{relation1} is also quadratic. Moreover, the argmax $\bm{a}_{t|n}$ of $Z_t(\cdot)$ can be expressed in terms of the argmax $\bm{a}_{t+1|n}$ of $Z_{t+1}(\cdot)$ as follows:
\begin{align}
\bm{a}_{t|n} &=\bm{a}_{t|t} +\bm{P}_{t|t} \bm{T}' \bm{I}_{t+1|t} (\bm{a}_{t+1|n}-\bm{c}-\bm{T}\bm{a}_{t|t}),
\label{RTS1}
\\
\bm{P}_{t|n}&=\bm{P}_{t|t}-\bm{P}_{t|t}\bm{T}' \bm{I}_{t+1|t} ( \bm{P}_{t+1|t} - \bm{P}_{t+1|n} ) \bm{I}_{t+1|t} \bm{T} \bm{P}_{t|t},
\label{RTS2}
\end{align}
where $\bm{I}_{t+1|t}:=(\bm{T}\bm{P}_{t|t}\bm{T}'+\bm{Q})^{-1}>\bm{0}$ and $\bm{I}_{t|n}=\bm{P}_{t|n}^{-1}>\bm{0}$ for $t=1,\ldots,n$ is the negative Hessian of $Z_t(\cdot)$. Expressions~\eqref{RTS1} and~\eqref{RTS2} are identical to the classic RTS smoother expressions, but in a more general---i.e.\ possibly approximate---context. 
\end{proposition}

The proof, presented in Supplement~\ref{app:RTS}, employs only standard matrix algebra, including a simple lemma on multivariate quadratic functions in Supplement~\ref{app:lemma}. Exact solubility of equation~\eqref{relation1} is clear given that all functions on its right-hand side are assumed to be quadratic; the crucial step is to relate the properties of $Z_t(\cdot)$ to those of $Z_{t+1}(\cdot)$ to obtain a backward recursion. The resulting RTS smoother~\eqref{RTS1} requires us to store the output of the filter for all time steps and subsequently to compute the smoothed state, $\bm{a}_{t|n}$, as a linear combination of the filtered state, $\bm{a}_{t|t}$, and the adjacent smoothed state, $\bm{a}_{t+1|n}$. The backward recursion can be initialised using the final filtered state, $\bm{a}_{n|n}$. The output of the backward matrix recursion~\eqref{RTS2}, which provides a measure of uncertainty, is not required if one is merely interested in the smoothed state estimates~\eqref{RTS1}.

\section{Parameter estimation by likelihood approximation}
\label{section6}

\noindent This section presents a heuristic approach to the static-parameter estimation problem, as distinct from the filtering problem, in that we aim to estimate both the time-varying states and the static (hyper)parameter~$\bm{\psi}$. 
I deviate from the literature by decomposing the log-likelihood function of the data in terms of the `fit' generated by the Bellman filter, penalised by a nonnegative term that resembles a `realised' version of the  Kullback-Leibler (KL, \citeyear{kullback1951information}) divergence between filtered and predicted states. Intuitively, this decomposition illustrates that we wish to maximise the congruence of the Bellman-filtered states and the data, while minimising the distance between the filtered and predicted states to prevent over-fitting. 

The proposed pseudo log-likelihood decomposition has the advantage that all terms can be evaluated or approximated using the output of the Bellman filter; no sampling techniques or numerical integration methods are required. While no formal guarantees of convergence are provided, I analyse the statistical properties of the proposed static-parameter estimator in extensive simulation studies (see section~\ref{section7}) and find that it performs on par with simulation-based methods at a fraction of the computational cost. The development of an asymptotic theory remains unresolved.

To introduce the proposed decomposition, I focus on the log-likelihood contribution of a single observation, $\ell(\bm{y}_t|\mathcal{F}_{t-1}):=\log p(\bm{y}_t|\mathcal{F}_{t-1})$. 
The equalities below follow immediately from the definition of conditional densities and the assumption of the state-space model~\eqref{DGP0}:
\begin{eqnarray}
\ell(\bm{y}_t|\mathcal{F}_{t-1})=\ell(\bm{y}_t,\bm{\alpha}_t|\mathcal{F}_{t-1})-\ell(\bm{\alpha}_t|\bm{y}_t,\mathcal{F}_{t-1})=\ell(\bm{y}_t|\bm{\alpha}_t)+\ell(\bm{\alpha}_t|\mathcal{F}_{t-1})-\ell(\bm{\alpha}_t|\mathcal{F}_{t}).
\end{eqnarray}
While the above decomposition is valid for any $\bm{\alpha}_t\in \mathbb{R}^m$, the resulting expression is not a computable quantity, as the true latent state $\bm{\alpha}_t$ remains unknown. It is practical to evaluate the expression at the Bellman-filtered state  $\bm{a}_{t|t}$ and swap the order of the last two terms, such that
\begin{eqnarray}
\label{KL decomposition}
\ell(\bm{y}_t|\mathcal{F}_{t-1})=\ell(\bm{y}_t|\bm{\alpha}_t)\Big|_{\bm{\alpha}_t=\bm{a}_{t|t}}\;-\;\underbrace{\Big\{\ell(\bm{\alpha}_t|\mathcal{F}_{t})- \ell(\bm{\alpha}_t|\mathcal{F}_{t-1})\Big\} \Big| _{\bm{\alpha}_t=\bm{a}_{t|t}}}_{\text{`realised' KL divergence}}.
\end{eqnarray}
The first term on the right-hand side, $\ell(\bm{y}_t|\bm{\alpha}_t)$ evaluated at $\bm{\alpha}_t=\bm{a}_{t|t}$, quantifies the congruence (or `fit') between the Bellman-filtered state $\bm{a}_{t|t}$ and the observation $\bm{y}_t$, which we wish to maximise. We simultaneously aim to minimise the term in curly brackets, i.e.\ the difference $\ell(\bm{\alpha}_t|\mathcal{F}_{t})-\ell(\bm{\alpha}_{t}|\mathcal{F}_{t-1})$ evaluated at $\bm{\alpha}_t=\bm{a}_{t|t}$. This difference can be viewed as a `realised' version of the KL divergence between the filtered and predicted densities; intuitively, it indicates the level of `surprise' associated with the filtered state $\bm{a}_{t|t}$. The standard KL divergence between filtered and predicted densities would have read $\mathbb{E}[\log(\bm{\alpha}_t|\mathcal{F}_{t}) - \log(\bm{\alpha}_t|\mathcal{F}_{t-1})]$, which involves an expectation operator that integrates out the state $\bm{\alpha}_t$ using the true density $p(\bm{\alpha}_t|\mathcal{F}_{t})$. Equation~\eqref{KL decomposition} contains no expectation but is simply evaluated at the filtered state $\bm{a}_{t|t}$; hence, it can be viewed as a realised version. The trade-off in equation~\eqref{KL decomposition} between maximising the fit while minimising the surprise gives rise to a meaningful optimisation problem.  

While decomposition~\eqref{KL decomposition} is exact, we do not generally have an exact expression for the terms in curly brackets. To ensure that the log-likelihood contribution~\eqref{KL decomposition} is computable, I now turn to approximating the realised KL divergence. In deriving the Bellman filter, I presumed that the researcher's knowledge, as measured in log-likelihood space for each time step, could be approximated by a multivariate quadratic function. Extending this line of reasoning, I consider the following approximations of the two terms that compose the realised KL divergence:
\begin{eqnarray}
\label{KL part 1}
\ell(\bm{\alpha}_t|\mathcal{F}_t)&\approx &
 \frac{1}{2} \log \det\{\bm{I}_{t|t}/(2\pi)\} \,-\,\frac{1}{2}(\bm{\alpha}_{t}-\bm{a}_{t|t})'\,\bm{I}_{t|t}\,(\bm{\alpha}_{t}-\bm{a}_{t|t}),\\
\label{KL part 2}
\ell(\bm{\alpha}_t|\mathcal{F}_{t-1})&\approx & \frac{1}{2} \log \det\{ \bm{I}_{t|t-1}/(2\pi) \} \,-\,\frac{1}{2}(\bm{\alpha}_{t}-\bm{a}_{t|t-1})'\,\bm{I}_{t|t-1}\,(\bm{\alpha}_{t}-\bm{a}_{t|t-1}).
\end{eqnarray}
Here the state $\bm{\alpha}_t$ is understood as a variable in $\mathbb{R}^m$, while $\bm{a}_{t|t-1}$, $\bm{a}_{t|t}$, $\bm{I}_{t|t-1}\geq \bm{0}$ and $\bm{I}_{t|t}\geq \bm{0}$ are known quantities determined by the Bellman filter in Table~\ref{table2} or~\ref{table3}, depending on the context.  If the model is linear and Gaussian, then the Bellman filter is exact (it is, in fact, the Kalman filter), as are equations~\eqref{KL part 1}--\eqref{KL part 2}. Based on approximations~\eqref{KL part 1} and~\eqref{KL part 2}, the approximation of the realised KL divergence reads
\begin{equation}
\label{KL approx}
 \ell(\bm{\alpha}_t|\mathcal{F}_{t})- \ell(\bm{\alpha}_t|\mathcal{F}_{t-1})\Big| _{\bm{\alpha}_t=\bm{a}_{t|t}} \; \approx 
\frac{1}{2} \log \frac{\det( \bm{I}_{t|t})}{\det( \bm{I}_{t|t-1})} +  \frac{1}{2}(\bm{a}_{t|t}-\bm{a}_{t|t-1})'\,\bm{I}_{t|t-1}\,(\bm{a}_{t|t}-\bm{a}_{t|t-1}),
\end{equation}
where all constants involving $\pi$ drop out. Nonnegativity of this quantity is guaranteed if $\bm{I}_{t|t}\geq \bm{I}_{t|t-1}$, which can be ensured in the implementation of the filter. Even when approximations~\eqref{KL part 1}--\eqref{KL part 2} are somewhat inaccurate, it may be that the approximation of their difference in equation~\eqref{KL approx} is quite accurate. Intuitively, the realised KL divergence between two densities can be approximated to second order by considering the difference between both argmaxes and the sharpness of both peaks.

To define the proposed approximate maximum-likelihood estimator (MLE) for the static parameters, I take the usual definition $ \widehat{\bm{\psi}}:= {\arg \max} \sum_t \ell(\bm{y}_t|\mathcal{F}_{t-1})$. Then I substitute the (exact) decomposition~\eqref{KL decomposition} and the KL approximation~\eqref{KL approx}, which gives
\begin{align}
\label{approximate estimator}
\widehat{\bm{\psi}}&:= \underset{\bm{\psi}}{\arg \max}
\sum_{t=t_0+1}^{n} \Bigg\{  \underbrace{\ell(\bm{y}_t|\bm{a}_{t|t})}_{\text{`fit' of the filter}} -\Big[ \underbrace{\frac{1}{2} \log \frac{\det( \bm{I}_{t|t})}{\det( \bm{I}_{t|t-1})}+  \frac{1}{2}(\bm{a}_{t|t}-\bm{a}_{t|t-1})'\,\bm{I}_{t|t-1}\,(\bm{a}_{t|t}-\bm{a}_{t|t-1})}_{\geq 0, \text{ KL-type penalty }}\Big] \Bigg\},
\end{align}
where all terms on the right-hand side implicitly or explicitly depend on the (hyper)parameter~$\bm{\psi}$. Time $t_0\geq 0$ is long enough to ensure the mode exists at time $t_0$.  If model~\eqref{DGP0.3} is stationary and $\bm{\alpha}_{0}$ is drawn from the unconditional distribution, as in the simulation studies in section~\ref{section7}, then $t_0=0$. The case $t_0>0$ is analogous to that for the Kalman filter when the first $t_0$ observations are used to  construct a `proper' prior (see~\citealp[p.~123]{harvey1990forecasting}). The first term inside curly brackets, involving the observation density, is given by model~\eqref{DGP0.3}.  The remaining terms can be computed based on the output of the Bellman filter in Table~\ref{table2} or~\ref{table3}. Expression~\eqref{approximate estimator} can be viewed as an alternative to the prediction-error decomposition for linear Gaussian state-space models (see e.g.\ \citealp[p.\ 126]{harvey1990forecasting}), the advantage being that estimator~\eqref{approximate estimator} remains applicable---albeit as an approximation---outside the classic context of linear Gaussian state-space models. 

\begin{corollary} Take the linear Gaussian state-space model specified in Corollary~\ref{corol1}. Assume that the Kalman-filtered covariance matrices $\{\bm{P}_{t|t}\}$ are positive definite. Estimator~\eqref{approximate estimator} then equals the MLE.
\end{corollary}

Estimator~\eqref{approximate estimator} is only slightly more computationally demanding than static-parameter estimation using the Kalman filter. The sole source of additional computational complexity derives from the fact that the Bellman filter in Table~\ref{table2} or~\ref{table3} may perform several optimisation steps for each time step, while the Kalman filter performs only one. However, because each optimisation step is straightforward and few steps are typically required, the additional computational burden is negligible. 

\section{Simulation studies}
\label{section7}

\subsection{Design}

\noindent This section contains an extensive Monte Carlo study to investigate the performance of the Bellman filter for a range of data-generating processes (DGPs).  I consider $10$ DGPs with linear Gaussian state dynamics~\eqref{DGP0.3}. (The empirical sections~\ref{section8} and~\ref{section9} consider high-dimensional and non-linear state dynamics, respectively.) 
The observation densities for this simulation study are listed in Supplement~\ref{sec:simulation}, which also includes link functions, scores and other quantities used by the Bellman filter. 
To avoid selection bias, these DGPs have  been taken from \citet{koopman2016predicting}. While the numerically accelerated importance-sampling (NAIS) method in~\cite{koopman2015numerically,koopman2016predicting} has been shown to produce highly accurate results, the Bellman filter turns out to be equally (if not more) accurate at a fraction of the computational cost. 

I add one DGP to the nine considered in~\citet{koopman2016predicting}: a local-level model with heavy-tailed observation noise. While a local-level model with additive Gaussian observation noise would be solved exactly by the Kalman filter, the latter does not adjust for heavy-tailed observation noise. Although the Kalman filter remains the best linear unbiased estimator of the state, the results below show that the (nonlinear) Bellman filter fares better.

The static (hyper)parameters for the first nine DGPs are taken from \citet[Table 3]{koopman2016predicting}. In particular, the state-transition equation (i.e.\ $\alpha_{t}=c+T \alpha_{t-1}+\eta_{t}$ with $\eta_{t}\sim \mathrm{N}(0,\sigma_\eta^2)$) has parameters $c=0,T=\phi=0.98$ and $\sigma_\eta=0.15$, except for both dependence models, in which case $c=0.02,T=\phi=0.98$ and $\sigma_\eta=0.10$. In the observation densities (provided in Supplement~\ref{sec:simulation}), the
Student's~\emph{t} distributions have $10$ degrees of freedom, i.e.\ $\nu=10$, except for the local-level model, in which case $\nu=3$. The remaining shape parameters are $\kappa=4$ for the negative binomial distribution, $\kappa=1.5$ for the Gamma distribution, $\kappa=1.2$ for the Weibull distribution and $\sigma=0.45$ for the local-level model.

For each of the $10$ DGPs, I simulate $1{,}000$ time series of length $5{,}000$. I take the first $2{,}500$ observations to represent the `in-sample' period. For the purpose of static-parameter estimation, I use either (a) all $2{,}500$ in-sample observations (long estimation window), (b) the last $1{,}000$ in-sample observations (medium estimation window), or (c) the last $250$ in-sample observations (short estimation window). 
Based on these parameter estimates, I run the Bellman filter and smoother in Table~\ref{table3} on the entire dataset, including the out-of-sample period from $t=2{,}501$ through $t=5{,}000$.  For the Bellman filter, I also produce out-of-sample `smoothed' state estimates $a_{t|n}$ using parameters estimated from in-sample period, but including out-of-sample data for the purpose of smoothing. 

I compute mean absolute errors (MAEs) and root mean squared errors (RMSEs) by comparing filtered and smoothed states against their true (simulated) counterparts.\footnote{The Bellman filter, being based on the mode, is technically suboptimal for both loss functions.} For each DGP and each method, the reported average loss is based on $2{,}500\times 1{,}000=2.5$~million filtered states. I consider five methods:

\begin{enumerate}
\item[1.] {\bf Infeasible mode estimator:} For filtering, I compute the mode using the true static parameters and a moving window of the most recent $250$ observations; hence, $250$ first-order conditions are solved for each time step (larger windows result in excessive computational times). The final state estimate $a_{t|t}$ for each time $t$ represents the filtered state. For smoothing, I use the mode estimator~\eqref{mode estimator} based on the true parameters with $t=n$ (i.e.\ based on the full sample). 

\item[2.] {\bf Bellman filter (BF):} The algorithm in Table~\ref{table3} is initialised using the unconditional distribution. Optimisation steps are performed until the estimated state is stable up to a tolerance of $0.0001$ (on average, ${\sim}5$ iterations are needed). The logarithmic observation density is smooth and concave for the first seven DPGs, in which case optimisation~\eqref{updatingrule v2} is strongly concave; quasi-Newton methods then quickly find the optimum (e.g.\ \citealp{nocedal2006numerical}). For simplicity, I pick Newton's method which proved fast and stable. For the last three DGPs, the logarithmic observation density fails to be concave; in this case, I amend Newton's method by replacing the Hessian of the logarithmic density by a weighted average of the Hessian and its expectation to ensure that the resulting expression is negative with probability one.\footnote{For the dependence model with the Gaussian distribution, the weight placed on the expectation should weakly exceed $1/2$. For the Student's \emph{t} distribution, this generalises to $1/2\times(\nu+4)/(\nu+3)$. For the local-level model with heavy-tailed noise, the weight given to the expectation should weakly exceed $(1+\nu/3)/(1+3\nu)$.}  
For these DGPs, the same weighting scheme ensures $I_{t|t}\geq I_{t|t-1}$ as desired for the static-parameter estimator~\eqref{approximate estimator}. Smoothed states are obtained as stated in Table~\ref{table3}.

\item[3.] {\bf Particle filter (PF):} I follow \citeauthor{malik2011particle}'s (\citeyear{malik2011particle}) implementation of the continuous sampling importance resampling (CSIR) particle filter, as it allows static parameters to be estimated using the same numerical optimisers employed for other methods. Experimentation suggests that using $1{,}000$ particles is necessary to achieve a performance similar to that of the other methods. The seed that controls randomness is fixed beforehand, after which new random variates are drawn for each of the $1{,}000$ times series; variations on this setup make no noticeable difference. The mean and the median of the particles at each time step are stored to compute RMSEs and MAEs, respectively. 

\item[4.] {\bf Numerically accelerated importance sampler (NAIS):} I follow  \cite{koopman2016predicting}, whose code  is available online, deviating slightly by computing not only the weighted mean but also the weighted median of the (simulated) states. The resulting filtered states are used to compute RMSEs and MAEs, respectively. 

\item[5.] {\bf Kalman filter (KF):} I follow \cite{ruiz1994quasi} and \cite{harvey1996estimation}
in using quasi maximum-likelihood estimation (QMLE) to estimate the static parameters of both stochastic-volatility (SV) models. For both SV models, the observations are squared and taking the logarithm produces a linear state-space model, albeit with biased and non-Gaussian observation noise (for details, see~\citealp{ruiz1994quasi} or \citealp{harvey1994multivariate}). For the local-level model with heavy-tailed observation noise, the Kalman filter is applied directly, i.e.\ without adjustments, and estimated by QMLE. For all three models, filtered and smoothed states are obtained, respectively, by the familiar Kalman filter and Rauch, Tung and Striebel smoother.  
\end{enumerate}

\begin{table}[tb]
\center
\caption{\label{table4} Average computing time (in seconds per sample) for parameter estimation and filtering}
\begin{footnotesize}
\renewcommand{\arraystretch}{1}
\begin{threeparttable}
\begin{tabular}{l@{\hspace{0.4cm}}l@{\hspace{1cm}}c@{\hspace{0.2cm}}c@{\hspace{0.2cm}}c@{\hspace{1cm}}c@{\hspace{0.2cm}}c@{\hspace{0.3cm}}c}
  \toprule
\multicolumn{2}{l}{\bf{DGP}}  & \multicolumn{3}{l}{\bf Parameter estimation}  & \multicolumn{3}{l}{\bf Filtering} \\
Type & Distribution  &   PF  &  NAIS  & BF &  NAIS  &  PF  & BF    \\
    \cmidrule(l{5pt}r{5pt}){1-2}  \cmidrule(l{5pt}r{5pt}){3-5} \cmidrule(l{5pt}r{5pt}){6-8}  
Count &	Poisson &	\hphantom{1}51& 1.1 & 0.25   &		4.0	& 0.7 &0.0024\\
Count &	Negative binomial & 146 &	3.1	 & 0.64  &		5.2&	1.0 &   0.0024\\
Intensity	& Exponential	& \hphantom{1}43 & 1.1 &	0.24 & 		3.4 &	0.6 & 0.0022\\
Duration  &	Gamma	 &138 & 3.8 	& 0.55  &	 4.8 &	1.0 & 0.0026\\
Duration & 	Weibull	& 162 & 8.4  &	0.84 & 		9.4&	1.4 & 0.0060\\
Volatility &	Gaussian &	\hphantom{1}48 &	1.3 & 0.28 &		3.7 & 0.7 	& 0.0023\\
Volatility &	Student's \emph{t} &\hphantom{1}95 &	2.7	 &  0.70	& 	5.2& 1.0 &  0.0027 \\
Dependence &	Gaussian & \hphantom{1}69&	2.4  &	0.57 & 		5.5 &	0.8 & 0.0050\\
Dependence	& Student's \emph{t} & 129 &	6.4 &	1.21   &	7.1 & 1.1 & 0.0060\\
Local level	& Student's \emph{t}	& 176 &n/a  & 1.01 & 	n/a &0.9 &	0.0029
\\
\bottomrule
\end{tabular}
\begin{tablenotes}
\item \emph{Note}: BF = Bellman filter. PF = particle filter. NAIS = numerically accelerated importance sampler. Computation times are measured on a computer running 64-bit Windows 8.1 Pro with an Intel(R) Core(TM) i7-4810MQ CPU @ 2.80GHz. Average parameter estimation times are based on the first 2{,}500 observations across 1{,}000 repetitions for each DGP. Average filtering times are based on filtering the entire sample of 5{,}000 observations across 1{,}000 repetitions for each DGP.
\end{tablenotes}
\end{threeparttable}
\end{footnotesize}
\end{table}

\vspace{-0.5cm}
\subsection{Results}

This section compares (a) computational complexity, (b) quality of estimated (hyper)parameters, (c) quality of filtered and (d) smoothed state estimates, and (e) coverage (and length) of predicted, filtered and smoothed confidence intervals.
\begin{enumerate}
\item[a.] \textbf{Computational complexity:}  Table~\ref{table4} shows  average computation times (in seconds per sample) required for parameter estimation (based on the long estimation window) and filtering (based on all data) for three methods (BF, PF and NAIS). The BF is considerably faster than both simulation-based methods for the purposes of both parameter estimation and filtering. Compared to the NAIS method, parameter estimation by the BF is faster by a factor $4$ to $10$, while filtering is faster by a factor between ${\sim}1{,}000$ and ${\sim}2{,}000$. Compared to the PF, parameter estimation by the BF is faster by a factor between ${\sim}100$ and ${\sim}250$, while filtering is faster by a factor between ${\sim}160$ and ${\sim}400$. 

\begin{table}[t!]
\center
\caption{\label{table5}  Average parameter estimates and RMSEs based on the long estimation window}
\begin{footnotesize}
\renewcommand{\arraystretch}{1}
\begin{threeparttable}
\begin{tabular}{
l@{\hspace{0.3cm}} 
l@{\hspace{0.8cm}} 
c@{\hspace{0.2cm}} 
c@{\hspace{0.6cm}} 
r@{\hspace{0.2cm}} 
c@{\hspace{0.6cm}} 
r@{\hspace{0.2cm}} 
c@{\hspace{0.6cm}} 
r@{\hspace{0.2cm}} 
c@{\hspace{0.2cm}} 
}
\toprule
\multicolumn{2}{l}{\bf{DGP}} 	&	\multicolumn{2}{c}{ }	& 	\multicolumn{2}{c}{\bf{BF}$\qquad$} 	&	\multicolumn{2}{c}{\bf{PF}$\qquad$} &	\multicolumn{2}{c}{\bf{NAIS}$\qquad$}		\\ 
Type & Distribution & \multicolumn{2}{c}{Truth $\quad$} & Average & RMSE & Average & RMSE & Average & RMSE  
\\
  \cmidrule(l{0pt}r{5pt}){1-2}   \cmidrule(l{5pt}r{5pt}){3-4}   \cmidrule(l{5pt}r{5pt}){5-6} \cmidrule(l{5pt}r{5pt}){7-8}  \cmidrule(l{5pt}r{0pt}){9-10}
Count&	Poisson&	\multicolumn{1}{c}{$c$}&	$0.000$&	$-0.007$&	﻿ $[0.008]$&	$0.000$&	﻿ $[0.003]$&	$0.000$&	﻿ $[0.003]$	\\ 	
&	&	\multicolumn{1}{c}{$\phi$}&	$0.980$&	$0.977$&	﻿ $[0.007]$&	$0.978$&	﻿ $[0.006]$&	$0.978$&	﻿ $[0.006]$	\\ 	
&	&	\multicolumn{1}{c}{$\sigma_\eta$}&	$0.150$&	$0.153$&	﻿ $[0.014]$&	$0.152$&	﻿ $[0.014]$&	$0.149$&	﻿ $[0.013]$	\\ 	\midrule
Count&	Negative Bin.&	\multicolumn{1}{c}{$c$}&	$0.000$&	$-0.004$&	﻿ $[0.005]$&	$0.000$&	﻿ $[0.003]$&	$0.000$&	﻿ $[0.003]$	\\ 	
&	&	\multicolumn{1}{c}{$\phi$}&	$0.980$&	$0.979$&	﻿ $[0.006]$&	$0.977$&	﻿ $[0.007]$&	$0.979$&	﻿ $[0.006]$	\\ 	
&	&	\multicolumn{1}{c}{$\sigma_\eta$}&	$0.150$&	$0.149$&	﻿ $[0.015]$&	$0.152$&	﻿ $[0.016]$&	$0.145$&	﻿ $[0.015]$	\\ 	
&	&	\multicolumn{1}{c}{$1/\kappa$}&	$0.250$&	$0.239$&	﻿ $[0.036]$&	$0.248$&	﻿ $[0.031]$&	$0.287$&	﻿ $[0.049]$	\\ 	\midrule
Intensity&	Exponential&	\multicolumn{1}{c}{$c$}&	$0.000$&	$-0.007$&	﻿ $[0.008]$&	$0.000$&	﻿ $[0.003]$&	$0.000$&	﻿ $[0.003]$	\\ 	
&	&	\multicolumn{1}{c}{$\phi$}&	$0.980$&	$0.976$&	﻿ $[0.008]$&	$0.978$&	﻿ $[0.007]$&	$0.978$&	﻿ $[0.007]$	\\ 	
&	&	\multicolumn{1}{c}{$\sigma_\eta$}&	$0.150$&	$0.158$&	﻿ $[0.017]$&	$0.151$&	﻿ $[0.014]$&	$0.151$&	﻿ $[0.014]$	\\ 	\midrule
Duration&	Gamma&	\multicolumn{1}{c}{$c$}&	$0.000$&	$0.007$&	﻿ $[0.008]$&	$0.000$&	﻿ $[0.004]$&	$0.000$&	﻿ $[0.004]$	\\ 	
&	&	\multicolumn{1}{c}{$\phi$}&	$0.980$&	$0.976$&	﻿ $[0.007]$&	$0.977$&	﻿ $[0.006]$&	$0.977$&	﻿ $[0.006]$	\\ 	
&	&	\multicolumn{1}{c}{$\sigma_\eta$}&	$0.150$&	$0.158$&	﻿ $[0.015]$&	$0.152$&	﻿ $[0.013]$&	$0.152$&	﻿ $[0.013]$	\\ 	
&	&	\multicolumn{1}{c}{$\kappa$}&	$1.500$&	$1.507$&	﻿ $[0.043]$&	$1.501$&	﻿ $[0.043]$&	$1.501$&	﻿ $[0.043]$	\\ 	\midrule
Duration&	Weibull&	\multicolumn{1}{c}{$c$}&	$0.000$&	$0.009$&	﻿ $[0.010]$&	$0.000$&	﻿ $[0.003]$&	$0.000$&	﻿ $[0.003]$	\\ 	
&	&	\multicolumn{1}{c}{$\phi$}&	$0.980$&	$0.975$&	﻿ $[0.008]$&	$0.978$&	﻿ $[0.006]$&	$0.978$&	﻿ $[0.006]$	\\ 	
&	&	\multicolumn{1}{c}{$\sigma_\eta$}&	$0.150$&	$0.160$&	﻿ $[0.018]$&	$0.152$&	﻿ $[0.013]$&	$0.152$&	﻿ $[0.013]$	\\ 	
&	&	\multicolumn{1}{c}{$\kappa$}&	$1.200$&	$1.207$&	﻿ $[0.023]$&	$1.200$&	﻿ $[0.021]$&	$1.200$&	﻿ $[0.021]$	\\ 	\midrule
Volatility&	Gaussian&	\multicolumn{1}{c}{$c$}&	$0.000$&	$0.007$&	﻿ $[0.008]$&	$0.000$&	﻿ $[0.004]$&	$0.000$&	﻿ $[0.004]$	\\ 	
&	&	\multicolumn{1}{c}{$\phi$}&	$0.980$&	$0.975$&	﻿ $[0.010]$&	$0.977$&	﻿ $[0.008]$&	$0.977$&	﻿ $[0.008]$	\\ 	
&	&	\multicolumn{1}{c}{$\sigma_\eta$}&	$0.150$&	$0.166$&	﻿ $[0.026]$&	$0.152$&	﻿ $[0.018]$&	$0.152$&	﻿ $[0.018]$	\\ 	\midrule
Volatility&	Student's \emph{t}&	\multicolumn{1}{c}{$c$}&	$0.000$&	$0.005$&	﻿ $[0.006]$&	$0.000$&	﻿ $[0.004]$&	$0.000$&	﻿ $[0.004]$	\\ 	
&	&	\multicolumn{1}{c}{$\phi$}&	$0.980$&	$0.975$&	﻿ $[0.010]$&	$0.977$&	﻿ $[0.008]$&	$0.977$&	﻿ $[0.008]$	\\ 	
&	&	\multicolumn{1}{c}{$\sigma_\eta$}&	$0.150$&	$0.162$&	﻿ $[0.031]$&	$0.153$&	﻿ $[0.021]$&	$0.153$&	﻿ $[0.022]$	\\ 	
&	&	\multicolumn{1}{c}{$1/\nu$}&	$0.100$&	$0.089$&	﻿ $[0.030]$&	$0.100$&	﻿ $[0.010]$&	$0.097$&	﻿ $[0.023]$	\\ 	\midrule
Dependence&	Gaussian&	\multicolumn{1}{c}{$c$}&	$0.020$&	$0.021$&	﻿ $[0.009]$&	$0.024$&	﻿ $[0.011]$&	$0.024$&	﻿ $[0.011]$	\\ 	
&	&	\multicolumn{1}{c}{$\phi$}&	$0.980$&	$0.979$&	﻿ $[0.008]$&	$0.977$&	﻿ $[0.010]$&	$0.977$&	﻿ $[0.010]$	\\ 	
&	&	\multicolumn{1}{c}{$\sigma_\eta$}&	$0.100$&	$0.095$&	﻿ $[0.020]$&	$0.103$&	﻿ $[0.024]$&	$0.103$&	﻿ $[0.024]$	\\ 	\midrule
Dependence&	Student's \emph{t}&	\multicolumn{1}{c}{$c$}&	$0.020$&	$0.022$&	﻿ $[0.010]$&	$0.025$&	﻿ $[0.013]$&	$0.025$&	﻿ $[0.014]$	\\ 	
&	&	\multicolumn{1}{c}{$\phi$}&	$0.980$&	$0.977$&	﻿ $[0.010]$&	$0.975$&	﻿ $[0.013]$&	$0.975$&	﻿ $[0.014]$	\\ 	
&	&	\multicolumn{1}{c}{$\sigma_\eta$}&	$0.100$&	$0.098$&	﻿ $[0.023]$&	$0.106$&	﻿ $[0.029]$&	$0.107$&	﻿ $[0.030]$	\\ 	
&	&	\multicolumn{1}{c}{$1/\nu$}&	$0.100$&	$0.103$&	﻿ $[0.012]$&	$0.100$&	﻿ $[0.006]$&	$0.098$&	﻿ $[0.025]$	\\ 	\midrule
Level&	Student's \emph{t}&	\multicolumn{1}{c}{$c$}&	$0.000$&	$0.000$&	﻿ $[0.004]$&	$0.000$&	﻿ $[0.003]$&	&		\\ 	
&	&	\multicolumn{1}{c}{$\phi$}&	$0.980$&	$0.979$&	﻿ $[0.005]$&	$0.978$&	﻿ $[0.005]$&	&		\\ 	
&	&	\multicolumn{1}{c}{$\sigma_\eta$}&	$0.150$&	$0.139$&	﻿ $[0.013]$&	$0.151$&	﻿ $[0.008]$&	&		\\ 	
&	&	\multicolumn{1}{c}{$\sigma$}&	$0.450$&	$0.453$&	﻿ $[0.025]$&	$0.451$&	﻿ $[0.027]$&	&		\\ 	
&	&	\multicolumn{1}{c}{$1/\nu$}&	$0.333$&	$0.277$&	﻿ $[0.066]$&	$0.332$&	﻿ $[0.024]$&	&		\\ 	
\bottomrule
 \end{tabular}
\begin{tablenotes}
\item \emph{Note}: BF = Bellman  filter. PF = Particle filter. NAIS = Numerically accelerated importance sampler. RMSE = root mean squared error. I simulated $1{,}000$ time series each of length $5{,}000$ for $10$ data-generating processes with linear Gaussian state dynamics~\eqref{DGP0.3}, i.e.\ $\smash{\alpha_{t+1}=c+\phi \alpha_t + \eta_{t+1}}$ with $\smash{\eta_{t+1}\sim \mathrm{N}(0,\sigma_\eta^2)}$. The observation densities are listed in Supplement~\ref{sec:simulation}. 
The estimation of static parameters is based on the long estimation window, which consists of $2{,}500$ observations. Parameter estimation is performed as follows: Bellman filter: based on estimator~\eqref{approximate estimator}; Particle filter: as in \cite{malik2011particle}; Importance sampler: as in \cite{koopman2015numerically,koopman2016predicting}.
\end{tablenotes}
\end{threeparttable}
\end{footnotesize}
\end{table}

\item[b.] \textbf{(Hyper)parameter estimates:} Table~\ref{table5} displays average (hyper)parameter estimates and root mean squared errors (RMSEs) versus the true parameters for three methods (BF, PF and NAIS) for  the long estimation window. Parameter estimates for the short and medium windows are presented in Supplement~\ref{S:parameter}.  The BF is about as accurate as both simulation-based methods for all three window sizes in terms of both average parameters and RMSEs relative to the true parameters. The average parameters are close to the true values and tend to be drawn even closer as the estimation window is increased, while the RMSEs decrease rapidly. These simulation results suggest that, for these models and sample sizes, any potential bias or loss of efficiency compared to the simulation-based methods under investigation is negligible.

\begin{table}[t!]
\center
\caption{\label{table6} MAEs of filtered states in out-of-sample period}
\renewcommand{\arraystretch}{1}
\begin{footnotesize}
\begin{threeparttable}
\begin{tabular}{
l@{\hspace{0.1cm}} 
l@{\hspace{0.1cm}} 
c@{\hspace{0.1cm}} 
c@{\hspace{0.2cm}} 
c@{\hspace{0.2cm}} 
c@{\hspace{0.2cm}} 
c@{\hspace{0.2cm}} 
c@{\hspace{0.2cm}} 
c@{\hspace{0.2cm}} 
c@{\hspace{0.2cm}} 
c@{\hspace{0.2cm}} 
c@{\hspace{0.2cm}} 
c@{\hspace{0.2cm}} 
c@{\hspace{0.2cm}} 
c 
}
\toprule
& & & 	\multicolumn{4}{c}{\bf Short estimation} & 	\multicolumn{4}{c}{\bf  Medium estimation }& 	\multicolumn{4}{c}{\bf  Long estimation}
\\
 & &	 {\bf Infeasible} & 	\multicolumn{4}{c}{\bf  window (250 obs.)} & 	\multicolumn{4}{c}{\bf  window (1{,}000 obs.) }& 	\multicolumn{4}{c}{\bf  window (2{,}500 obs.)}
\\
{\bf DGP} & & {\bf estimator}&BF &	PF &	NAIS &	KF &	BF &	PF &	NAIS &	KF &	BF &	PF &	NAIS &	KF
\\
\cmidrule(r{5pt}l{5pt}){3-3}
\cmidrule(r{5pt}l{5pt}){4-7}\cmidrule(r{5pt}l{5pt}){8-11}\cmidrule(r{5pt}l{5pt}){12-15}
Type &	Distribution & MAE & \multicolumn{4}{c}{Relative MAE}	& \multicolumn{4}{c}{Relative MAE}	 & \multicolumn{4}{c}{Relative MAE}		\\
\cmidrule(r{5pt}l{5pt}){1-2} \cmidrule(r{5pt}l{5pt}){3-3}\cmidrule(r{5pt}l{5pt}){4-7}\cmidrule(r{5pt}l{5pt}){8-11}\cmidrule(r{5pt}l{5pt}){12-15}
Count & 	Poisson &	$0.283$&	$1.145$&	$1.141$&	$1.140$&	&	$1.015$&	$1.015$&	$1.016$&	&	$1.001$&	$1.002$&	$1.003$&		\\
Count &	Neg. Bin.  &	$0.300$&	$1.159$&	$1.154$&	$1.155$&	&	$1.018$&	$1.019$&	$1.020$&	&	$1.005$&	$1.006$&	$1.007$&		\\
Intensity  &	Exponential  &	$0.286$&	$1.128$&	$1.130$&	$1.128$&	&	$1.013$&	$1.014$&	$1.014$&	&	$1.002$&	$1.003$&	$1.003$&		\\
Duration  &	Gamma  &	$0.259$&	$1.158$&	$1.156$&	$1.154$&	&	$1.023$&	$1.024$&	$1.023$&	&	$1.007$&	$1.007$&	$1.007$&		\\
Duration  &	Weibull  &	$0.264$&	$1.117$&	$1.115$&	$1.114$&	&	$1.012$&	$1.012$&	$1.012$&	&	$1.001$&	$1.001$&	$1.001$&		\\
Volatility  &	Gaussian  &	$0.337$&	$1.198$&	$1.200$&	$1.200$&	$1.473$&	$1.023$&	$1.023$&	$1.023$&	$1.230$&	$1.005$&	$1.005$&	$1.005$&	$1.230$	\\
Volatility  &	Student's \emph{t}  &	$0.352$&	$1.231$&	$1.213$&	$1.217$&	$1.574$&	$1.038$&	$1.029$&	$1.030$&	$1.336$&	$1.012$&	$1.009$&	$1.010$&	$1.275$	\\
Dependence  &	Gaussian  &	$0.288$&	$1.291$&	$1.296$&	$1.290$&	&	$1.056$&	$1.056$&	$1.055$&	&	$1.018$&	$1.016$&	$1.016$&		\\
Dependence  &	Student's \emph{t}  &	$0.295$&	$1.301$&	$1.313$&	$1.291$&	&	$1.063$&	$1.065$&	$1.067$&	&	$1.022$&	$1.022$&	$1.022$&		\\
Level  &	Student's \emph{t}  &	$0.159$&	$1.059$&	$1.045$&	 &	$1.196$&	$1.014$&	$1.004$&	 &	$1.128$&	$1.003$&	$1.000$&	 &	$1.122$	\\\bottomrule
 \end{tabular}
\begin{tablenotes}
\item \emph{Note}: MAE = mean absolute error. BF = Bellman  filter. PF = particle filter. NAIS = numerically accelerated importance sampler. KF = Kalman filter. I simulated $1{,}000$ time series each of length $5{,}000$ for $10$  data-generating processes of type~\eqref{DGP0.3}; the observation densities are listed in  Supplement~\ref{sec:simulation}. The data is split in an `in-sample' period (first $2{,}500$ observations) and an `out-of-sample' period (last $2{,}500$ observations). The short, medium and long estimation windows consist of the $250$, $1{,}000$ or $2{,}500$ observations, respectively, of the in-sample period. 
Filtered states based on simulation-based methods (importance sampler and particle filter) are computed by taking the median of the simulated states. In all cases, MAEs are computed by comparing  the last $2{,}500$ filtered states with their true (simulated) counterparts. MAEs are reported relative to the MAE of the infeasible mode estimator.
\end{tablenotes}
\end{threeparttable}
\end{footnotesize}
\end{table}

\item[c.] \textbf{Filtered state estimates:}  Table~\ref{table6} shows mean absolute errors (MAEs) of filtered states in the out-of-sample period, reported relative to the MAEs of the infeasible mode estimator, for four methods: BF, PF, NAIS and KF.  The infeasible estimator uses true parameters and the same information set as the filtering methods. The main finding is that the BF,  PF and NAIS perform near identically, while the KF, when applicable, lags substantially behind.\footnote{This difference is not due to the choice of loss function; the relative performance of the KF deteriorates further when reporting RMSEs (see Supplement~\ref{app:F}).} The out-of-sample performance of the BF based on the long estimation window falls within ${\sim}2\%$ of that of the infeasible state estimator across all DGPs. For this estimation window, the BF marginally outperforms the PF and NAIS for three DGPs (for the Poisson, negative binomial and exponential distributions). It performs on par with both these methods for four DGPs (with the Gamma/Weibull distributions and for the Gaussian volatility and Student's \emph{t} dependence models), but is marginally outperformed for three DGPs (for the Student's \emph{t} volatility, Gaussian dependence and local-level models), albeit by max ${\sim}0.3\%$. Filtering results deteriorate by a few percentage points for the medium estimation window, and by ${\sim}10{-}30\%$ for the short estimation window, in paricular for both dependence models. Even for the short estimation window, the results for the BF, PF and NAIS are virtually identical with the KF lagging behind. 
The robustness of the BF means that it compares favourably with the KF for both the SV and local-level models: e.g.\ for the local-level model, the maximum absolute error in the out-of-sample period, averaged across $1{,}000$ samples, is $1.80$ for the KF; double that for the BF ($0.90$). The BF is thus more robust in the face of heavy-tailed observation noise, while having only a single additional parameter to estimate (the degrees of freedom of the observation noise, $\nu$).

\begin{table}[t!]
\center
\caption{\label{table7} MAEs of smoothed states in out-of-sample period}
\begin{footnotesize}
\renewcommand{\arraystretch}{1}
\begin{threeparttable}
\begin{tabular}{
l@{\hspace{0.2cm}} 
l@{\hspace{0.2cm}} 
c@{\hspace{0.2cm}} 
c@{\hspace{0.1cm}} 
c@{\hspace{0.2cm}} 
r@{\hspace{0.2cm}} 
c@{\hspace{0.2cm}} 
c@{\hspace{0.2cm}} 
c 
}
\toprule
& & & 	\multicolumn{2}{c}{\bf Short estimation} & 	\multicolumn{2}{c}{\bf  Medium estimation }& 	\multicolumn{2}{c}{\bf  Long estimation}
\\
 & &	 {\bf Infeasible} & 	\multicolumn{2}{c}{\bf  window (250 obs.)} & 	\multicolumn{2}{c}{\bf  window (1{,}000 obs.) }& 	\multicolumn{2}{c}{\bf  window (2{,}500 obs.)}
\\
{\bf DGP} & & {\bf estimator}&BF &	KF &BF &	KF &	BF &	KF 
\\
\cmidrule(r{5pt}l{5pt}){3-3}
\cmidrule(r{5pt}l{5pt}){4-5}\cmidrule(r{5pt}l{5pt}){6-7}\cmidrule(r{5pt}l{5pt}){8-9}
Type &	Distribution & MAE & \multicolumn{2}{c}{Relative MAE}	& \multicolumn{2}{c}{Relative MAE}	 & \multicolumn{2}{c}{Relative MAE}		 \\
\cmidrule(r{5pt}l{5pt}){1-2}
\cmidrule(r{5pt}l{5pt}){3-3}
\cmidrule(r{5pt}l{5pt}){4-5}\cmidrule(r{5pt}l{5pt}){6-7}\cmidrule(r{5pt}l{5pt}){8-9}
Count & 	Poisson &	0.222	&	1.118	&		&	1.020	&		&	1.013	&			\\
Count &	Neg. Bin.  &	0.236	&	1.139	&		&	1.018	&		&	1.009	&			\\
Intensity  &	Exponential  &	0.222	&	1.099	&		&	1.021	&		&	1.016	&				\\
Duration  &	Gamma  &	0.201	&	1.168	&		&	1.040	&		&	1.024	&			\\
Duration  &	Weibull  &	0.204	&	1.096	&		&	1.026	&		&	1.021	&			\\
Volatility  &	Gaussian  &	0.266	&	1.196	&	1.628	&	1.033	&	1.259	&	1.022	&	1.221		\\
Volatility  &	Student's \emph{t}  &	0.280	&	1.247	&	2.156	&	1.047	&	1.433	&	1.024	&	1.366			\\
Dependence  &	Gaussian  &	0.240	&	1.359	&		&	1.056	&		&	1.018	&			\\
Dependence  &	Student's \emph{t}  &	0.247	&	1.379	&		&	1.064	&		&	1.021	&				\\
Level  &	Student's \emph{t}  &	0.126	&	1.035	&	1.154	&	1.017	&	1.131	&	1.015	&	1.129			\\
\bottomrule
 \end{tabular}
\begin{tablenotes}
\item \emph{Note}: For the simulation setting, see the note to Table~\ref{table6}.  For the SV models, the static parameters in the Kalman filter are estimated by QMLE as in \cite{ruiz1994quasi}, after which the RTS smoother is applied (\citealp{rauch1965maximum}). MAEs are reported relative to the MAE of the infeasible estimator~\eqref{mode estimator}.
\end{tablenotes}
\end{threeparttable}
\end{footnotesize}
\end{table}

\item[d.] \textbf{Smoothed state estimates:} Table~\ref{table7} shows the MAEs of smoothed states in the out-of-sample period obtained by the Bellman filter/smoother combination in Table~\ref{table3}, where the static parameters are estimated  based on three different in-sample estimation windows. The results are reported relative to those of the infeasible state estimator~\eqref{mode estimator} with $t=n$, which similarly exploits all data and uses the true parameters. Where appropriate, results are also reported for the Kalman filter/smoother. The performance of the Bellman filter/smoother using the long estimation window lies within ${\sim}2\%$ of that of the infeasible state estimator across all DGPs. The performance compared with the filtering results in Table~\ref{table6} is improved by ${\sim}20\%$. This shows that smoothing has substantial benefits, which  the Bellman filter/smoother successfully exploits. The KF smoothing results are comparatively poor, especially for the short estimation window. Neither \cite{malik2011particle} nor \cite{koopman2016predicting} present smoothing methods; hence, no PF or NAIS smoothing results are reported.

\begin{table}[tb]
\center
\caption{\label{table8} Coverage (in $\%$) and  average length (in square brackets) of Bellman-predicted, -filtered and -smoothed confidence intervals for different parameter-estimation windows}
\begin{footnotesize}
\renewcommand{\arraystretch}{1}
\begin{threeparttable}
\begin{tabular}{l@{\hspace{0.2cm}}l@{\hspace{0.4cm}}c@{\hspace{0.2cm}}c@{\hspace{0.2cm}}c@{\hspace{0.4cm}}c@{\hspace{0.2cm}}c@{\hspace{0.2cm}}c@{\hspace{0.4cm}}c@{\hspace{0.2cm}}c@{\hspace{0.2cm}}c}
\toprule
& &  	\multicolumn{3}{c}{\bf Short estimation} & 	\multicolumn{3}{c}{\bf  Medium estimation }& 	\multicolumn{3}{c}{\bf  Long estimation}
\\
\multicolumn{2}{l}{\bf  DGP }&	  	\multicolumn{3}{c}{\bf  window (250 obs.)} & 	\multicolumn{3}{c}{\bf  window (1{,}000 obs.) }& 	\multicolumn{3}{c}{\bf  window (2{,}500 obs.)}
\\
Type & Distribution & Predict & Filter & Smooth 
 & Predict & Filter & Smooth & Predict & Filter & Smooth \\
\cmidrule(r{5pt}l{5pt}){1-2}  \cmidrule(r{5pt}l{5pt}){3-5} \cmidrule(r{5pt}l{5pt}){6-8}  \cmidrule(r{5pt}l{5pt}){9-11}
Count & 	Poisson &	$90.2$&	$90.6$&	$92.5$&	$94.7$&	$94.8$&	$94.7$&	$95.2$&	$95.3$&	$94.9$	\\
&	&	$[1.52]$&	$[1.41]$&	$[1.17]$&	$[1.51]$&	$[1.41]$&	$[1.11]$&	$[1.51]$&	$[1.41]$&	$[1.11]$	\\
Count &	Neg. Bin.  &	$89.5$&	$89.7$&	$91.7$&	$94.3$&	$94.3$&	$94.3$&	$94.9$&	$94.9$&	$94.6$	\\
&	&	$[1.61]$&	$[1.50]$&	$[1.24]$&	$[1.57]$&	$[1.48]$&	$[1.16]$&	$[1.57]$&	$[1.48]$&	$[1.16]$	\\
Intensity  &	Exponential  &	$90.8$&	$91.1$&	$93.4$&	$95.4$&	$95.4$&	$95.5$&	$95.8$&	$95.8$&	$95.5$	\\
&	&	$[1.56]$&	$[1.46]$&	$[1.20]$&	$[1.57]$&	$[1.47]$&	$[1.16]$&	$[1.57]$&	$[1.47]$&	$[1.15]$	\\
Duration  &	Gamma  &	$90.8$&	$90.9$&	$92.1$&	$95.2$&	$95.2$&	$94.9$&	$95.7$&	$95.7$&	$95.3$	\\
&	&	$[1.43]$&	$[1.31]$&	$[1.06]$&	$[1.44]$&	$[1.32]$&	$[1.04]$&	$[1.44]$&	$[1.33]$&	$[1.03]$	\\
Duration  &	Weibull  &	$92.4$&	$92.6$&	$94.3$&	$95.6$&	$95.6$&	$95.5$&	$96.0$&	$95.9$&	$95.5$	\\
&	&	$[1.50]$&	$[1.37]$&	$[1.12]$&	$[1.48]$&	$[1.36]$&	$[1.07]$&	$[1.48]$&	$[1.36]$&	$[1.06]$	\\
Volatility  &	Gaussian  &	$88.1$&	$88.4$&	$90.8$&	$95.3$&	$95.3$&	$95.5$&	$96.1$&	$96.0$&	$95.8$	\\
&	&	$[1.81]$&	$[1.73]$&	$[1.47]$&	$[1.84]$&	$[1.76]$&	$[1.42]$&	$[1.84]$&	$[1.77]$&	$[1.41]$	\\
Volatility  &	Student's \emph{t}  &	$88.4$&	$88.4$&	$90.5$&	$94.5$&	$94.5$&	$94.7$&	$95.4$&	$95.3$&	$95.2$	\\
&	&	$[1.98]$&	$[1.87]$&	$[1.61]$&	$[1.88]$&	$[1.81]$&	$[1.46]$&	$[1.87]$&	$[1.80]$&	$[1.44]$	\\
Dependence  &	Gaussian  &	$73.9$&	$74.0$&	$75.7$&	$90.5$&	$90.6$&	$91.2$&	$93.1$&	$93.1$&	$93.1$	\\
&	&	$[1.26]$&	$[1.23]$&	$[1.10]$&	$[1.37]$&	$[1.34]$&	$[1.14]$&	$[1.39]$&	$[1.36]$&	$[1.14]$	\\
Dependence  &	Student's \emph{t}  &	$71.9$&	$71.9$&	$73.5$&	$90.4$&	$90.4$&	$91.2$&	$93.0$&	$93.1$&	$93.4$	\\
&	&	$[1.28]$&	$[1.25]$&	$[1.13]$&	$[1.42]$&	$[1.40]$&	$[1.20]$&	$[1.43]$&	$[1.41]$&	$[1.19]$	\\
Level  &	Student's \emph{t}  &	$93.1$&	$93.5$&	$94.7$&	$94.9$&	$95.0$&	$95.2$&	$95.1$&	$95.1$&	$95.3$	\\
&	&	$[0.98]$&	$[0.80]$&	$[0.65]$&	$[0.99]$&	$[0.81]$&	$[0.64]$&	$[0.99]$&	$[0.81]$&	$[0.64]$	\\
\bottomrule
 \end{tabular}
\begin{tablenotes}
\item \emph{Note}: For the simulation setting, see the note to Table~\ref{table6}.
\end{tablenotes}
\end{threeparttable}
\end{footnotesize}
\end{table}

\item[e.] \textbf{Coverage of confidence intervals:} Table~\ref{table8} shows the coverage of approximate Bellman-predicted, -filtered and -smoothed confidence intervals with endpoints given by $\smash{a_{t|t-1}\pm 2/\sqrt{I_{t|t-1}}}$, $\smash{a_{t|t}\pm 2/\sqrt{I_{t|t}}}$ and $\smash{a_{t|n}\pm 2/\sqrt{I_{t|n}}}$, respectively, as well as the average length of these intervals, where the estimation of static parameters is based on three possible window sizes. These confidence intervals are based on the quadratic approximation of the value function and are analogous to those in the Kalman filter. For brevity, both simulation-based approaches are excluded. The Bellman-predicted, -filtered and -smoothed confidence intervals based on the medium and long estimation windows tend to be fairly accurate, containing the true states ${\sim} 93- 96\%$ of the time for most DGPs and ${\sim}90{-}96\%$ for both dependence models. Confidence intervals based on the short estimation window tend to be overly optimistic, especially for the two dependence models. Finally, the length of confidence intervals based on the smoothed states is substantially reduced, while the coverage remains good for the medium and long estimation windows, further highlighting the benefits of smoothing.
\end{enumerate}

\section{Application I: High-dimensional state space}
\label{section8}

This section considers the modelling of high-dimensional cloud-intensity data from 
a regional climate model as in \citet{katzfuss2020ensemble}. In a simulation study with realistic parameter values, I demonstrate that the performance of the Bellman filter is unaffected as the dimension of the state increases from $10$ to $150$, while the performance of the standard (bootstrap) particle filter deteriorates sharply---even when using very many particles. When predicting real data, I show that the Bellman filter substantially outperforms the particle-ensemble Kalman filter in \cite{katzfuss2020ensemble} and the exact approximation of the Rao-Blackwellised particle filter in \cite{johansen2012exact}.

\subsection{Model}

Following \citet[p.\ 868]{katzfuss2020ensemble}, I consider a multivariate overdispersed Poisson density that generates an integer number of clouds recorded at adjacent locations over a period of time, in combination with a linear Gaussian state equation for the logarithmic cloud intensities. The model for $t=1,\ldots,n$ reads
\begin{align}
\bm{y}_t &\sim \text{Poisson}(\exp(\bm{\beta_t})), && \bm{y}_t\in \mathbb{N}^m,\bm{\beta}_t\in \mathbb{R}^m,
\label{katzfuss1}
\\
\bm{\beta}_{t}&= \bm{\alpha}_{t}\, + \,\bm{\xi}_{t}, && \bm{\xi}_t \sim \text{i.i.d.}\,\mathrm{N}(\bm{0}_m,\sigma_{\xi}^2 \mathds{1}_{m\times m}),
\label{katzfuss2}
\\
\bm{\alpha}_t & =(\mathds{1}_{m\times m}-\bm{T})\,  \bm{c}\,+\, \bm{T} \bm{\alpha}_{t-1}\, +\, \bm{\eta}_t, && \bm{\eta}_t  \sim \text{i.i.d.}\,\mathrm{N}(\bm{0}_m,\bm{Q}),
\label{katzfuss3}
\end{align}
where $\bm{\alpha}_t\in \mathbb{R}^m$ is the latent state, $\bm{\beta}_t\in \mathbb{R}^m$ is an overdispersed (i.e.\ noisy) realisation of $\bm{\alpha}_t$ with overdispersion parameter $\sigma_{\xi}\geq 0$, and $\bm{y}_t\in \mathbb{N}^m$ is a vector of $m$ Poisson-generated counts with corresponding intensities $\exp(\bm{\beta}_t)$. The exponent of a vector in equation~\eqref{katzfuss1} is understood elementwise, i.e.\ observation $y_{i,t}$ is drawn independently from a Poisson density with intensity $\exp(\beta_{i,t})$ for each $i=1,\ldots,m$. When $\sigma_{\xi}=0$, such that $\bm{\alpha}_t=\bm{\beta}_t$ for all $t$, the model collapses to a standard state-space model with state vector $\bm{\alpha}_t$ of length $m$. For $\sigma_{\xi}>0$, the hierarchical structure~\eqref{katzfuss1}--\eqref{katzfuss3} can be cast in the standard state-space format as I show below, where the dimension of the state is $2m$. Models with $\sigma_{\xi}=0$ and $\sigma_{\xi}>0$ are referred to as the `standard' and `overdispersed' versions of the model, respectively.  

The system vectors and matrices in the state-transition equation are  $\bm{c}\in \mathbb{R}^m$ and $\bm{T},\bm{Q}\in \mathbb{R}^{m\times m}$. Following \cite{katzfuss2020ensemble}, I assume that $\bm{T}$ is  tridiagonal with $\gamma_1$ on the main diagonal, $\gamma_2$ above the main diagonal, and $\gamma_3$ below the main diagonal. Intuitively, these parameters govern the probability of cloud intensities staying in place or drifting left or right. As in \cite{katzfuss2020ensemble}, I assume new cloud formation to be more highly correlated at shorter distances.  Specifically,  the covariance matrix $\bm{Q}$ is assumed to be a spatial Mat{\`e}rn covariance matrix, with a smoothness of $1.5$, spatial dependence parameter $\lambda>0$, and overall scale governed by $\tau>0$, i.e.\ $(\bm{Q})_{ij} = \tau^2 (1+\sqrt{3}|i-j|/\lambda)\exp(-\sqrt{3}|i-j|/\lambda)$ for $i,j=1,\ldots,m$. While \cite{katzfuss2020ensemble} set $\bm{c}=\bm{0}_m$, I consider the more general case $\bm{c}\neq\bm{0}_m$, where $\bm{c}$ can be interpreted as the long-run average of $\bm{\alpha}_t$ if the eigenvalues of $\bm{T}$ lie inside the unit circle. For simplicity I set $\bm{c}=c \mathds{1}_m$, where a single parameter $c\in \mathbb{R}$ controls the overall level. Static parameters are collected in the vector $\bm{\psi}=(c,\gamma_1,\gamma_2,\gamma_3,\tau,\lambda,\sigma_{\xi})'$.

\subsection{State-space formulation and Bellman-filter implementation}

For $\sigma_{\xi}>0$, a standard state-space model can be obtained by writing the dynamics of $\bm{\alpha}_t$ and $\bm{\beta}_t$  jointly as
\begin{align}
\label{katzfuss4}
\begin{bmatrix}
\bm{\beta}_t \\
\bm{\alpha}_{t+1}
\end{bmatrix}
&=
\begin{bmatrix}
\bm{0}_m
\\
(\mathds{1}_{m\times m}-\bm{T})  \bm{c}
\end{bmatrix} 
+
\begin{bmatrix}
\bm{0}_{m\times m} & \mathds{1}_{m\times m}
\\
\bm{0}_{m\times m} & \bm{T} 
\end{bmatrix} 
\begin{bmatrix}
\bm{\beta}_{t-1}\\
\bm{\alpha}_{t}
\end{bmatrix} + \begin{bmatrix} \bm{\xi}_t \\ \bm{\eta}_{t+1} 
\end{bmatrix},
\end{align}
where $\{\bm{\xi}_t\}$ and $\{\bm{\eta}_t\}$ are series of i.i.d.\ disturbances with characteristics specified in equations~\eqref{katzfuss2}--\eqref{katzfuss3}. The state vector in the overdispersed model is $(\bm{\beta}_t',\bm{\alpha}_{t+1}')' \in \mathbb{R}^{2m}$, which is $120$-dimensional when $m=60$ (as in \citealp{katzfuss2020ensemble}). The Bellman filter in Table~\ref{table3} is directly applicable after appropriate redefinitions; e.g.\ $\bm{c}$ in Table~\ref{table3} should be identified with the first vector on the right-hand side of equation~\eqref{katzfuss4}.

The Bellman filter solves a high-dimensional optimisation problem at each time step. The logarithmic Poisson density is jointly concave in all elements of $\bm{\beta}_t$.  The Bellman-filtered state in equation~\eqref{updatingrule v2} then is unique; it can typically be found using e.g.\ Newton steps. To avoid the need for repeated large-matrix inversions, however, I opted for the Broyden-Fletcher-Goldfarb-Shanno (BFGS) algorithm (e.g.\ \citealp[{\S}6.1]{nocedal2006numerical}), which proved both fast and stable. Indeed, at the estimated parameter values, executing the Bellman filter for the standard (overdispersed) model using data from  \citeauthor{katzfuss2020ensemble} (\citeyear{katzfuss2020ensemble}), involving a $60$-dimensional ($120$-dimensional) optimisation problem for each of $80$ time steps, takes about ${\sim}0.25$  (${\sim}0.60$) seconds. In both cases, convergence with a tolerance of $10^{-5}$ at each time step is reached within ${\sim}12$ BFGS optimisation steps.

\subsection{Simulation study with high-dimensional state space}

\begin{figure}
\caption{MAE of filtered states and filtering times (in seconds per sample) }
\begin{threeparttable}
\begin{tabular}{l@{\hspace{1cm}}l}
\subfloat[Mean absolute error]{\includegraphics[width = 7.5cm]{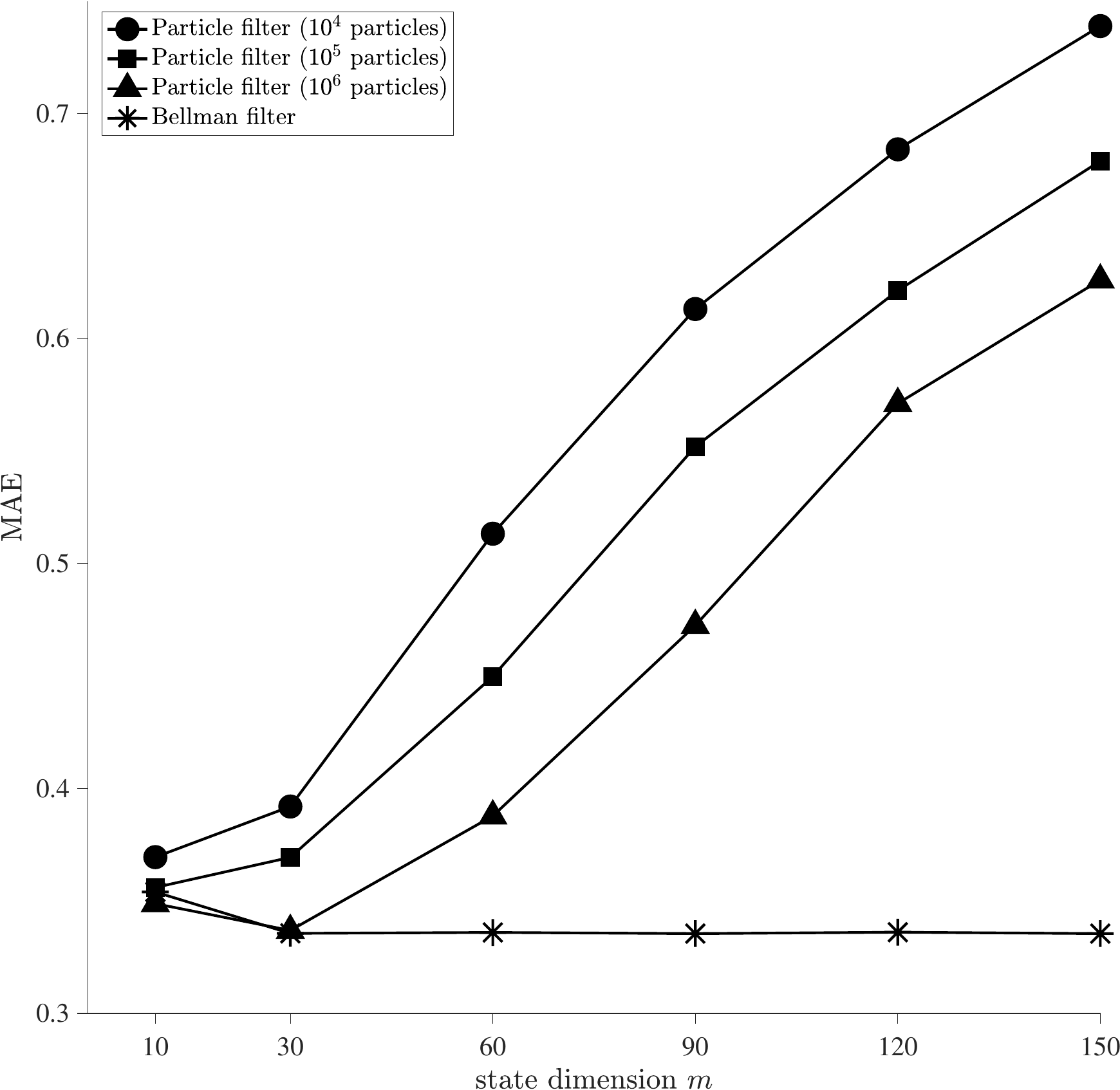}} &
\subfloat[Filtering time (seconds/sample)]{\includegraphics[width = 7.5cm]{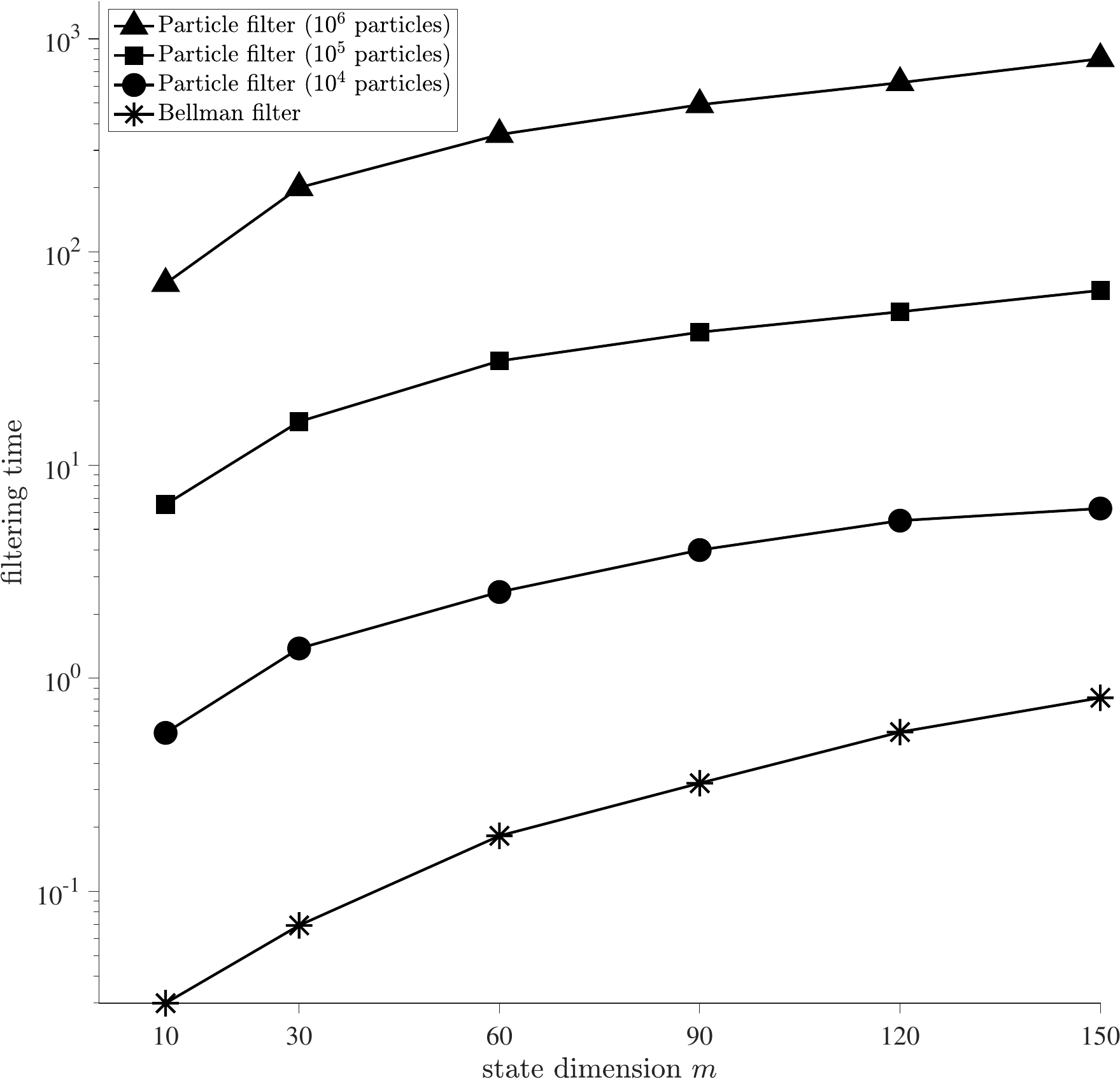}} \\
\end{tabular}
\begin{tablenotes}
\footnotesize
\emph{Note}: MAE = mean absolute error. I simulated $100$ instances of the model~\eqref{katzfuss1}--\eqref{katzfuss3} with $n=80$ time steps and static parameters $\bm{\psi}=(0,0.4,0,0.4,0.8,5,0)'$ for various values of the state dimension $m$. Using the true static parameter for the purpose of filtering, I recorded the MAE of the filtered states $\bm{a}_{t|t}$ relative to the true (simulated) states $\bm{\alpha}_{t}$ and runtime in seconds per sample for the Bellman filter and particle filter, where the latter was implemented with $10^4$,  $10^5$ and $10^6$ particles.
\end{tablenotes}
\end{threeparttable}
\label{fig2}
\end{figure}

This section investigates the performance of the Bellman filter in high-dimensional state spaces by performing a simulation study for the model~\eqref{katzfuss1}--\eqref{katzfuss3} with varying spatial dimension $m$. I compare the Bellman filter's performance against that of the standard (bootstrap) particle filter.  For simplicity, the static parameter $\bm{\psi}$ is considered known and taken as $\bm{\psi}=(c,\gamma_1,\gamma_2,\gamma_3,\tau,\lambda,\sigma_{\xi})'=(0,0.4,0,0.4,0.8,5,0)'$, which is similar to the empirical parameter estimates obtained from real data. As in the real data, the relatively large value of $\gamma_3=0.4$ reflects the fact that logarithmic cloud intensities tend to float from lower to higher location numbers, which may be due to a fixed wind direction during the observation period. The overdispersion parameter $\sigma_{\xi}$ is set to zero, as my empirical study contains no evidence to suggest otherwise. For $\sigma_{\xi}=0$, the state-augmentation procedure~\eqref{katzfuss4} is not required; hence, the dimension of the state space is simply $m$. I investigate cases where $m$ equals $10$, $30$, $60$, $90$, $120$ or $150$, thus exploring different spatial dimensions beyond that of the real data set considered in \cite{katzfuss2020ensemble}, where $m=60$. For each $m$, I simulate $100$ datasets with $80$ time steps, matching the time dimension of the real data. 

The particle filter is subject to the curse of dimensionality and may struggle in higher dimensions (e.g.\ \citealp{surace2019avoid}). Hence, I experiment with $10^4$, $10^5$ and $10^6$ particles;  increasing this number further turns out to be computationally infeasible (see further discussion below). I compute the median of the particles as the filtered state.  For both methods, mean absolute errors (MAEs) of filtered states are computed by taking the one-norm of the vector $\bm{a}_{t|t}-\bm{\alpha}_t\in \mathbb{R}^m$, dividing this norm by $m$, and averaging the resulting quantity across $80$ time steps and $100$ simulated data sets. 

Figure~\ref{fig2} (Panel~A) shows that the MAE of the Bellman filter is almost entirely flat at ${\sim}0.34$, independently of the dimension $m$. In fact, the MAE appears to improve slightly as the dimension $m$ increases,  possibly because the filter benefits from improved predictions: cloud observations even in distant locations may, due to wind conditions, be informative as to the possible future presence of clouds at other locations. In contrast, the MAE of the particle filter increases sharply with $m$ and substantially exceeds that of the Bellman filter even at $m=60$ or $m=90$. This heightened inaccuracy in higher dimensions materialises for any (fixed) number of particles. Even with $10^6$ particles, the particle filter at $m=150$ produces an MAE of ${\sim}0.63$, a factor ${\sim}1.8$ higher than that of the Bellman filter.

Figure~\ref{fig2} (Panel~B) shows that using $10^6$ particles in $m=150$ dimensions necessitates a filtering time of ${\sim}800$ seconds per simulated dataset, such that the total runtime for the particle filter across $100$ simulations is $100\times 800$ seconds $={\sim} 22$ hours. The BFGS implementation of the Bellman filter required between $0.03$ seconds (for $m=10$) and $0.80$ seconds (for $m=150$), translating in the latter case to a total runtime across $100$ simulations of only ${\sim}1.3$ minutes. Panel~B also shows that the computational complexity of the particle filter scales with the number of particles employed: for $10^6$ particles, the difference with the Bellman filter is around three orders of magnitude for any $m$. The relative accuracy and speed of the Bellman filter as demonstrated in this section can largely be attributed to its approach to optimisation, which is simpler than the sampling/integration approach used in the particle filter---especially in higher dimensions.

\subsection{Real-data application with artificially missing data}

For the real-data application, I take the  cloud-motion data investigated by \cite{katzfuss2020ensemble}, which contains $m=60$ locations along a spatial transect (i.e.\ a line), where the number of visible clouds is recorded at each of $n=80$ time steps. Following their procedure, I artificially introduce `missing data' by assuming that at each time step only $90\%$ of the locations, i.e.\ $54$ randomly selected locations, deliver a measurement that the researcher  can use for parameter estimation and state filtering. The remaining $80\times 6=480$ observations are declared `missing', but remain available for testing. For reproducibility, the same missing data  are considered as in \cite{katzfuss2020ensemble}, whose code is available online. The aim is to `nowcast' the (same) missing data by running the Bellman filter on the available data.

\begin{table}[t!]
\center
\caption{\label{table9} Full-sample-with-missing-data parameter estimates for model~\eqref{katzfuss1}--\eqref{katzfuss3}}
\renewcommand{\arraystretch}{1}
\begin{small}
\begin{threeparttable}
\begin{tabular}{l@{\hspace{0.2cm}}r@{\hspace{0.2cm}}r@{\hspace{0.2cm}}r@{\hspace{0.2cm}}r@{\hspace{0.2cm}}r@{\hspace{0.2cm}}r@{\hspace{0.2cm}}r@{\hspace{0.5cm}}r@{\hspace{0.2cm}}r}
\toprule
&  \multicolumn{1}{l}{$c$} 
&   \multicolumn{1}{l}{$\gamma_1$} &  \multicolumn{1}{l}{$\gamma_2$} &  \multicolumn{1}{l}{$\gamma_3$} &    \multicolumn{1}{l}{$\tau$} &  \multicolumn{1}{l}{$\lambda$} &  \multicolumn{1}{l}{$\sigma_{\xi}$} & \multicolumn{1}{l}{MSE} & \multicolumn{1}{l}{CRPS}   \\
     \cmidrule(r{5pt}){1-1}    \cmidrule(r{5pt}){2-8} \cmidrule(r{5pt}){9-10}
           Standard model  
     &  $-3.656\phantom{]}$ &  $0.254\phantom{]} $ &   $0.050\phantom{]}$  &  $ 0.372\phantom{]}    $ &   $ 1.749\phantom{]} $ &  $ 7.040\phantom{]} $  &  & $ 0.513$ & $  0.185$
   \\ 
   &    $[  0.242]$   & $[ 0.053 ]$ &  $[ 0.040  ]$  &   $[ 0.056  ]$   &  $[ 0.100 ]$   &  $[ 0.471  ]$  
  \\ 
 Standard model ($c=0$) &   &$0.260\phantom{]}$  &$0.127\phantom{]}$  &$0.482\phantom{]}$  & $1.771\phantom{]}$	 & $8.295\phantom{]}$ & 	 & 	$0.547$    & $  0.192$       
   \\
&       &  $[0.060]$ &   $[0.047]$  &   $[0.055]$   & $[0.108]$   &  $[0.561]$  &  
\\
\midrule
\\
Overdispersed model & $-4.236\phantom{]}$  &$0.245\phantom{]}$  &$0.055\phantom{]}$  &$0.384\phantom{]}$  & $1.839\phantom{]}  $	 & $7.249\phantom{]} $ & 	$0.000\phantom{]}$ & 	 $      0.509$ & $ 0.185$   
\\ &   $[0.072 ]$   &  $[ 0.025 ]$ &    $[0.033 ]$  &   $[0.027 ]$   &  $[ 0.053   ]$   &  $[   0.053 ]$  &  $[ 0.018]$ &
 \\
Overdispersed model ($c=0$)  &  &$ 0.230\phantom{]}  $  &$0.142\phantom{]} $  &$0.494\phantom{]} $  & $1.791\phantom{]} $	 & $ 8.301\phantom{]} $ & 	$0.000\phantom{]}$ & 	$      0.556$  & $  0.197$ 
   \\&  &  $[0.055  ]$   & $[ 0.045 ]$ &   $[ 0.047  ]$  &  $[ 0.102 ]$   &   $[ 0.346 ]$   &  $[  0.035]$  
 \\
  \bottomrule
\end{tabular}
\begin{tablenotes}
\item \emph{Note}: MSE = mean squared error. CRPS = continuously ranked probability score. The standard model has $\sigma_{\xi}=0$, while the overdispersed model has $\sigma_{\xi}>0$. Numerical standard errors in square brackets are computed by taking the square root of diagonal elements of the inverse of the negative finite-difference Hessian matrix. Using the output of the Bellman filter at times and locations where observations were declared missing, I produce `nowcasts' of missing data, the quality of which can be judged on the basis of MSE and CRPS values in the right-most columns.
\end{tablenotes}
\end{threeparttable}
\end{small}
\end{table}

To implement the Bellman filter with missing data, I write the logarithm of the observation density at time $t$ used in the Bellman-filter update~\eqref{updatingrule v2} as 
\begin{equation}
\label{missing data}
\log \text{Poisson}(\bm{y}_t |\exp(\bm{\beta}_t)) \;=\; \sum_{i \in \mathcal{O}_t}\;  \log \text{Poisson}(y_{i,t} |\exp(\beta_{i,t}) ) ,
\end{equation}
where $\mathcal{O}_t$ is the set of available observations at time $t$; i.e.\ log-likelihood contributions of missing data are excluded. The Bellman filter in Table~\ref{table3} remains applicable as long as the score and (realised) information quantities are computed by taking derivatives of the logarithmic density on the right-hand side of equation~\eqref{missing data}. This implies that elements of the score vector corresponding to missing observations are set to zero. Nevertheless, the Bellman-filtered states at times and locations for which observations are declared missing remain non-trivial, because the filtered state---representing the solution to an optimisation problem---is affected by \emph{all} available observations at a given time step. The Bellman filter in Table~\ref{table3} is initialised with $\bm{I}_{1|0}$ equal to a small multiple of the identity. The static parameter~$\bm{\psi}$ is estimated using the approximate maximum-likelihood estimator~\eqref{approximate estimator}, employing equation~\eqref{missing data} to exclude data declared missing.

\subsection{Results: Full sample with missing data}

\begin{figure}
\caption{Expanding-window parameter estimation results for model~\eqref{katzfuss1}--\eqref{katzfuss3}}
\begin{threeparttable}
\begin{tabular}{l@{\hspace{1cm}}l}
\subfloat[Estimates of $\gamma_1,\gamma_2,\gamma_3$]{\includegraphics[width = 7.5cm]{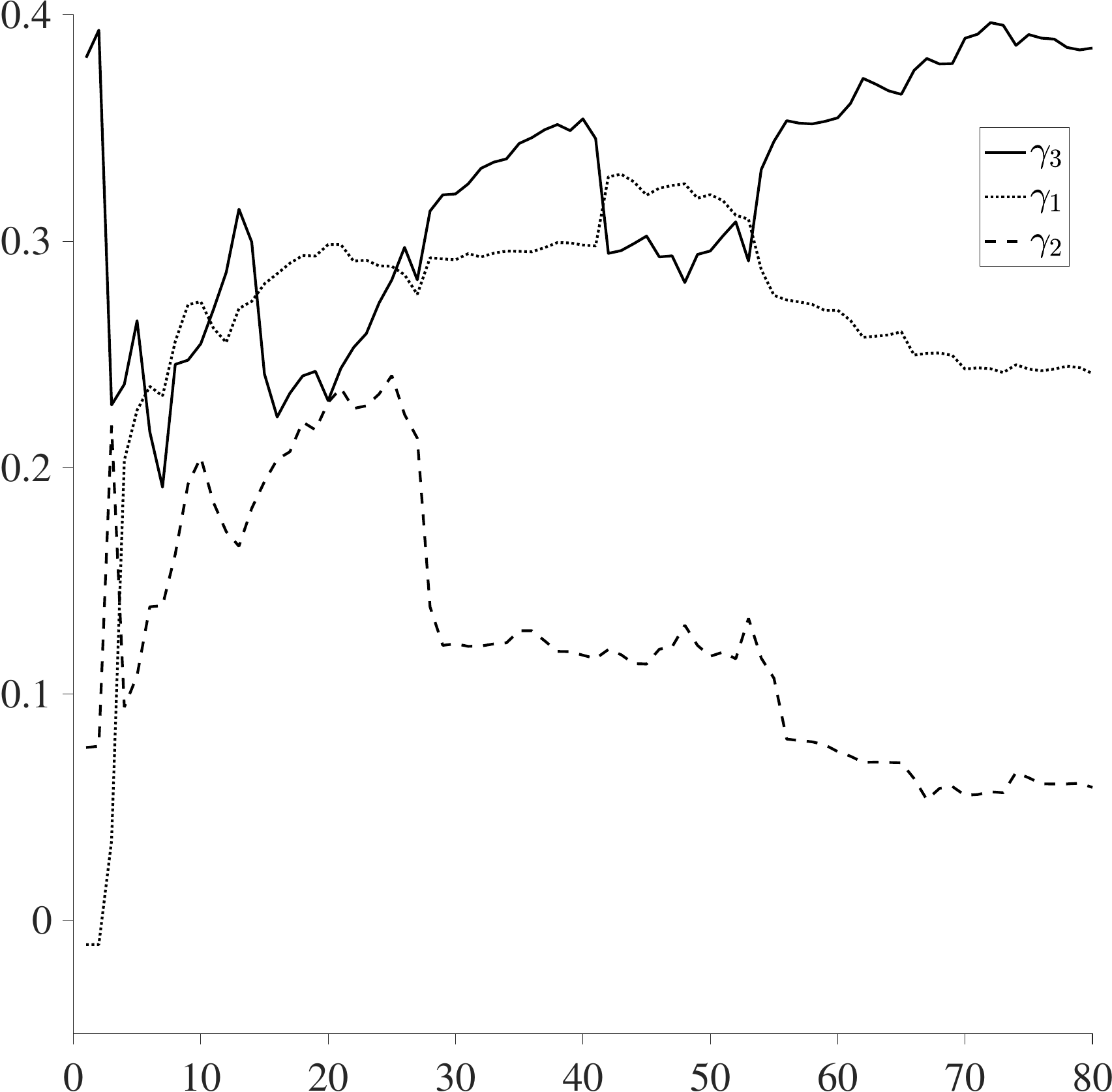}} &
\subfloat[Estimates of $\lambda,\tau,\sigma_{\xi},c$]{\includegraphics[width = 7.5cm]{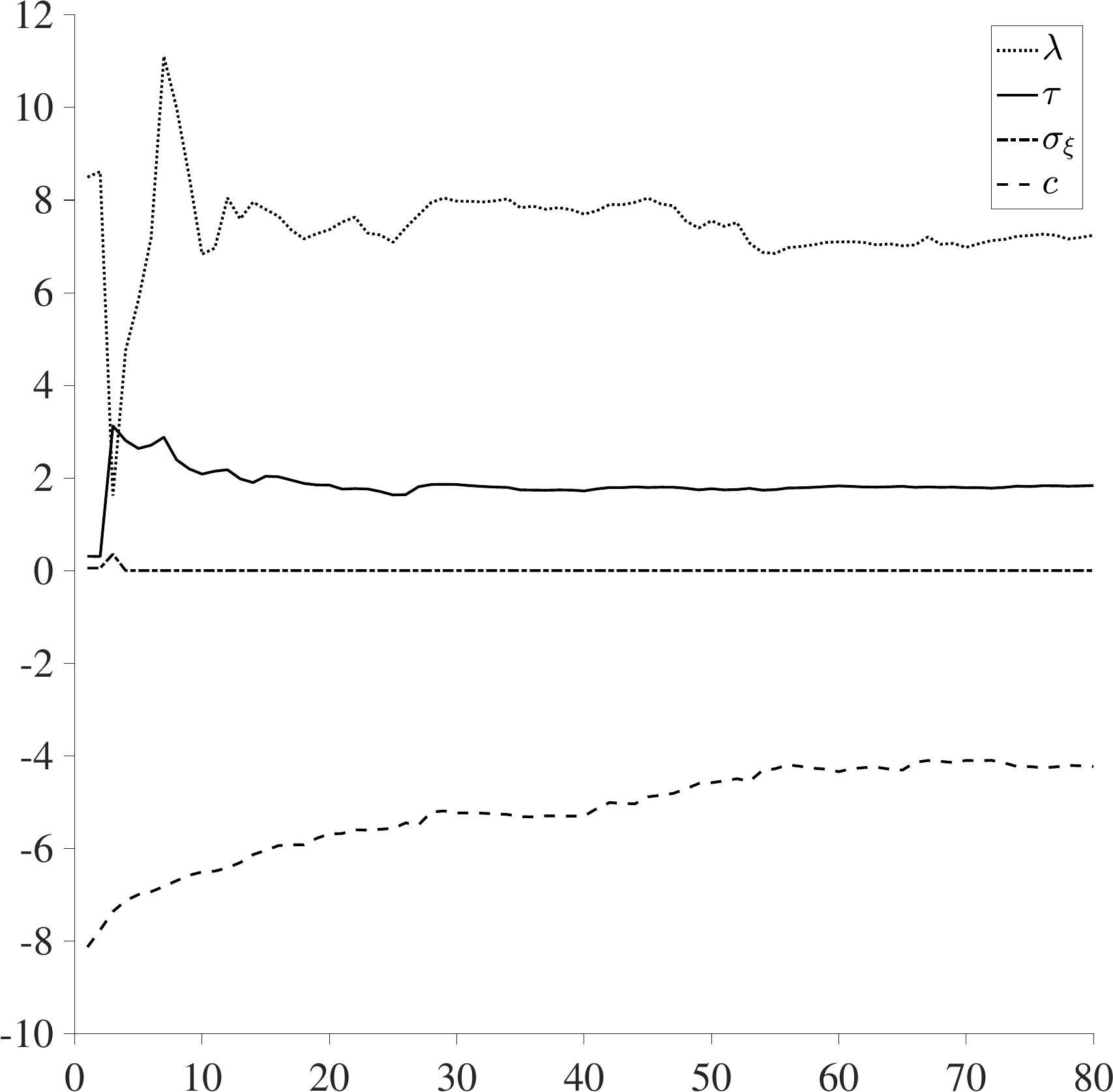}} \\
\end{tabular}
\begin{tablenotes}
\footnotesize
\emph{Note}: Parameters estimated by an expanding window using cloud data from \cite{katzfuss2020ensemble}.
\end{tablenotes}
\end{threeparttable}
\label{fig3}
\end{figure}

Table~\ref{table9} contains the resulting parameter estimates for various model specifications, where the parameter-estimation procedure used all data deemed available.  Consistent with \cite{katzfuss2020ensemble}, in all specifications the relatively large estimate of $\gamma_3$ picks up the drift of clouds along the spatial transect, indicating that clouds tend to float from lower to higher location numbers. While \cite{katzfuss2020ensemble} investigated only the overdispersed model, our comparison of the overdispersed model and the standard model yields no evidence that the former is preferable to the latter: estimates of the overdispersion parameter $\sigma_{\xi}$ are practically zero. On the other hand, the inclusion of an additional parameter~$c$ governing the overall level appears to be beneficial.

Running the Bellman filter on the entire sample with missing data produces filtered states at times and locations for which observations were declared missing. By taking the exponent, a filtered state translates to an intensity, which in turn equals the expected value of a draw from the relevant Poisson distribution. This allows us to produce both point and density `nowcasts' of missing data conditional on the available data up to and including the relevant time step. Following \cite{katzfuss2020ensemble}, these point and density nowcasts can be compared with the actual observations using the mean squared error (MSE) and continuously ranked probability score (CRPS), respectively, which are reported in the right-most columns of Table~\ref{table9}. Depending on the model specification, the MSEs of the Bellman filter lie in the range ${\sim}0.51{-}0.56$, the CRPS in ${\sim}0.18{-}0.20$. These numbers are not (yet) directly comparable with those in \cite{katzfuss2020ensemble}, who use an expanding window for the purpose of parameter estimation. This is addressed in the next section.

\subsection{Results: Expanding window with missing data}

The highly parametrised model~\eqref{katzfuss1}--\eqref{katzfuss3} allows us to estimate the static parameters in an expanding-window-with-missing-data setting, starting with a window of one time step. For the most general (i.e.\ overdispersed) version of model, Figure~\ref{fig3} shows the parameter estimates over time. At the end of the sample, the parameter estimates match the results in Table~\ref{table9}. For all time steps, the estimate of $\sigma_{\xi}$ is practically zero. After some variation at the start of the sample, the estimates of $\lambda,\tau$ and $c$  converge relatively quickly. The estimates of $\gamma_1,\gamma_2,\gamma_3$, however, show considerable time variation even towards the end of the sample, indicating that these parameters may not in fact be static. This may explain why the expanding-window results, discussed below, appear to be no worse than the full-sample results. 

For the purpose of nowcasting missing data, Table~\ref{table10} shows that 
both the standard ($\sigma_{\xi}=0$) and overdispersed ($\sigma_{\xi}>0$) versions of the model with $c\neq 0$ achieve MSEs of ${\sim}0.52$, with the particle ensemble Kalman filter and Rao-Blackwellised particle filter lagging behind by ${\sim}45\%$ and ${\sim}140\%$, respectively.  Irrespective of the exact specification, the Bellman filter achieves CRPS values of ${\sim}0.19$, with the corresponding numbers for both particle-filtering methods inflated by ${\sim}30\%$ and ${\sim}75\%$. This demonstrates that Bellman filter can outperform state-of-the-art particle filtering methods in high-dimensional settings, while the computational burden remains low. 

\begin{table}[t!]
\center
\caption{\label{table10} Quality of nowcasts using an expanding window for parameter estimation and filtering}
\renewcommand{\arraystretch}{1}
\begin{footnotesize}
\begin{threeparttable}
\begin{tabular}{l@{\hspace{4mm}}l@{\hspace{4mm}}r@{\hspace{4mm}}r}
  \toprule
Model & Method &  
  MSE 
   &
CRPS 
   \\
   \midrule
 Overdispersed $(\sigma_{\xi}>0)$     &   Rao-Blackwellised particle filter ($c=0$, \citealp{johansen2012exact})   & 1.26 & 0.33
         \\
       & Particle ensemble Kalman filter ($c=0$, \citealp{katzfuss2020ensemble}) & 0.75 & 0.25
\\
& Bellman filter ($c=0$) & $  0.554$ & $    0.194$ 
\\
& Bellman filter ($c\neq0$) & $0.519$ & $0.188$ 
\\
\midrule
Standard $(\sigma_{\xi}=0)$  
& Bellman filter ($c=0$) & $0.556$ & $ 0.196$ 
\\
& Bellman filter ($c\neq 0$) & $0.525$ & $0.190$ 
\\
  \bottomrule
\end{tabular}
\begin{tablenotes}
\item \emph{Note}: MSE = mean squared error. CRPS = continuously ranked probability score. The data (including the classification of training and test data) are available from~\cite{katzfuss2020ensemble}. The first two rows are copied from~\cite{katzfuss2020ensemble}, who consider only the overdispersed model with $c=0$.
\end{tablenotes}
\end{threeparttable}
\end{footnotesize}
\end{table}

\section{Application II: Nonlinear and degenerate state dynamics}
\label{section9}

This section considers a recent state-space model in financial econometrics featuring multidimensional, nonlinear and degenerate state dynamics.  A simulation study demonstrates that the Bellman filter outperforms the particle filter for the purposes of both parameter estimation and filtering, while an empirical application using real data yields similar results for both methods.

\subsection{Model}

\citet[eq.\ 1]{catania2022stochastic} considers a stochastic-volatility model with a general leverage specification:
\begin{align}
y_t & = \mu + \exp(h_t/2) \, \varepsilon_t, & \varepsilon_t \sim \text{i.i.d.}\,\mathrm{N}(0,1),\label{catania eqn 1}
\\[1.2ex]
h_t & = c+ \varphi \, h_{t-1} + \sigma_\eta \, \eta_t , & \label{catania eqn 2}
\\
\eta_t & = \sum_{j=0}^{k} \rho_j \, \varepsilon_{t-j} +\sigma_\xi\,\xi_t,  & \xi_t \sim \text{i.i.d.}\, \mathrm{N}(0,1). \label{catania eqn 3}
\end{align}
Here, $y_t$ is a financial log return, with median (but not mean, as we shall see) $\mu$.  The dynamics for the log-volatility process $\{h_t\}$ feature the intercept $c$, persistence parameter $|\varphi|<1$  and variability $\sigma_\eta>0$. The volatility shock $\eta_t$ is a linear function of current and lagged return shocks, i.e.\ $\varepsilon_t,\ldots,\varepsilon_{t-k}$, where $k\geq 0$ represents the maximum lag length. Unlike in standard volatility models, the return shock $\varepsilon_t$ and log-volatility $h_t$ are generally dependent; both are related to $\eta_t$ whenever $\rho_0\neq 0$. When $\rho_0<0$, as is typical for financial returns, a negative return shock $\varepsilon_t$ tends to coincide, contemporaneously, with a positive volatility shock $\eta_t$. This is known as the `volatility-feedback effect' (e.g.\ \citealp{carr2017leverage}) and implies that the distribution of $y_t$ is negatively skewed, explaining why $\mu$ is the median but not generally the mean. While \cite{catania2022stochastic} sets $\mu=0$, the introduction of $\mu$ enables a more accurate estimation of $\rho_0$ by disentangling the location and scale.
Parameters $\rho_j\in (-1,1)$ for $j=1,\ldots,k$ quantify a generalised `leverage effect': the impact of multiple lagged return shocks $\varepsilon_{t-j}$ on the volatility shock $\eta_t$.  \cite{catania2022stochastic} sets $\sigma_\xi^2= 1-\sum_{j=0}^k \rho_j^2$ with $\sum_{j=0}^k \rho_j^2<1$ to ensure that the unconditional variance of $\eta_t$ is unity; this is required for the identification of $\sigma_\eta$.

\subsection{State-space formulation}

Model~\eqref{catania eqn 1} through~\eqref{catania eqn 3} can be written in the general state-space format~\eqref{DGP0} if the latent state is identified as $\bm{a}_{t}=(h_{t},h_{t-1},\ldots,h_{t-k})'\in \mathbb{R}^{k+1}$, which contains the log volatility $h_t$ as well as $k$ lags. As shown in Supplement~\ref{app:Catania}, the probability density of $y_t\in \mathbb{R}$ conditional on the (now multidimensional) state~$\bm{a}_t$ and the information set at time $t-1$ is Gaussian with mean $\mu_{y,t}$ and standard deviation $\sigma_{y,t}$ as follows:
\allowdisplaybreaks
\begin{align}
\allowdisplaybreaks
&p(y_t|\bm{a}_t,\mathcal{F}_{t-1}) = \frac{1}{\sigma_{y,t}\sqrt{2\pi}} \exp\left(-\frac{(y_t-\mu_{y,t})^2}{2\sigma_{y,t}^2} \right),\quad \sigma_{y,t}  = \exp(h_t/2) \sqrt{1-\frac{\rho_0^2}{1-\sum_{j=1}^k \rho_j^2}},
\label{Catania observation density}
\\
\allowdisplaybreaks
&\mu_{y,t} = \mu + \frac{\rho_0}{1-\sum_{j=1}^k \rho_j^2}\,\exp(h_t/2) \left[ \frac{h_t-c-\varphi\, h_{t-1}}{\sigma_\eta}- \sum_{j=1}^k \rho_j\, \frac{y_{t-j}-\mu}{\exp(h_{t-j}/2)}  \right].
\notag
\end{align}
The mean $\mu_{y,t}$ depends on the log volatility $h_t$ as well as $k$ of its lags (except when $\rho_0=0$), such that $y_t$ provides information about the entire state vector $\bm{a}_t=(h_t,\ldots,h_{t-k})'$. This implies that, at each time step, $k+1$ logarithmic volatilities must be estimated; this insight will be important for the choice of estimation method. The density of the state vector $\bm{a}_t$ conditional on the previous state and the information set $\mathcal{F}_{t-1}$ is a degenerate Gaussian (for details, see Supplement~\ref{app:Catania}). The first element of $\bm{a}_t$ (i.e.\ $h_t$) has a proper distribution, while lagged versions of $h_t$ are not random when the conditioning set includes the previous state $\bm{a}_{t-1}$:
\begin{align}
&p(\bm{a}_t|\bm{a}_{t-1},\mathcal{F}_{t-1})  = \frac{1}{\sigma_{h,t}\sqrt{2\pi}} \exp\left(-\frac{(h_t-\mu_{h,t})^2}{2\sigma_{h,t}^2} \right) \; \times\; \prod_{j=1}^{k} \delta(a_{j+1,t} - a_{j,t-1}),
\label{degenerate state dynamics}
 \\
&\mu_{h,t} = c +\varphi\, h_{t-1} + \sigma_\eta\,\sum_{j=1}^k \rho_j\, \frac{y_{t-j}-\mu}{\exp(h_{t-j}/2)}  ,\qquad
\sigma_{h,t} = \sigma_\eta \sqrt{1-\sum_{j=1}^k \rho_j^2}.
\notag
\end{align}
Here, $a_{j,t}$ denotes the $j$-th element of the state vector $\bm{a}_{t}=(h_{t},h_{t-1},\ldots,h_{t-k})'$, and  $\delta(\cdot)$ denotes the Dirac delta function. The product of Dirac deltas ensures that the second element of $\bm{a}_t$ equals the first element in $\bm{a}_{t-1}$, and so on. The resulting state dynamics are multidimensional, nonlinear and degenerate. This is problematic, as parameter estimation for multidimensional states (\citealp[p.\ 335]{kantas2015particle}) and/or degenerate state dynamics (\citealp[p.\ 1396]{kunsch2013particle}) using  particle-filtering methods remains a challenge that has not yet been fully resolved in the literature. For the same reasons, approximate filters such as that in \cite{koyama2010approximate} are ruled out. 

\subsection{Parameter-estimation methods}

\cite{catania2022stochastic} estimates the static parameters of the state-space model~\eqref{Catania observation density} and \eqref{degenerate state dynamics} using a univariate implementation of \citeauthor{malik2011particle}'s (\citeyear{malik2011particle}) continuous sampling importance resampling (CSIR) method. The effect of this univariate approach on parameter estimation and model selection is a priori unclear. Moreover, this approach comes with three potential disadvantages. First, the univariate approach means that only the first element of the state vector $\bm{a}_{t}=(h_{t},h_{t-1},\ldots,h_{t-k})'$ is estimated at time $t$, while the other elements remain fixed at previously estimated values. However, the observation $y_t$ contains information about the entire state vector $\bm{a}_t$, as can be seen from the observation density~\eqref{Catania observation density}. While actual (i.e.\ true) lags of $h_t$ are constant over time, the researcher's estimates need not be. Even when focusing purely on the real-time estimation of $h_t$, the decision not to re-estimate the lags at each point in time may lead to an efficiency loss. Second, while the CSIR method guarantees a continuous approximation of the log-likelihood function, this approximation need not be smooth, potentially causing standard gradient-based optimisers to fail. I employ a grid search to identify promising areas of the parameter space, followed by a simplex-based optimisation algorithm that does not utilise gradients. Third, numerical standard errors derived from the inversion of negative Hessian matrices may be misleading when the objective function is nonsmooth. For a piecewise linear approximation as in the CSIR method, finite-difference Hessian matrices may be badly scaled when evaluated near kinks, or identically zero when evaluated on linear pieces. This may explain the exceedingly small standard errors reported in \cite{catania2022stochastic}, as well as my finding that Hessian matrices based on the CSIR method frequently fail to be invertible. 

\begin{table}[t!]
\center
\caption{\label{table11} Average parameter estimates across $100$ samples, standard deviations (in parentheses) and the average of numerical standard errors (in square brackets).}
\renewcommand{\arraystretch}{1}
\begin{footnotesize}
\begin{threeparttable}
\begin{tabular}{l@{\hspace{1cm}}r@{\hspace{0.6cm}}r@{\hspace{0.6cm}}r@{\hspace{0.6cm}}r@{\hspace{0.6cm}}r@{\hspace{0.6cm}}r@{\hspace{0.6cm}}r@{\hspace{1cm}}r}
  \toprule
 \multicolumn{8}{c}{Parameter estimates} &  \multicolumn{1}{c}{MAE}
 \\
  &  \multicolumn{1}{l}{$\mu$} 
   &   \multicolumn{1}{l}{$c$} &  \multicolumn{1}{l}{$\varphi$} &  \multicolumn{1}{l}{$\sigma_\eta$} &  \multicolumn{1}{l}{$\rho_0$} &  \multicolumn{1}{l}{$\rho_1$} &  \multicolumn{1}{l}{$\rho_2$} & \multicolumn{1}{l}{$h_{t|t-1}$}  \\
     \cmidrule(r{5pt}){1-1}    \cmidrule(r{5pt}){2-8} \cmidrule(r{0pt}){9-9}

True value $\rightarrow$  & $0.0015$ & $-0.200$ & $0.980$ & $0.250$ & $-0.700$ & $-0.400$ & $0.300$ &  \multicolumn{1}{c}{$h_t$}   \\
    \cmidrule(r{5pt}){1-1}   \cmidrule(r{5pt}){2-8} \cmidrule(r{0pt}){9-9} 
 Bellman filter &   $0.0015$  & $-0.207$ &    $0.979$ &  $0.252$ & $-0.651$ & $-0.438$ &   $0.294$ & 
 $0.358$
\\
 &   $(0.0001 )$   &     $(0.038)$    &    $(0.004 )$    &   $(0.024)$   &   $ (0.089 )$   &   $ (0.115 )$    &    $(0.101)$    
\\   
 &   $[0.0001]$   &    $[0.033]$    &   $[0.003]$    &   $[0.026]$   &   $[0.094]$   &   $[0.107]$    &    $[0.102]$
\\                             
Particle filter   &$0.0016$   &$-0.262$    &$0.974$    &$0.279$   &$-0.739$   &$-0.109$    & $0.095$& $0.382$  \\
                  &   $(0.0002)$   &   $(0.155)$    &   $(0.016)$    &   $(0.051)$   &   $(0.110)$   &   $(0.293)$    &    $(0.203)$\\ 
                    &   $[0.0001]$   &   $[  0.004 ]$    &   $[   0.001   ]$    &   $[  0.004   ]$   &   $[0.004]$   &   $[ 0.005]$    &    $[   0.005]$\\

   \cmidrule(r{5pt}){1-1}    \cmidrule(r{5pt}){2-8} \cmidrule(r{0pt}){9-9}

True value $\rightarrow$ & $0.0015$ & $-0.200$ & $0.980$ & $0.250$ & $-0.400$ & $-0.700$ & $0.300$ & \multicolumn{1}{c}{$h_t$} \\
    \cmidrule(r{5pt}){1-1} \cmidrule(r{5pt}){2-8} \cmidrule(r{0pt}){9-9} 
Bellman filter &    $0.0015$ &  $-0.208$  &  $0.979$  &  $0.265$ &   $-0.355$ &  $-0.715 $    & $0.306$ & $0.335$
\\
    &   $(0.0001)$   &   $(0.033)$  &    $(0.004)$ &      $( 0.034)$ &    $(0.083)$  &    $(0.062)$ &    $(0.084)$   
    \\
  &    $[0.0001]$ &     $[0.034]$  &    $[0.004]$  &    $[0.033]$ &     $[0.088]$  &    $[0.064]$  &    $[0.089]$    
   
  \\
 Particle filter &   $0.0015$  & $-0.242$  &   $0.976 $ &   $0.250$  & $-0.471  $  & $-0.441$  & $0.061$ & $  0.358 $  \\   
    
&     $( 0.0003)$  &    $(0.099)$ &     $(0.010)$  &    $(0.062)$  &    $(0.207   )$  &    $(0.347 )$  &    $(0.258)$ \\ 
&   $[0.0001]$   &   $[ 0.005]$    &   $[0.001]$    &   $[ 0.006   ]$   &   $ [0.007]$   &   $[ 0.007]$    &    $[   0.008]$\\   
\bottomrule
\end{tabular}
\begin{tablenotes}
\item \emph{Note}: MAE = mean absolute error. For both sets of true parameter values, I simulate $100$ samples of length $5{,}000$ and compute parameter estimates based on the first $2{,}500$ observations. For the Bellman filter, the proposed approximate estimator~\eqref{approximate estimator} is used. For the particle filter, I follow \cite{catania2022stochastic} in using \citeauthor{malik2011particle}'s (\citeyear{malik2011particle}) continuous sampling importance resampling (CSIR) particle filter with $5{,}000$ particles. For each sample I compute, in addition to parameter estimates, numerical standard errors by inverting the negative Hessian matrix evaluated at the peak and taking the square root of the diagonal. I exclude standard errors based on non-invertible Hessian matrices, which were encountered in ${\sim}40\%$ of samples based on the CSIR method. Using estimated parameters, I make out-of-sample predictions by running the filter on the entire data set, computing mean absolute errors (MAEs) by comparing out-of-sample predictions $h_{t|t-1}$ with actual (simulated) values $h_t$ for $t>2{,}500$. 
\end{tablenotes}
\end{threeparttable}
\end{footnotesize}
\end{table}

In addition to the particle filter, I employ the general version of the Bellman filter (section~\ref{section31}) extended to account for degenerate state dynamics (section~\ref{section32}). The Bellman filter is implemented using closed-form expressions (given in Supplement~\ref{app:Catania2}) for derivatives of the observation and state-transition log densities with respect to the entire state vector $\bm{a}_{t}=(h_t,h_{t-1},\ldots,h_{t-k})$; hence, the entire $(k+1)$-dimensional state is estimated at each time $t$. I allow up to $k_{\max}=10$ lags, implying that the Bellman filter solves an optimisation problem with up to $11$ dimensions at each time step.  To estimate the static parameters, I identify promising starting values using a grid search, after which I implement estimator~\eqref{approximate estimator} using a gradient-based numerical optimiser. In the Bellman-filtering procedure, at each time step I execute Newton or Fisher optimisation steps when the search direction is well-defined; otherwise, the optimisation is skipped and the update is set equal to the prediction. This somewhat crude approach ensures that the filter runs smoothly even when using flawed parameter values, which may be encountered during the black-box estimation routine~\eqref{approximate estimator}. At the optimal parameter values identified using this routine, the filter is convergent at every time step.

\subsection{Simulation results}

To investigate the difference between the multivariate approach and the (one-dimensional) CSIR method, a simulation study is performed. Two sets of realistic parameter values are shown in Table~\ref{table11}. I generate $100$ series of length $5{,}000$, using the first half for parameter estimation. The results in Table~\ref{table11} show that average parameter estimates of $\rho_0,\rho_1$ and $\rho_2$ obtained by the CSIR particle filter are inaccurate, while those based on the Bellman filter are relatively accurate. For example, the average estimate of $\rho_2$ by the Bellman filter differs from the true value by no more than $0.01$, compared to at least $0.20$ for the particle filter. While \cite{catania2022stochastic} demonstrated that the CSIR method may produce accurate parameter estimates, this finding may partly be explained by the fact that the parameter-optimisation routine there was initialised using the true parameters, in which case the CSIR estimates typically remain close to the starting point. 
 The results also show that the parameter estimates based on the particle filter vary greatly across samples, as can be seen from the large standard deviations in parentheses in Table~\ref{table11}, while parameter estimates based on the Bellman filter are relatively stable. Additionally, the average of numerically computed standard errors, in square brackets, indicates that standard errors are somewhat reliable for the Bellman filter, closely matching the actual variation across samples, but not for the CSIR method, where they are several orders of magnitude too small. This may be due to the nonsmooth approximation of the log-likelihood function in the CSIR method, and casts doubt on the validity of similarly small standard errors in \cite{catania2022stochastic}. Finally, the right-most column shows that the improved parameter estimates lead to out-of-sample forecasting gains, which are consistent across samples (the Bellman filter produces better forecasts for each sample) and overwhelmingly statistically significant according to a standard Diebold-Mariano test (not shown). 

\begin{table}[t!]
\center
\caption{\label{table12} Parameter estimates for preferred model specifications and numerical standard errors in square brackets}
\renewcommand{\arraystretch}{1}
\begin{footnotesize}
\begin{threeparttable}
\begin{tabular}{l@{\hspace{0.5cm}}r@{\hspace{0.3cm}}r@{\hspace{0.3cm}}r@{\hspace{0.3cm}}r@{\hspace{0.3cm}}r@{\hspace{0.3cm}}r@{\hspace{0.3cm}}r@{\hspace{0.3cm}}r}
  \toprule
  &  \multicolumn{1}{l}{$\mu$} 
   &   \multicolumn{1}{l}{$c$} &  \multicolumn{1}{l}{$\varphi$} &  \multicolumn{1}{l}{$\sigma_\eta$} &  \multicolumn{1}{l}{$\rho_0$} &  \multicolumn{1}{l}{$\rho_1$} &  \multicolumn{1}{l}{$\rho_2$} &  \multicolumn{1}{l}{$\rho_3$}   \\
     \cmidrule(r{5pt}){1-1}    \cmidrule(r{5pt}){2-9}      Bellman filter  &      $0.051$	 & $-0.001$ &	$0.982$ &	$0.258$ &	$-0.377$ & 	$-0.583$ & 	$-0.091$ & $	0.463$     
   \\
&      $[0.008]$   &      $[0.002]$ &      $[0.003]$  &     $[0.016]$   &    $[0.049]$   &    $[0.066]$  &    $[ 0.099]$ &    $[0.060]$ 
 \\
 Particle filter &  $0.052$ &	$-0.006$ & 	$0.983$ & 	$0.239$ & 	$-0.398$ & $	-0.571$ & 	$-0.114$ & 	$0.459$
   \\ 
&       $[ 0.004 ]$   &    $[0.002]$ &      $[0.002  ]$  &     $[0.005 ]$   &    $[0.009   ]$   &    $[0.007 ]$  &    $[ 0.007]$ &    $[ 0.005 ]$  
\\
  \bottomrule
\end{tabular}
\begin{tablenotes}
\item \emph{Note}: For both parameter-estimation methods, the preferred model determined by the Bayesian information criterion (BIC) has three lags. Full parameter-estimation results with up to ten lags are available in Supplement~\ref{additional estimation results}. 
The data are log returns of the S$\&$P500 (multiplied by $100$) from $3$ Jan $1990$ to $31$ Dec $2019$ ($7{,}558$ observations).
\end{tablenotes}
\end{threeparttable}
\end{footnotesize}
\end{table}

\subsection{Empirical results}

For the empirical application, I take log returns of the S$\&$P500 from $3$ Jan $1990$ to $31$ Dec $2019$ ($7{,}558$ observations). Table~\ref{table12} shows  preferred models when using the Bayesian information criterion, which suggests setting $k=3$ lags for both parameter-estimation methods when up to $10$ lags are allowed (full results are available in Supplement~\ref{additional estimation results}). Parameter estimates for both methods  are similar, perhaps due to the comparatively long dataset. Both methods indicate that volatility feedback and leverage play important roles, with the positive estimate of $\rho_3$ suggesting that the leverage effect is temporary: upward volatility shocks following negative returns may be partially reversed on day three. 
The small standard errors for the particle filter, similar to those reported in~\citet[table 2]{catania2022stochastic}, may underestimate the true uncertainty surrounding the parameter estimates. Standard errors based on the Bellman filter, which are up to an order of magnitude higher for the parameters of interest, were in simulation studies found to be reasonably accurate.

\section{Conclusion}
\label{section10}

\noindent The Bellman filter for state-space models as developed in this article generalises the Kalman filter and is equally computationally inexpensive in high-dimensional state spaces, but robust in the case of heavy-tailed observation noise and applicable to a wider range of (nonlinear and non-Gaussian) models. Under suitable conditions, the Bellman-filtered states are globally contractive to a small region around the true state at every time step, while filtering errors remain uniformly bounded over time. A second contribution is the development of a Bellman smoother that is mathematically equivalent to the classic Rauch, Tung and Striebel (\citeyear{rauch1965maximum}) smoother, but applicable more generally---as an approximation---to state-space models with nonlinear and/or non-Gaussian observation equations.
Third, the approximate static-parameter estimation procedure developed here is straightforward to implement and, again, computationally inexpensive; the resulting parameter estimates for various sample sizes appear to be no less accurate or efficient than those of (asymptotically exact) simulation-based methods.

In a simulation study involving a wide range of univariate models, the performance of the Bellman filter  is near identical to those of state-of-the art simulation-based methods in terms of parameter estimation and filtering, while additionally enabling smoothing. Filtering speeds are improved by factors up to ${\sim}160$ (compared to particle filters) and ${\sim}2{,}000$ (cf.\ importance samplers). Likewise, computation times for estimating the static parameters are reduced by factors up to ${\sim}10$ (cf.\ importance samplers) and ${\sim}400$ (cf.\ particle filters). In an application with a high-dimensional climate model, the tracking performance of the Bellman filter remains virtually unchanged as the dimension of the state space is increased from $10$ to $150$, while that of the particle filter deteriorates sharply---due to the curse of dimensionality---even when employing very many particles: e.g.\ with $10^6$ particles in $150$ spatial dimensions, the Bellman filter is both faster (by a factor ${\sim}1{,}000$) and more accurate (by a factor ${\sim}1.8$ in terms of mean absolute filtering error). In a second application with highly nonlinear and degenerate state dynamics, the Bellman filter outperforms the particle filter for the purposes of both parameter estimation and filtering. 

\clearpage
{
\small
\setlength{\bibsep}{0pt plus 1.0ex} 
\bibliographystyle{rss}
\bibliography{mybibfile}
}

\clearpage
\appendix
\small
\setstretch{1}
\numberwithin{equation}{section}
\numberwithin{table}{section}

\section{Proof of Proposition~\ref{prop1}}
\label{A:proof} 

To understand how a recursive approach may be feasible, we start by noting that the joint log-likelihood function~\eqref{loglik} satisfies a straightforward recursive relation for $2\leq t\leq n$ as follows:
\begin{eqnarray}
\label{recursive}
L_{1:t}(\bm{a}_{1},\ldots,\bm{a}_t)\;=\; \ell(\bm{y}_{t}|\bm{a}_{t})\,+\,\ell(\bm{a}_{t}|\bm{a}_{t-1}) \, +\,L_{1:t-1}(\bm{a}_{1},\ldots,\bm{a}_{t-1}) .
\end{eqnarray}
That is, in transitioning from time $t-1$ to time $t$, two terms are added: one representing the state-transition density, $\ell(\bm{a}_{t}|\bm{a}_{t-1})$; the other representing the observation density, $\ell(\bm{y}_{t}|\bm{a}_{t})$. Next, standard dynamic-programming arguments imply
\allowdisplaybreaks
\begin{eqnarray}
\label{RML}
V_{t}(\bm{a}_{t}) \;&:=& \underset{(\bm{a}_{1},\ldots,\bm{a}_{t-1})\in \mathbb{R}^{m\times (t-1)}}{\max} \; L_{1:t}(\bm{a}_1,\ldots,\bm{a}_{t}),\qquad \text{  by definition~\eqref{value function},}
\\
&=& \underset{\bm{a}_{1:t-1}\in \mathbb{R}^{m\times (t-1)}}{\max} \; \big\{ \ell(\bm{y}_{t}|\bm{a}_{t})+\ell(\bm{a}_{t}|\bm{a}_{t-1})  + L_{1:t-1}(\bm{a}_{1},\ldots,\bm{a}_{t-1})\big\}, \text{  by recursion \eqref{recursive},}\notag
\\
&=& \underset{\bm{a}_{t-1} \in \mathbb{R}^m}{\max}\;\;\Big\{\ell(\bm{y}_{t}|\bm{a}_{t})+\ell(\bm{a}_{t}|\bm{a}_{t-1})   + \underset{(\bm{a}_{1},\ldots,\bm{a}_{t-2})\in \mathbb{R}^{m\times(t-2)}}{\max}  L_{1:t-1}(\bm{a}_{1},\ldots,\bm{a}_{t-1}) \Big\},\notag\\
& & \qquad  \text{ by moving all but one maximisation inside curly brackets,}\notag
\\
&=& \underset{\bm{a}_{t-1} \in \mathbb{R}^m}{\max}\;\;\big\{\ell(\bm{y}_{t}|\bm{a}_{t})+\ell(\bm{a}_{t}|\bm{a}_{t-1})   + V_{t-1}(\bm{a}_{t-1}) \big\}, \text{  again by definition~\eqref{value function},}\notag
\\
&=& \ell(\bm{y}_{t}|\bm{a}_{t})+\underset{\bm{a}_{t-1} \in \mathbb{R}^m}{\max}\;\;\big\{\ell(\bm{a}_{t}|\bm{a}_{t-1})   + V_{t-1}(\bm{a}_{t-1}) \big\}.\notag
\end{eqnarray}
Further, it is evident that
\begin{equation}
\bm{a}_{t|t}=\underset{ \bm{a}_t \in \mathbb{R}^m }{\arg \max }\; V_t(\bm{a}_t) = \underset{\bm{a}_t \in \mathbb{R}^m}{ \arg\max}
\underset{(\bm{a}_{1},\ldots,\bm{a}_{t-1})\in \mathbb{R}^{m\times(t-1)}}{\max}
 L_{1:t}(\bm{a}_{1},\ldots,\bm{a}_t).
\end{equation}

\section{Block-matrix inversion}
\label{app:block matrix}

Consider the second diagonal block of the negative Hessian matrix in equation~\eqref{Newton general}. Define this block as  $\bm{D}_t\in \mathbb{R}^{m\times m}$ and define its Schur complement $\bm{S}_t\in \mathbb{R}^{m\times m}$ as follows:
\begin{align}
\bm{D}_t:=\bm{I}_{t-1|t-1}+\bm{J}_{t}^{22}, \hspace{1cm} \bm{S}_t:=  \bm{J}^{11}_{t} -\bm{J}^{12}_{t}\,\bm{D}_t^{-1}\,\bm{J}^{21}_{t} -\frac{\dd^2 \ell(\bm{y}_t|\bm{a}_t)}{\dd\bm{a}_t\dd\bm{a}_t'}.
\end{align}
As is standard (e.g.\  \citealp[p.\ 108]{bernstein2009matrix}), the required block-matrix inverse can then be expressed as 
\begin{equation}
\left[\begin{array}{cc}\bm{J}^{11}_{t} -\frac{\dd^2\ell(\bm{y}_{t}|\bm{a}_{t})}{\dd\bm{a}_t\dd\bm{a}_t' } & \bm{J}^{12}_{t}\\ \bm{J}^{21}_{t} & \bm{I}_{t-1|t-1} + \bm{J}^{22}_{t} \end{array} \right]^{-1} =  \left[\begin{array} {cc} \bm{S}_t^{-1} & -\bm{S}_{t}^{-1} \bm{J}_{t}^{12} \bm{D}_t^{-1}
 \\  -\bm{D}_t^{-1} \bm{J}_{t}^{21}  \bm{S}_{t}^{-1}  & \bm{D}_t^{-1}+ \bm{D}_t^{-1} \bm{J}_{t}^{21} \bm{S}_{t}^{-1} \bm{J}_{t}^{12} \bm{D}_{t}^{-1}
 \end{array} \right] ,
\end{equation}
as long as the required inverses exist.

\section{Derivation of equation~\eqref{general information update}}
\label{app:envelope}

Here we compute the negative Hessian of  the value function, i.e.\
\begin{align}
V_t(\bm{a}_t)&=\ell(\bm{y}_t|\bm{a}_t)+\underset{\bm{a}_{t-1}\in \mathbb{R}^m}{ \max } \Big\{ \ell(\bm{a}_t|\bm{a}_{t-1} )-\frac{1}{2}(\bm{a}_{t-1}-\bm{a}_{t-1|t-1})'\bm{I}_{t-1|t-1}(\bm{a}_{t-1}-\bm{a}_{t-1|t-1})  \Big\}
,\notag
\\
&=\ell(\bm{y}_t|\bm{a}_t)+\ell(\bm{a}_t|\bm{a}_{t-1}^\ast )-\frac{1}{2}(\bm{a}^\ast_{t-1}-\bm{a}_{t-1|t-1})'\bm{I}_{t-1|t-1}(\bm{a}_{t-1}^\ast-\bm{a}_{t-1|t-1}),
\end{align}
where the second line employs the definition
\begin{equation}
\bm{a}^\ast_{t-1} := \underset{\bm{a}_{t-1}\in \mathbb{R}^m}{\arg \max } \Big\{ \ell(\bm{a}_t|\bm{a}_{t-1} )-\frac{1}{2}(\bm{a}_{t-1}-\bm{a}_{t-1|t-1})'\bm{I}_{t-1|t-1}(\bm{a}_{t-1}^\ast-\bm{a}_{t-1|t-1})  \Big\} .
\end{equation}
We must keep in mind that $\bm{a}_{t-1}^\ast$ depends on $\bm{a}_t$; we could have written $\bm{a}^\ast_{t-1}(\bm{a}_t)$. Indeed, to compute the negative Hessian of $V_t(\bm{a}_t)$, we must account for the change in $\bm{a}^\ast_{t-1}(\bm{a}_t)$ using the chain rule. The first-order condition satisfied by $\bm{a}^\ast_{t-1}$, i.e.
\begin{equation}
\bm{0}\;=\;\frac{\dd\ell(\bm{a}_t|\bm{a}^\ast_{t-1})}{\dd\bm{a}^\ast_{t-1}}-\bm{I}_{t-1|t-1}(\bm{a}^\ast_{t-1}-\bm{a}_{t-1|t-1}),
\end{equation}
can be differentiated with respect to $\bm{a}_t$ to obtain
\begin{equation}
\bm{0}\;=\;\left[-\bm{J}^{21}_{t}-\bm{J}^{22}_{t}\frac{\dd\bm{a}^\ast_{t-1}}{\dd\bm{a}_{t}'}  -\bm{I}_{t-1|t-1} \frac{\dd\bm{a}^\ast_{t-1}}{\dd\bm{a}_{t}'}\right]_{\bm{a}_{t-1}=\bm{a}^\ast_{t-1}},
\end{equation}
where $\bm{J}^{21}_{t}$ and $\bm{J}^{22}_{t}$ are as in equation~\eqref{J}. Solving for the sensitivity of $\bm{a}^\ast_{t-1}$ with respect to $\bm{a}_t$, we obtain
\begin{equation}
\frac{\dd\bm{a}^\ast_{t-1}}{\dd\bm{a}_{t}'} \;=\; \left[-(\bm{I}_{t-1|t-1}+\bm{J}^{22}_{t}
) ^{-1}\bm{J}^{21}_{t}\right]_{\bm{a}_{t-1}=\bm{a}^\ast_{t-1}}.
\end{equation}
Next, the chain rule tells us that the Hessian with respect to $\bm{a}_t$ can be computed as
\begin{equation}
\frac{\dd^2 \, \cdot }{\dd \bm{a}_{t} \dd\bm{a}_{t}'} = 
\left[ 
\begin{array}{c}
\mathds{1}_{m\times m}
\\
\displaystyle\frac{\dd\bm{a}^\ast_{t-1}}{\dd\bm{a}_t'}
\end{array}
\right]'
\left[ 
\begin{array}{cc}
\displaystyle \frac{\partial^2 \, \cdot }{\partial\bm{a}_{t}\partial\bm{a}_{t}'} & 
\displaystyle \frac{\partial^2  \, \cdot }{\partial \bm{a}_{t} \partial {\bm{a}^\ast_{t-1}}' }
\\
 \displaystyle \frac{\partial^2  \, \cdot}{\partial\bm{a}^\ast_{t-1}\partial\bm{a}_{t}'}
   & 
\displaystyle \frac{\partial ^2 \, \cdot }{\partial\bm{a}^\ast_{t-1}\partial{\bm{a}^\ast_{t-1}}'}
\end{array}
\right]
\left[ 
\begin{array}{c}
\mathds{1}_{m \times m}
\\
\displaystyle \frac{\dd\bm{a}^\ast_{t-1}}{\dd\bm{a}_t'}
\end{array}
\right],
\end{equation}
where instances of $\partial$ and $\dd$ denote `partial' and `total' derivatives, respectively, while $\mathds{1}_{m \times m}$ denotes an identity matrix of size $m\times m$. By the first-order envelope theorem, no first order derivative with respect to $\bm{a}^\ast_{t-1}$ appears. The negative Hessian of $V_t(\bm{a}_t)$
becomes
\begin{eqnarray}
-\frac{\dd^2 V_t(\bm{a}_t)
}{\dd\bm{a}_t \dd\bm{a}_t'}&=&\left. \left[ 
\begin{array}{c}
\mathds{1}_{m \times m}
\\
\displaystyle\frac{\dd\bm{a}^\ast_{t-1}}{\dd\bm{a}_t'}
\end{array}
\right]'
\left[
\begin{array}{cc}
 \displaystyle \bm{J}^{11}_{t} -\frac{\dd^2 \ell(\bm{y}_t|\bm{a}_t)}{\dd\bm{a}_t\dd\bm{a}_t'} & \bm{J}^{12}_{t} \\
\bm{J}^{21}_{t}   &\bm{I}_{t-1,t-1}+\bm{J}^{22}_{t} 
\end{array}
\right]
\left[ 
\begin{array}{c}
\mathds{1}_{m \times m}
\\
\displaystyle \frac{\dd\bm{a}^\ast_{t-1}}{\dd\bm{a}_t'}
\end{array}
\right]
\right |_{\bm{a}_{t-1}=\bm{a}^\ast_{t-1}},
\notag
\\
&=&
\left.
\bm{J}^{11}_{t} -\frac{\dd^2 \ell(\bm{y}_t|\bm{a}_t)}{\dd\bm{a}_t\dd\bm{a}_t'}-2\bm{J}^{12}_{t}(\bm{I}_{t-1|t-1}+\bm{J}^{22}_{t}
) ^{-1}\bm{J}^{21}_{t} +\frac{\dd\bm{a}_{t-1}^\ast}{\dd\bm{a}_t} (\bm{I}_{t-1,t-1}+\bm{J}^{22}_{t}) \frac{\dd\bm{a}_{t-1}^\ast}{\dd\bm{a}_t'}
\right |_{\bm{a}_{t-1}=\bm{a}^\ast_{t-1}},
\notag
\\
&=&
\left.
\bm{J}^{11}_{t} -\frac{\dd^2 \ell(\bm{y}_t|\bm{a}_t)}{\dd\bm{a}_t\dd\bm{a}_t'}-\bm{J}^{12}_{t}(\bm{I}_{t-1|t-1}+\bm{J}^{22}_{t}
) ^{-1}\bm{J}^{12}_{t}
\right |_{\bm{a}_{t-1}=\bm{a}^\ast_{t-1}}
.
\end{eqnarray}
Finally $\bm{a}^\ast_{t-1}(\bm{a}_{t|t})=\bm{a}_{t-1|t}$, such that
\begin{equation}
\left.-\frac{\dd^2 V_t(\bm{a}_t)
}{\dd\bm{a}_t \dd\bm{a}_t'} \right|_{\bm{a}_{t|t}} \;=\; \left[ \bm{J}^{11}_{t} -\frac{\dd^2 \ell(\bm{y}_t|\bm{a}_t)}{\dd\bm{a}_t\dd\bm{a}_t'}-\bm{J}^{12}_{t}(\bm{I}_{t-1|t-1}+\bm{J}^{22}_{t}
) ^{-1}\bm{J}^{21}_{t}\right]_{\bm{a}_t=\bm{a}_{t|t},\bm{a}_{t-1}=\bm{a}_{t-1|t}},
\end{equation}
which confirms equation~\eqref{general information update}.

\section{Kalman information update as a special case of~\eqref{general information update}}
\label{app: Kalman info}

For the linear Gaussian model in Corollary~\ref{corol0}, we have $\bm{J}^{11}_{t}=\bm{Q}^{-1}$, $\bm{J}^{12}_{t}=\bm{Q}^{-1}\bm{T}$, $\bm{J}^{21}_{t}=\bm{T}'\bm{Q}^{-1}$, $\bm{J}^{22}_{t}=\bm{T}'\bm{Q}^{-1}\bm{T}$ and $\dd^2 \ell(\bm{y}_t|\bm{a}_t)/(\dd\bm{a}_t\dd\bm{a}_t')=-\bm{Z}' \bm{H}^{-1}\bm{Z}$. Substituting these equalities into the  information update~\eqref{general information update}, we obtain
\begin{align}
\bm{I}_{t|t}&=\bm{Q}^{-1} - \bm{Q}^{-1}\bm{T} (\bm{I}_{t-1|t-1}+\bm{T}'\bm{Q}^{-1}\bm{T})^{-1} \bm{T}' \bm{Q}^{-1} + \bm{Z}'\bm{H}^{-1}\bm{Z},\notag
\\
&= \bm{I}_{t|t-1} + \bm{Z}'\bm{H}^{-1}\bm{Z},
\end{align}
where $\bm{I}_{t|t-1}$ is defined as
\begin{equation}
\bm{I}_{t|t-1}:=\bm{Q}^{-1} - \bm{Q}^{-1}\bm{T} (\bm{I}_{t-1|t-1}+\bm{T}'\bm{Q}^{-1}\bm{T})^{-1} \bm{T}' \bm{Q}^{-1}=(\bm{T}\bm{I}_{t-1|t-1}^{-1}\bm{T}' +\bm{Q})^{-1},
\end{equation}
and where the second equality follows by the Woodbury matrix equality (e.g.\ \citealp[eq.\ 1]{henderson1981deriving}). Next, assuming the inverses $\bm{P}_{t|t-1}:=\bm{I}_{t|t-1}^{-1}$ and $\bm{P}_{t|t}:=\bm{I}_{t|t}^{-1}$ exist, using again \citet[eq.\ 1]{henderson1981deriving},  we find 
\begin{equation}
\bm{P}_{t|t}=\bm{I}_{t|t}^{-1}=( \bm{I}_{t|t-1} + \bm{Z}'\bm{H}^{-1}\bm{Z})^{-1}=\bm{P}_{t|t-1}-\bm{P}_{t|t-1}\bm{Z}' (\bm{Z}\bm{P}_{t|t-1}\bm{Z}'+\bm{H})^{-1}\bm{Z}\bm{P}_{t|t-1}, 
\end{equation}
which is exactly the Kalman filter covariance matrix updating step (again, see \citealp[p.\ 106]{harvey1990forecasting}).

\section{Derivation of equation~\eqref{alphastarmain}}
\label{app:inner maximisation}

The first-order condition for the maximisation over $\bm{a}_{t-1}$ in equation~\eqref{approximate Bellman} can be usefully manipulated as follows:
\begin{eqnarray}
\bm{a}_{t-1}^\ast &=&\big(\bm{I}_{t-1|t-1}\,+\,\bm{T}' \bm{Q}^{-1}\bm{T}\big)^{-1}\,\big(\bm{I}_{t-1|t-1}\,\bm{a}_{t-1|t-1}+ \bm{T}' \bm{Q}^{-1}(\bm{a}_{t}-\bm{c})\big),
\notag
\\
&=& \bm{a}_{t-1|t-1} + (\bm{I}_{t-1|t-1}\,+\,\bm{T}' \bm{Q}^{-1}\bm{T}\big)^{-1} \bm{T}'\bm{Q}^{-1}\,\big(\bm{a}_t-\bm{c}-\bm{T} \bm{a}_{t-1|t-1}\big),
\notag
\\
&=& \bm{a}_{t-1|t-1} + \bm{I}_{t-1|t-1}^{-1}\, \bm{T}'
\, \big(\bm{T}\bm{I}_{t-1|t-1}^{-1} \bm{T}'+\bm{Q}\big)^{-1}
\,\big(\bm{a}_t-\bm{c}-\bm{T} \bm{a}_{t-1|t-1}\big),
\notag
\\
&=& \bm{a}_{t-1|t-1} +  \bm{I}_{t-1|t-1}^{-1}\, \bm{T}'
\, \bm{I}_{t|t-1} \,  \big(\bm{a}_t-\bm{a}_{t|t-1}\big),
\end{eqnarray}
which confirms equation~\eqref{alphastarmain} in the main text.
This second line expresses $\bm{a}_{t-1}^\ast$ as the sum of $\bm{a}_{t-1|t-1}$ and a correction that is linear in the `innovation' $\bm{a}_t -\bm{c}- \bm{T}\bm{a}_{t-1|t-1}$. The third line uses matrix-inversion formulas by \citet[eqns. 9--11]{henderson1981deriving} to ensure that $\bm{Q}^{-1}$ no longer appears, such that by a limiting argument the result remains valid even when $\bm{Q}$ is singular. The last line employs the definitions of $\bm{a}_{t|t-1}$ and $\bm{I}_{t|t-1}$ in Table~\ref{table3}.

\section{Derivation of equation~\eqref{approximate Bellman v2}}
\label{app:B}

\noindent 
Computing the first-order condition in equation~\eqref{approximate Bellman v2}, 
with respect to $\bm{a}_{t-1}$, we obtain
\begin{equation}
\bm{0} = \bm{T}'\bm{Q}^{-1}(\bm{a}_{t}-\bm{c}-\bm{T}\bm{a}_{t-1})-\bm{I}_{t-1|t-1}(\bm{a}_{t-1}-\bm{a}_{t-1|t-1}), 
\end{equation}
the solution of which reads
\begin{eqnarray}
\label{alphastar2}
\bm{a}_{t-1}^\ast =\big(\bm{I}_{t-1|t-1}\,+\,\bm{T}' \bm{Q}^{-1}\bm{T}\big)^{-1}\,\big\{\bm{I}_{t-1|t-1}\,\bm{a}_{t-1|t-1}+ 
\bm{T}' \bm{Q}^{-1}(\bm{a}_{t}-\bm{c})\big\},
\end{eqnarray}
which depends linearly on $\bm{a}_t$. In principle, equation~\eqref{approximate Bellman v2} in the main text can be obtained by substituting equation~\eqref{alphastar2} into equation~\eqref{approximate Bellman} and performing algebraic manipulations. The desired result can be obtained more elegantly by `completing the square' as follows. First, we replace $\bm{a}_{t-1}$ with $\bm{a}^\ast_{t-1}$ in equation~\eqref{approximate Bellman}, which then contains the following terms:
\begin{equation}
\label{C1}
-\frac{1}{2}(\bm{a}_{t}-\bm{c}-\bm{T}\bm{a}^\ast_{t-1})'\;\bm{Q}^{-1}\;(\bm{a}_{t}-\bm{c}-\bm{T}\bm{a}^\ast_{t-1})
- \frac{1}{2}(\bm{a}^\ast_{t-1}-\bm{a}_{t-1|t-1})'\;\bm{I}_{t-1|t-1}\;(\bm{a}^\ast_{t-1}-\bm{a}_{t-1|t-1}).
\end{equation}
Then we recall from equation~\eqref{alphastar2} that $\bm{a}^\ast_{t-1}$ is linear in $\bm{a}_{t}$, such that the collection of terms in equation \eqref{C1} above is at most multivariate quadratic in $\bm{a}_{t}$. Hence, we should be able to rewrite equation~\eqref{C1} as a quadratic function (i.e., by completing the square) as follows:
\begin{equation}
\label{C2}
-\frac{1}{2}(\bm{a}_{t}-\bm{a}_{t|t-1})'\,\bm{I}_{t|t-1}\bm(\bm{a}_{t}-\bm{a}_{t|t-1}) \; + \; \text{ constants,}
\end{equation}
for some vector $\bm{a}_{t|t-1}$ to be found and some matrix $\bm{I}_{t|t-1}$ to be determined. 

To do this, we note that $\bm{a}_{t|t-1}$ represents the argmax of equation~\eqref{C2}, which can most readily be found by differentiating equation~\eqref{C1} with respect to $\bm{a}_{t}$ and setting the result to zero. Using the envelope theorem, we need not account for the fact that $\bm{a}^\ast_{t-1}$ depends on $\bm{a}_{t}$ (the first derivative with respect to $\bm{a}^\ast_{t-1}$ is zero because $\bm{a}^\ast_{t-1}$ is optimal). Thus we set the derivative of equation~\eqref{C1} with respect to $\bm{a}_{t}$ equal to zero, which gives $\bm{0}=\displaystyle \bm{a}_{t}-\bm{c}-\bm{T}\bm{a}^\ast_{t-1}$, or, by substituting $\bm{a}^\ast_{t-1}$ from equation~\eqref{alphastar2}, we obtain
\begin{eqnarray}
\bm{0}&=& \displaystyle\bm{a}_{t}-\bm{c}-\bm{T}[\bm{I}_{t-1|t-1}\,+\,\bm{T}' \bm{Q}^{-1}\bm{T}]^{-1} \bm{I}_{t-1|t-1}\bm{a}_{t-1|t-1}\\
&&
\hspace{5cm}
-\bm{T}[\bm{I}_{t-1|t-1}\,+\,\bm{T}' \bm{Q}^{-1}\bm{T}]^{-1} 
\bm{T}' \bm{Q}^{-1}(\bm{a}_{t}-\bm{c}).\notag
\end{eqnarray}
The solution to this equation reads $
\bm{a}_{t|t-1} := \bm{T} \bm{a}_{t-1|t-1}+\bm{c}$, which confirms the expression in Table~\ref{table3}.

Next, we compute the negative second derivative of equation~\eqref{C1} with respect to $\bm{a}_{t}$, which should give us $\bm{I}_{t|t-1}$. To account for the dependence of $\bm{a}_{t-1}^\ast$ on $\bm{a}_{t}$, we use the chain rule. Specifically, in equation~\eqref{alphastar2}, $\bm{a}^\ast_{t-1}$ is linear in $\bm{a}_{t}$, with the following Jacobian matrix:
\begin{equation}
\label{C4}
\bm{J}:=\frac{\dd \bm{a}^\ast_{t-1}} {\dd\bm{a}'_{t}} \;=\;[\bm{I}_{t-1|t-1}\,+\,\bm{T}' \bm{Q}^{-1}\bm{T}]^{-1}\,\bm{T}'\,\bm{Q}^{-1}.
\end{equation}
Next, the chain rule tells us that
\begin{equation}
\label{block}
\frac{\dd^2 \, \cdot }{\dd \bm{a}_{t} \dd\bm{a}_{t}'} = 
\left[ 
\begin{array}{c}
\mathds{1}_{m \times m}
\\
\bm{J}
\end{array}
\right]'
\left[ 
\begin{array}{cc}
\displaystyle \frac{\partial^2 \, \cdot }{\partial\bm{a}_{t}\partial\bm{a}_{t}'} & 
\displaystyle \frac{\partial^2  \, \cdot }{\partial \bm{a}_{t} \partial {\bm{a}^\ast_{t-1}}' }
\\
 \displaystyle \frac{\partial^2  \, \cdot}{\partial\bm{a}^\ast_{t-1}\partial\bm{a}_{t}'}
   & 
\displaystyle \frac{\partial ^2 \, \cdot }{\partial\bm{a}^\ast_{t-1}\partial{\bm{a}^\ast_{t-1}}'}
\end{array}
\right]
\left[ 
\begin{array}{c}
\mathds{1}_{m \times m}
\\
\bm{J}
\end{array}
\right],
\end{equation}
where instances of $\partial$ and $\dd$ denote `partial' and `total' derivatives, respectively, while $\mathds{1}_{m \times m}$ denotes an identity matrix. As before, the envelope theorem ensures that no \emph{first} derivative with respect to $\bm{a}_t^\ast$ appears. When applying equation~\eqref{block}, we find that the negative second derivative of equation~\eqref{C1} becomes
\begin{eqnarray}
&&\left[ 
\begin{array}{c}
\mathds{1}_{m \times m}
\\
\bm{J}
\end{array}
\right]'
\left[
\begin{array}{cc}
 \bm{Q}^{-1} & - \bm{Q}^{-1}\bm{T} \\
- \bm{T}'\bm{Q}^{-1}   &\bm{I}_{t-1|t-1}+\bm{T}'\bm{Q}^{-1}\bm{T}
\end{array}
\right]
\left[ 
\begin{array}{c}
\mathds{1}_{m \times m}
\\
\bm{J}
\end{array}
\right]
\notag
\\
&=&\bm{Q}^{-1}-\underbrace{\bm{Q}^{-1}\bm{T}\bm{J}}-\underbrace{\bm{J}'\bm{T}'\bm{Q}^{-1}}+\underbrace{\bm{J}'[\bm{I}_{t-1|t-1}+\bm{T}'\bm{Q}^{-1}\bm{T}]\bm{J}},
\notag
\\
&=&
\bm{Q}^{-1}-\bm{Q}^{-1}\bm{T}[\bm{I}_{t-1|t-1}+\bm{T}' \bm{Q}^{-1}\bm{T}]^{-1}\bm{T}'\bm{Q}^{-1}.
\end{eqnarray}
In the last line, we have used the fact that all three terms with curly brackets equal $\bm{Q}^{-1}\bm{T}[\bm{I}_{t|t}+\bm{T}' \bm{Q}^{-1}\bm{T}]^{-1}\bm{T}'\bm{Q}^{-1}$, such that two terms with curly brackets and opposite signs cancel, leaving only one term with a negative sign, which confirms the expression for $\bm{I}_{t|t-1}$ in Table~\ref{table3}.

\section{Kalman filter as a special case}
\label{app:D}
\label{comparison}

Consider the linear Gaussian state-space model in Corollary~\ref{corol1}. Suppose the inverse of the Kalman-filtered covariance matrix exists, i.e.\ $\bm{P}^{-1}_{t-1|t-1}:=\bm{I}_{t-1|t-1}$ exists. In Table~\ref{table3}, take the starting point $\bm{a}_{t|t}^{(0)}=\bm{a}_{t|t-1}$, and use Newton or Fisher optimisation steps. Given that the observation density is Gaussian, the log likelihood $\ell(\bm{y}_t|\bm{a}_t)$ is multivariate quadratic in $\bm{a}_t$, such that the entire objective function~\eqref{approximate Bellman v2} turns out to be multivariate quadratic in $\bm{a}_t$. The matrix of second derivatives is constant, such that Newton and Fisher optimisation steps are identical. Moreover, given the quadratic nature of the objective function, both methods find the location of the optimum in a single step. Indeed, the result is the classic Kalman filter, albeit written in the information form.

More explicitly, take $\bm{y}_t =  \bm{d} +\bm{Z}\,\bm{\alpha}_t + \bm{\varepsilon}_t$ with $\bm{\varepsilon}_t\sim\text{i.i.d. } \mathrm{N}(\bm{0},\bm{H})$. Then 
\begin{equation}
\ell(\bm{y}_t|\bm{a}_t)\;=\; - 1/2 (\bm{y}_t-\bm{d}-\bm{Z}\bm{a}_t)' \bm{H}^{-1}(\bm{y}_t-\bm{d}-\bm{Z}\bm{a}_t)+\text{constants}.
\end{equation}
The score and realised information are 
\begin{equation}
\frac{\dd\, \ell\big(\bm{y}_{t}|\bm{a}_t\big)}{\dd\bm{a}_t}=\bm{Z}'\,\bm{H}^{-1}\,(\bm{y}_{t}-\bm{d}-\bm{Z}\bm{a}_t),\qquad\qquad-
\frac{\dd^2\, \ell\big(\bm{y}_{t}|\bm{a}_t\big)}{\dd\bm{a}_t\,\dd\bm{a}_t'}=  \bm{Z}'\bm{H}^{-1}\bm{Z}.
\end{equation}
As the realised information is constant, it equals the (expected) marginal information. Taking the starting point $\bm{a}_{t|t}^{(0)}=\bm{a}_{t|t-1}$ for Newton's optimisation method, the estimate after a single Newton iteration reads
\begin{equation}
\bm{a}^{(1)}_{t|t} =\bm{a}_{t|t-1}+\left(\bm{I}_{t|t-1}+\bm{Z}'\bm{H}^{-1}\bm{Z} \right)^{-1}\,\bm{Z}' \bm{H}^{-1} (\bm{y}_t-\bm{d}-\bm{Z}\bm{a}_{t|t-1}),
\end{equation}
which is exactly the Kalman filter level update written in information form. To see the equivalence with the covariance form of the Kalman filter, suppose that $\bm{P}_{t|t-1}:=\bm{I}_{t|t-1}^{-1}$ exists. Then, using a standard matrix-inversion formula (see e.g.\ \citealp[eqns. 9--10]{henderson1981deriving}), the expression above is equivalent to
\begin{equation}
\bm{a}^{(1)}_{t|t} 
=\bm{a}_{t|t-1}+\bm{P}_{t|t-1}\bm{Z}'(\bm{Z}\bm{P}_{t|t-1}\bm{Z}'+\bm{H})^{-1} (\bm{y}_t-\bm{d}-\bm{Z}\bm{a}_{t|t-1}),
\end{equation}
which is exactly the Kalman filter updating step (see e.g.\ \citealp[p.\ 106]{harvey1990forecasting}). For the information matrix update we have
\begin{equation}
\bm{I}_{t|t}=\bm{I}_{t|t-1}-\left.\frac{\dd^2\, \ell\big(\bm{y}_{t}|\bm{a}\big)}{\dd\bm{a}\,\dd\bm{a}'}\right|_{\bm{a}=\bm{a}_{t|t}} = \bm{I}_{t|t-1} + \bm{Z}'\bm{H}^{-1}\bm{Z}.
\end{equation}
If the inverses $\bm{P}_{t|t-1}:=\bm{I}_{t|t-1}^{-1}$ and $\bm{P}_{t|t}:=\bm{I}_{t|t}^{-1}$ exist, then, again using \citet[eq. 1]{henderson1981deriving},  we find 
\begin{equation}
\bm{P}_{t|t}=\bm{I}_{t|t}^{-1}=( \bm{I}_{t|t-1} + \bm{Z}'\bm{H}^{-1}\bm{Z})^{-1}=\bm{P}_{t|t-1}-\bm{P}_{t|t-1}\bm{Z}' (\bm{Z}\bm{P}_{t|t-1}\bm{Z}'+\bm{H})^{-1}\bm{Z}\bm{P}_{t|t-1}, 
\end{equation}
which is exactly the Kalman filter covariance matrix updating step (again, see \citealp[p.\ 106]{harvey1990forecasting}).

\section{Iterated extended Kalman filter as a special case}
\label{app:E}

Consider the linear Gaussian state-space model in Corollary~\ref{corol1}, except let $\bm{y}_t =\bm{d}+ \bm{Z}(\bm{\alpha}_t) + \bm{\varepsilon}_t$ for some nonlinear vector function $\bm{Z}(\cdot)$ and $\bm{\varepsilon}_t\sim\text{i.i.d. } \mathrm{N}(\bm{0},\bm{H})$. In Table~\ref{table3}, take the starting point $\bm{a}_{t|t}^{(0)}=\bm{a}_{t|t-1}$ and perform Fisher optimisation steps, ignoring (i.e.\ setting to zero) all second-order derivatives of $\bm{Z}(\cdot)$. The iterated extended Kalman filter is then obtained as a special case. 

More explicitly, take $\bm{y}_t= \bm{d}+\bm{Z}(\bm{\alpha}_t)+\bm{\varepsilon}_t$ with $\bm{\varepsilon}_t\sim \text{ i.i.d. } \mathrm{N}(\bm{0},\bm{H})$. Here, $\bm{Z}_t:=\bm{Z}(\bm{\alpha}_t)$ is a column vector of the same size as $\bm{y}_t$, where each element of $\bm{Z}_t$ depends on the elements of $\bm{\alpha}_t$.
Then
\begin{equation}
\ell(\bm{y}_t|\bm{a}_t)\;=\; - 1/2 (\bm{y}_t-\bm{d}-\bm{Z}(\bm{a}_t))' \bm{H}^{-1}(\bm{y}_t-\bm{d}-\bm{Z}(\bm{a}_t))+\text{constants}.
\end{equation}
The score and marginal information are similar to those in Appendix~\ref{app:D}, as long as $\bm{Z}$ there is replaced by the Jacobian of the transformation from $\bm{\alpha}_t$ to $\bm{Z}_t$, i.e.\ $\dd\bm{Z}(\bm{a}_t)/\dd\bm{a}_t'$. Hence
\begin{eqnarray}
\frac{\dd\, \ell\big(\bm{y}_{t}|\bm{a}_t\big)}{\dd\bm{a}_t}&=& \frac{\dd\bm{Z}'}{\dd\bm{a}_t}\,\bm{H}^{-1}\,(\bm{y}_t-\bm{d}-\bm{Z}(\bm{a}_t)),
\\
\frac{\dd^2\, \ell\big(\bm{y}_{t}|\bm{a}_t\big)}{\dd\bm{a}_t\,\dd\bm{a}_t'}&=& - \frac{\dd\bm{Z}'}{\dd\bm{a}_t}\bm{H}^{-1}\frac{\dd\bm{Z}}{\dd\bm{a}_t'}+\text{second-order derivatives}.
\end{eqnarray}
The iterated extended Kalman filter (IEKF) is obtained from the Bellman filter by choosing Newton's method and by making one further simplifying approximation: namely that all second-order derivatives of elements of $\bm{Z}_t$ with respect to the elements of $\bm{\alpha}_t$ are zero. It is not obvious under what circumstances this approximation is justified, but here we are interested only in showing that the IEKF is a special case of the Bellman filter. Higher-order IEKFs may be obtained by retaining the second-order derivatives. If the observation noise $\bm{\varepsilon}_t$ is heavy tailed, however, the Bellman filter in Table~\ref{table3} suggests a `robustified' version of the Kalman filter and its extensions, in which case the tail behaviour of $p(\bm{y}_t|\bm{a}_t)$ is accounted for in the optimisation step by using the score $\dd\ell(\bm{y}_t|\bm{a}_t)/\dd\bm{a}_t$. 

\section{Fahrmeir's approximate mode estimator as a special case}
\label{app fahrmeir}

When considering an observation density $p(\bm{y}_t|\bm{a}_t)$ from the exponential family and taking just one optimisation step, we recover \citeauthor{fahrmeir1992posterior}'s (\citeyear{fahrmeir1992posterior}) approximate mode estimator. 
Our analysis differs from \citeauthor{fahrmeir1992posterior}'s in that (a) we show that online mode estimation can in theory be performed exactly by solving Bellman's equation, (b) we consider a general (rather than exponential) observation distribution, and (c) we allow more than one optimisation step. 

\section{Laplace Gaussian filter as a special case}
\label{app LGF}

When the state-transition density is linear and Gaussian, step 4 in the algorithm of \cite{koyama2010approximate} can be performed in closed form. The first-order Laplace Gaussian filter in step three of their algorithm is then equivalent to maximisation~\eqref{updatingrule v2}. Both algorithms differ when the state transition is nonlinear and/or non-Gaussian.

\section{Implicit stochastic gradient method as a special case}
\label{app ISG}

In model~\eqref{DGP0.3}, suppose that $\bm{c}=\bm{0}$, $\bm{Q}=\bm{0}$ and $\bm{T}=\mathds{1}_{m\times m}$, where $\mathds{1}_{m\times m}$ is an $m\times m$ identity matrix. The (constant) state $\bm{\alpha}_t=\bm{\alpha}_1$ for all $t=1,2,\ldots$ now represents an unknown parameter to be estimated recursively over time. The prediction step of the Bellman filter simplifies to $\smash{\bm{a}_{t|t-1}=\bm{a}_{t-1|t-1}}$ and $\smash{\bm{I}_{t|t-1}=\bm{I}_{t-1|t-1}}$, while update~\eqref{updatingrule v2} equates to an implicit stochastic gradient method (e.g.\ \citealp{toulis2015scalable},  \citealp{toulis2016towards},  \citealp{toulis2017asymptotic},  \citealp{toulis2021proximal}). In this case, the Bellman filter with BHHH updating steps becomes an implicit version of the (explicit) stochastic gradient methods in~\citet[eq.\ 2.14]{amari2000adaptive} or \citet[eq.\ 11]{toulis2017asymptotic}. While such methods are asymptotically convergent to the true parameter value, the Bellman filter typically remains perpetually responsive.

\section{Proof of Theorem~\ref{thrm1}}
\label{proof of theorem 1}

\begin{enumerate}
\item The objective function $V_t(\bm{a}):= \ell(\bm{y}_t|\bm{a})-1/2\|\bm{a}-\bm{a}_{t|t-1}\|^2_{\bm{I}_{t|t-1}}$ is strongly concave with probability one because $\ell(\bm{y}_t|\cdot)$ is concave with probability one (Assumption~1a), while $-1/2\|\bm{a}-\bm{a}_{t|t-1}\|^2_{\bm{I}_{t|t-1}}$ is strongly concave. Because the objective function is also real valued, $\bm{a}_{t|t}$ is well defined. Moreover, $V_t(\bm{a}_{t|t}) \geq V_t(\bm{a}_{t|t-1})=\ell(\bm{y}_t|\bm{a}_{t|t-1})$, i.e.
\begin{equation}
\label{bound on stepsize 1}
0\leq   V_t(\bm{a}_{t|t}) - V_t(\bm{a}_{t|t-1}) \; = \ell(\bm{y}_t|\bm{a}_{t|t}) -\frac{1}{2}\left\|\bm{a}_{t|t}-\bm{a}_{t|t-1} \right\|_{\bm{I}_{t|t-1}}^2 - \ell(\bm{y}_t|\bm{a}_{t|t-1}).
\end{equation}
Re-arranging gives
\begin{equation}
\frac{1}{2}\left\|\bm{a}_{t|t}-\bm{a}_{t|t-1} \right\|_{\bm{I}_{t|t-1}}^2 \; \leq \;  \ell(\bm{y}_t|\bm{a}_{t|t}) -  \ell(\bm{y}_t|\bm{a}_{t|t-1}).
\end{equation}
The right-hand side is bounded because the set $\{\bm{a}\in \mathbb{R}^m:V_t(\bm{a}) \geq V_t(\bm{a}_{t|t-1})\}$ is bounded.

\item Assuming that $\bm{a}\mapsto \ell(\bm{y}_t|\bm{a})$ is twice continuously differentiable (Assumption~2b), the following first- and second-order conditions must hold at the Bellman-filtered state $\bm{a}_{t|t}\in \mathbb{R}^m$: 
\begin{align}
\text{first-order condition:} \hspace{-2cm} &&\nabla \, \ell(\bm{y}_t|\bm{a}_{t|t})-\bm{I}_{t|t-1}(\bm{a}_{t|t}-\bm{a}_{t|t-1})&=\bm{0}_m,
\\
\text{second-order condition:}\hspace{-2cm}  &&
\nabla^2\, \ell(\bm{y}_t|\bm{a}_{t|t})-\bm{I}_{t|t-1} &\leq \bm{0}_{m\times m},\label{second order condition}
\end{align}
where the weak inequality in the second line means the matrix on the left-hand side is negative semi-definite. 
Differentiating the first-order condition with respect to $\bm{a}_{t|t-1}$, we obtain
\begin{equation}
\nabla^2\, \ell(\bm{y}_t|\bm{a}_{t|t})\, \frac{\dd \bm{a}_{t|t}}{\dd \bm{a}_{t|t-1}'} \; =\; \bm{I}_{t|t-1}\, \left[\frac{\dd \bm{a}_{t|t}}{\dd \bm{a}_{t|t-1}'}- \mathds{1}_{m\times m} \right],
\end{equation}
which can be re-written as
\begin{equation}
\label{equation with X and Y}
\frac{\dd \bm{a}_{t|t}}{\dd \bm{a}_{t|t-1}'} \;=\; \left[ \bm{I}_{t|t-1}  \,-\, \nabla^2\, \ell(\bm{y}_t|\bm{a}_{t|t}) \right]^{-1} \,\bm{I}_{t|t-1}  ,
\end{equation}
where the required inverse exists because $\bm{I}_{t|t-1} - \nabla^2\, \ell(\bm{y}_t|\bm{a}_{t|t})$ is positive definite by assumption.

Next, we use a result of \citet[eq.\ 2]{wang1993some}, which says that $\lambda_{\min}(\bm{A}) \lambda_{\min}(\bm{B}) \leq \lambda_{\min}(\bm{A}\bm{B})$ for two square, symmetric and positive semidefinite matrices $\bm{A}$ and $\bm{B}$, where $\lambda_{\min}(\cdot)$ denotes the smallest eigenvalue of a matrix. Denoting $\bm{H}_t:=-\nabla^2 \ell(\bm{y}_t|\bm{a}_{t|t})$ and applying this result to $(\bm{I}_{t|t-1}+\bm{H}_t)^{-1}\bm{I}_{t|t-1}$ yields
\begin{equation}
\label{ev1}
0 < \frac{\lambda_{\min}(\bm{I}_{t|t-1})}{\lambda_{\max}(\bm{I}_{t|t-1}+\bm{H}_t)}= \lambda_{\min}[(\bm{I}_{t|t-1}+\bm{H}_t)^{-1}]\lambda_{\min}(\bm{I}_{t|t-1})\leq \lambda_{\min}[(\bm{I}_{t|t-1}+\bm{H}_t)^{-1}\bm{I}_{t|t-1}] .
\end{equation}
Hence, the eigenvalues of $(\bm{I}_{t|t-1}+\bm{H}_t)^{-1}\bm{I}_{t|t-1}$ are strictly positive. 
To show that the eigenvalues of $(\bm{I}_{t|t-1}+\bm{H}_t)^{-1}\bm{I}_{t|t-1}$ are bounded above by one, we note that
\begin{align}
\lambda_{\max}[(\bm{I}_{t|t-1}+\bm{H}_t)^{-1}\bm{I}_{t|t-1} ] &= \lambda_{\max}[\mathds{1}_{m\times m} - (\bm{I}_{t|t-1}+\bm{H}_t)^{-1}\bm{H}_{t}],
\notag
\\
&=1 -\lambda_{\min}[(\bm{I}_{t|t-1}+\bm{H}_t)^{-1}\bm{H}_t ] ,
\notag
\\
&\leq 1 -\lambda_{\min}[(\bm{I}_{t|t-1}+\bm{H}_t)^{-1}] \lambda_{\min}(\bm{H}_t) ,
\notag
\\
&=  1 -\frac{ \lambda_{\min}(\bm{H}_t)}{\lambda_{\max}(\bm{I}_{t|t-1}+\bm{H}_t)} \;\leq\; 1 -\frac{ \lambda_{\min}(\bm{H}_t)}{\lambda_{\max}(\bm{I}_{t|t-1})+\lambda_{\max}(\bm{H}_t)} ,
\label{ev2}
\end{align}
which does not exceed (is strictly smaller than) than unity if $\bm{H}_t\geq 0$ ($\bm{H}_t>0$). The conditions $\bm{H}_t\geq 0$ or $\bm{H}_t>0$ are ensured, respectively, if the observation log density is concave (Assumption 1a) or strictly concave (Assumption 1b).

Next, we use the well known fact (e.g.\ \citealp[p.\ 39]{jungers2009joint}) that the induced matrix norm satisfies
$$
\| \bm{M}\|_{\bm{W}} =\| \bm{W}^{1/2} \bm{M} \bm{W}^{-1/2}\| =\sqrt{ \lambda_{\max}\left(   \bm{W}^{1/2} \bm{M} \bm{W}^{-1}  \bm{M}' \bm{W}^{1/2} \right)}=\sqrt{ \lambda_{\max}\left(   \bm{M} \bm{W}^{-1}  \bm{M}' \bm{W} \right)},
$$
where the last equality follows by cyclically rotating inside the $\lambda_{\max}(\cdot)$ operator. Here $\bm{M},\bm{W}\in \mathbb{R}^{m\times m}$ and $\bm{W}>\bm{0}$ is the positive definite weight matrix. Using this fact along with the symmetry of $\bm{I}_{t|t-1}$ and $\bm{H}_t$, we then obtain
\begin{align}
\left \| \frac{\dd \bm{a}_{t|t}}{\dd \bm{a}'_{t|t-1} } \right\|_{\bm{I}_{t|t-1}} & =
\Big \| (\bm{I}_{t|t-1}+\bm{H}_t)^{-1}\bm{I}_{t|t-1} \Big \|_{\bm{I}_{t|t-1}} ,
\notag
\\
&= \sqrt{ \lambda_{\max}\left\{  (\bm{I}_{t|t-1}+\bm{H}_t)^{-1}\bm{I}_{t|t-1} \bm{I}_{t|t-1}^{-1} \bm{I}_{t|t-1}  (\bm{I}_{t|t-1}+\bm{H}_t)^{-1} \bm{I}_{t|t-1} \right\}},
\notag
\\
&= \sqrt{ \lambda_{\max}\left\{ \Big[ (\bm{I}_{t|t-1}+\bm{H}_t)^{-1}\bm{I}_{t|t-1}\Big]^2 \right\} } \; \leq \; 1 -\frac{ \lambda_{\min}(\bm{H}_t)}{\lambda_{\max}(\bm{I}_{t|t-1})+\lambda_{\max}(\bm{H}_t)},
\label{useful result for thrm2}
\end{align}
where we have used equation~\eqref{ev2} along with the fact that the eigenvalues of the square of a matrix are equal to the squares of the eigenvalues of the original matrix. If additionally Assumption~1a (1b) holds, then we have $\lambda_{\min}(\bm{H}_t)\geq 0$ ($\lambda_{\min}(\bm{H}_t)> 0$), such that the right-hand side does not exceed (is strictly less than) unity.

\item Assuming that $\bm{a}\mapsto \ell(\bm{y}_t|\bm{a})$ is strongly concave with parameter $\epsilon>0$ (Assumption~1c) and once
continuously differentiable (Assumption~2a), standard arguments (e.g.\ \citealp[eq.\ 2.1.17]{nesterov2003introductory}) give
\begin{equation}
\label{strong concavity}
\braket{\,\bm{a}_t - \bm{\alpha}_t\,,\,\nabla \ell(\bm{y}_t|\bm{a}_t) - \nabla \ell(\bm{y}_t|\bm{\alpha}_t) \, } \; \leq \; - \epsilon \, \cdot\, \|\bm{a}_t - \bm{\alpha}_t \|^2, \qquad \forall \bm{a}_t,\bm{\alpha}_t\in \mathbb{R}^m.
\end{equation}
Strong concavity means that equation~\eqref{strong concavity} holds for all pairs $\bm{a}_t,\bm{\alpha}_t\in \mathbb{R}^m$, but we shall need it only when $\bm{\alpha}_t$ is the true state. Assuming differentiability (Assumption~2a), the first-order condition $\bm{I}_{t|t-1}( \bm{a}_{t|t} - \bm{a}_{t|t-1} )= \nabla\ell(\bm{y}_t|\bm{a}_{t|t})$ is rewritten by pre-multiplying the equation by $\bm{I}_{t|t-1}^{-1/2}$ and subtracting $\bm{I}_{t|t-1}^{1/2}\bm{\alpha}_t-\bm{I}_{t|t-1}^{-1/2}\nabla \ell(\bm{y}_t|\bm{\alpha}_t) $ from both sides to obtain
\begin{equation}
\bm{I}_{t|t-1}^{1/2}(\bm{a}_{t|t}-\bm{\alpha}_t) - \bm{I}_{t|t-1}^{-1/2} \left\{\nabla \ell(\bm{y}_t|\bm{a}_{t|t})-\nabla \ell(\bm{y}_t|\bm{\alpha}_{t}) \right\}
 =\bm{I}_{t|t-1}^{1/2} (\bm{a}_{t|t-1}-\bm{\alpha}_t) + \bm{I}_{t|t-1}^{-1/2}\nabla \ell(\bm{y}_t|\bm{\alpha}_{t}).
\end{equation}
Computing the quadratic norm on both sides and ignoring one term on the left, we obtain an inequality as follows:
\begin{align}
&\left\| \bm{a}_{t|t}-\bm{\alpha}_t\right\|_{\bm{I}_{t|t-1}}^2 - 2 \braket{\bm{a}_{t|t}-\bm{\alpha}_t,\nabla \ell(\bm{y}_t|\bm{a}_{t|t})-\nabla \ell(\bm{y}_t|\bm{\alpha}_{t})}   \notag
\\
&\hspace{4cm}\leq
  \left\| \bm{a}_{t|t-1}-\bm{\alpha}_t\right\|_{\bm{I}_{t|t-1}}^2 +2\braket{\bm{a}_{t|t-1}-\bm{\alpha}_t,\nabla\ell(\bm{y}_t|\bm{\alpha}_t)} +\left\|\nabla \ell(\bm{y}_t|\bm{\alpha}_{t}) \right\|^2_{\bm{I}_{t|t-1}^{-1}}.\notag
\end{align}
By strong concavity \eqref{strong concavity}, we have
\begin{align}
&\left\| \bm{a}_{t|t}-\bm{\alpha}_t\right\|_{\bm{I}_{t|t-1}}^2 + 2 \epsilon\cdot \left\| \bm{a}_{t|t}-\bm{\alpha}_t\right\|^2   \notag
  \\
 &\hspace{4cm}\leq
  \left\| \bm{a}_{t|t-1}-\bm{\alpha}_t\right\|_{\bm{I}_{t|t-1}}^2+2\braket{\bm{a}_{t|t-1}-\bm{\alpha}_t,\nabla\ell(\bm{y}_t|\bm{\alpha}_t)}
+ \left\|\nabla \ell(\bm{y}_t|\bm{\alpha}_{t}) \right\|^2_{\bm{I}_{t|t-1}^{-1}}.
\end{align}
Taking expectations yields
\begin{align}
&\mathbb{E}\left( \left\| \bm{a}_{t|t}-\bm{\alpha}_t\right\|_{\bm{I}_{t|t-1}}^2 \right) +2 \epsilon\;
\mathbb{E}\left(\left\| \bm{a}_{t|t}-\bm{\alpha}_t\right\|^2\right)\notag
\\
 &\hspace{5cm}\leq \mathbb{E}\left( \left\| \bm{a}_{t|t-1}-\bm{\alpha}_t\right\|_{\bm{I}_{t|t-1}}^2\right)+\mathbb{E}\left( \left\|\nabla \ell(\bm{y}_t|\bm{\alpha}_{t}) \right\|^2_{\bm{I}_{t|t-1}^{-1}}\right).
\end{align}
where we have used $\mathbb{E}\braket{\bm{a}_{t|t-1}-\bm{\alpha}_t,\nabla\ell(\bm{y}_t|\bm{\alpha}_t)}=0$, which is obvious from the expectation of the score being zero, i.e.\ $\mathbb{E}[\nabla\ell(\bm{y}_t|\bm{\alpha}_t)|\bm{\alpha}_t]=0$. Finally, the theorem is proved by noting that the left-hand side is $\mathbb{E}\left( \left\| \bm{a}_{t|t}-\bm{\alpha}_t\right\|_{\bm{I}_{t|t-1}+2\epsilon\mathds{1}_{m\times m}}^2 \right)$, where $\mathds{1}_{m\times m}$ is an $m\times m$ identity matrix, while Assumption~3 together with the assumed positive definiteness of $\bm{I}_{t|t-1}$ implies that on the right-hand side we have
$$\mathbb{E}\left( \|
\nabla \ell(\bm{y}_t|\bm{\alpha}_{t}) \|^2_{\bm{I}_{t|t-1}^{-1}}\right)\leq \sigma^2/\lambda_{\min}.
$$
\end{enumerate}

\section{Comparison of Theorem~\ref{thrm1} with \cite{toulis2016towards}}
\label{comparison toulis}

This section casts light on the different definitions of strong concavity used in Theorem~\ref{thrm1} and in \cite{toulis2016towards}. Here we show  that Theorem~\ref{thrm1} applies to e.g.\ the Kalman filter, while the seemingly stronger result in \cite{toulis2016towards} does not.

By the combination of Assumptions 1c (strong concavity) and 2b (twice differentiability), part~3 of Theorem~\ref{thrm1} assumes that the negative Hessian  $-\nabla^2 \ell(\bm{y}_t|\bm{a})$ is strictly positive definite with smallest eigenvalue $\epsilon>0$.
Standard arguments (e.g.\ \citealp[eq.\ 2.1.17]{nesterov2003introductory})  imply that
\begin{equation}
\label{used in thrm1}
 \braket{\bm{a}_t - \bm{\alpha}_t,\nabla \ell(\bm{y}_t|\bm{a}_t) - \nabla \ell(\bm{y}_t|\bm{\alpha}_t)  } \; \leq \; - \epsilon \, \cdot\, \|\bm{a}_t - \bm{\alpha}_t \|^2, \qquad \forall \bm{a}_t,\bm{\alpha}_t\in \mathbb{R}^m.
\end{equation}
\citet{toulis2016towards} take a different view on strong concavity, defining a log-likelihood function to be strongly concave, for a typical observation $\bm{y}_t\in \mathbb{R}^l$, when
\begin{equation}
\label{toulis strong concavity}
\text{strong concavity in \cite{toulis2016towards}:}\quad 
 \braket{\bm{a}_t - \bm{\alpha}_t,\nabla \ell(\bm{y}_t|\bm{a}_t)  } \leq- \epsilon\, \cdot\, \|\bm{a}_t - \bm{\alpha}_t \|^2,\quad \forall \bm{a}_t,\bm{\alpha}_t\in \mathbb{R}^m,
\end{equation}
which differs from definition~\eqref{used in thrm1} in that the term~$\nabla \ell(\bm{y}_t|\bm{\alpha}_t)$ is no longer present. Inequality~\eqref{toulis strong concavity} appears in Remark~2 and equation~17 of the supplementary material to \cite{toulis2016towards}, where $\mu_t>0$ appears instead of our $\epsilon$, the random draw $\xi_t$ appears instead of our $\bm{y}_t$, $\theta_t$ appears instead of our $\bm{a}_t$, the true value $\theta_{\star}$ appears instead of our $\bm{\alpha}_t$, their $L$ is a negative log-likelihood function, and index $n$ is used instead of our $t$. \cite{toulis2016towards} permit the parameter of strong concavity to depend on the observation; for simplicity, we do not. The term $\nabla \ell(\bm{y}_t|\bm{\alpha}_t)$, which appears in equation~\eqref{used in thrm1} but not equation~\eqref{toulis strong concavity}, is the score function evaluated at the true parameter; hence, this term is zero on average. For many models of interest, however, realisations of the score are non-zero with probability one, such that definition~\eqref{toulis strong concavity} materially differs from~\eqref{used in thrm1}.

While definition~\eqref{used in thrm1} of strong concavity was used in the proof of Theorem~\ref{thrm1}, definition~\eqref{toulis strong concavity} allows a stronger result due to \cite{toulis2016towards} to be derived. First, the first-order condition corresponding to maximisation~\eqref{updatingrule v2}, i.e.\ $\bm{I}_{t|t-1}(\bm{a}_{t|t}-\bm{a}_{t|t-1})=\nabla \ell(\bm{y}|\bm{a}_{t|t})$, is rewritten as
\begin{equation}
\bm{I}_{t|t-1}^{1/2}(\bm{a}_{t|t}-\bm{\alpha}_t) - \bm{I}_{t|t-1}^{-1/2}\nabla \ell(\bm{y}_t|\bm{a}_{t|t})
 =\bm{I}_{t|t-1}^{1/2} (\bm{a}_{t|t-1}-\bm{\alpha}_t) .
\end{equation}
Computing the quadratic norm on both sides, we have 
\begin{equation}
 \left\| \bm{a}_{t|t}-\bm{\alpha}_t\right\|_{\bm{I}_{t|t-1}}^2 - 2  \braket{\bm{a}_{t|t}-\bm{\alpha}_t,\nabla \ell(\bm{y}_t|\bm{a}_{t|t})}  + \left \|\nabla \ell(\bm{y}_t|\bm{a}_{t|t}) \right \|_{\bm{I}^{-1}_{t|t-1}}^2
= \left\| \bm{a}_{t|t-1}-\bm{\alpha}_t\right\|_{\bm{I}_{t|t-1}}^2.
\end{equation}
By strong concavity~\eqref{toulis strong concavity}, it follows that
\begin{equation}
 \left\| \bm{a}_{t|t}-\bm{\alpha}_t\right\|_{\bm{I}_{t|t-1}}^2 + 2\,\epsilon\cdot \left\| \bm{a}_{t|t}-\bm{\alpha}_t\right\|^2  + \left \|\nabla \ell(\bm{y}_t|\bm{a}_{t|t}) \right \|_{\bm{I}^{-1}_{t|t-1}}^2
\leq \left\| \bm{a}_{t|t-1}-\bm{\alpha}_t\right\|_{\bm{I}_{t|t-1}}^2 .
\end{equation}
Ignoring the third term on the left-hand side and combining terms, we find
\begin{equation}
 \left\| \bm{a}_{t|t}-\bm{\alpha}_t\right\|_{\bm{I}_{t|t-1}+2\,\epsilon \mathds{1}_{m\times m}}^2  
\leq \left\| \bm{a}_{t|t-1}-\bm{\alpha}_t\right\|_{\bm{I}_{t|t-1}}^2 ,
\end{equation}
where $\mathds{1}_{m\times m}$ denotes an $m\times m$ identity matrix. In \citet[p.\ 1291]{toulis2016towards} it holds that $\bm{I}_{t|t-1}=\gamma^{-1}\mathds{1}_{m\times m}$, where $\mathds{1}_{m\times m}$ is an $m\times m$ identity matrix and $\gamma>0$ is a learning parameter, in which case we obtain
\begin{equation}
\label{result toulis}
\left\| \bm{a}_{t|t}-\bm{\alpha}_t\right\|^2  
\leq \frac{1}{1+2\gamma \epsilon}\left\| \bm{a}_{t|t-1}-\bm{\alpha}_t\right\|^2,
\end{equation}
as in \citet[p.\ 1291]{toulis2016towards}. This result is stronger than that in Theorem~\ref{thrm1}, because \eqref{result toulis} holds for all realisations~$\bm{y}_t$, without taking expectations. 
Inequality~\eqref{result toulis} implies that the update is `contracting almost surely' \cite[p.\ 1291]{toulis2016towards}. Unfortunately, this is desirable property is not observed in practice for e.g.\ the Kalman filter.

To explain why the Kalman filter fails to be almost surely contractive in the sense of \cite{toulis2016towards}, we observe that the Kalman filter satisfies our assumption~\eqref{used in thrm1} as used in Theorem~\ref{thrm1}, but \emph{not} assumption~\eqref{toulis strong concavity} as used by \cite{toulis2016towards}. To demonstrate this, we take the linear Gaussian state-space model in Corollary~\ref{corol1}, such that the observation density $p(\bm{y}_t|\bm{\alpha}_t)$ is Gaussian with mean $\bm{d}+\bm{Z}\bm{\alpha}_t$ and covariance matrix $\bm{H}$, which is assumed positive definite. The log-likelihood function and its gradient then read
\begin{align}
\ell(\bm{y}_t |\bm{\alpha}_t ) \; &= \; -\frac{1}{2} (\bm{y}_t - \bm{d} - \bm{Z}\bm{\alpha}_t)'\, \bm{H}^{-1} \,(\bm{y}_t - \bm{d} - \bm{Z}\bm{\alpha}_t)
+\text{constants},
\\
\nabla \ell(\bm{y}_t |\bm{\alpha}_t ) \; & = \bm{Z}'\bm{H}^{-1} \,(\bm{y}_t - \bm{d} - \bm{Z}\bm{\alpha}_t).
\end{align}
The multivariate Gaussian is strongly concave according to our definition~\eqref{used in thrm1}, because
\begin{align}
\braket{\bm{a}_t-\bm{\alpha}_t, \nabla \ell(\bm{y}_t|\bm{a}_t)-\nabla \ell(\bm{y}_t|\bm{\alpha}_t) } &= \braket{\bm{a}_t-\bm{\alpha}_t, \bm{Z}'\bm{H}^{-1} \,(\bm{y}_t - \bm{d} - \bm{Z}\bm{a}_t)-\bm{Z}'\bm{H}^{-1} \,(\bm{y}_t - \bm{d} - \bm{Z}\bm{\alpha}_t) },
\notag
\\
&= - \braket{\bm{a}_t-\bm{\alpha}_t, \bm{Z}'\bm{H}^{-1} \bm{Z}(\bm{a}_t-\bm{\alpha}_t) },\notag
\\
&= - \| \bm{a}_t-\bm{\alpha}_t\|_{ \bm{Z}'\bm{H}^{-1} \bm{Z}}^2 \,,\notag
\\
&
\leq - \lambda_{\min}\left( \bm{Z}'\bm{H}^{-1} \bm{Z}\right)\, \cdot\,\|\bm{a}_t-\bm{\alpha}_t \|^2,
\end{align}
where $\lambda_{\min}(\cdot)$ denotes the smallest eigenvalues of a matrix. Hence, condition~\eqref{used in thrm1} is satisfied with $\epsilon=\lambda_{\min}(\bm{Z}'\bm{H}^{-1} \bm{Z})>0$. Conversely, the multivariate Gaussian fails to be strongly concave when using the alternative definition~\eqref{toulis strong concavity} of \cite{toulis2016towards}, because
\begin{equation}
\braket{\bm{a}_t-\bm{\alpha}_t, \nabla \ell(\bm{y}_t|\bm{a}_t) } \;=\; \braket{\bm{a}_t-\bm{\alpha}_t, \bm{Z}'\bm{H}^{-1} \,(\bm{y}_t - \bm{d} - \bm{Z}\bm{a}_t) } \; \nleq\; -\,\text{positive scalar}\, \cdot\, \|\bm{a}_t-\bm{\alpha}_t\|^2.
\end{equation}

Stepping back, it is not too surprising that the almost sure contractive property of \cite{toulis2016towards} fails for the Kalman filter, because the Kalman filter can (and does) move in the wrong direction when confronted with atypical observations. The contribution of Theorem~\ref{thrm1} is to demonstrate that, in a general context, such `bad' behaviour does not dominate. Theorem~\ref{thrm1} allows for the fact that updates may be less accurate than predictions, while still ensuring that the updates are contractive in quadratic mean towards a noise-dominated region around the true state, which is the situation that is relevant in practice.

\section{Proof of Proposition~\ref{corol unit root}}
\label{S geometric convergence}

Repeated self-substitution of the recursions~\eqref{MSE2} yields:
\begin{align*}
\textnormal{MSE}_{t|t}
\; &\leq\;  \left(\frac{\gamma}{\gamma+2\epsilon}\right)^{t}  \textnormal{MSE}_{1|0}+
 \frac{\sigma^2}{\gamma^2} \sum_{i=1}^{t}
 \left(\frac{\gamma}{\gamma+2\epsilon}\right)^i+\sigma_\eta^2\sum_{i=1}^{t-1}
 \left(\frac{\gamma}{\gamma+2\epsilon}\right)^i ,
 \\ 
 \;&=\; \left(\frac{\gamma}{\gamma+2\epsilon}\right)^{t} \textnormal{MSE}_{1|0} +\frac{\sigma^2}{\gamma^2} \left(\frac{\gamma}{\gamma+2\epsilon}\right)\frac{1-\left(\frac{\gamma}{\gamma+2\epsilon}\right)^t}{1-\frac{\gamma}{\gamma+2\epsilon}}+\sigma_{\eta}^2 \left(\frac{\gamma}{\gamma+2\epsilon}\right)\frac{1-\left(\frac{\gamma}{\gamma+2\epsilon}\right)^{t-1}}{1-\frac{\gamma}{\gamma+2\epsilon}},
\end{align*}
where the second line employs $\sum_{i=1}^t x^{i-1} = (1-x^t)/(1-x)$ for $-1<x<1$. Using $\gamma,\epsilon>0$ and taking the the limit $t\to \infty$ yields equation~\eqref{upper bound}.

\section{Proof of Theorem~\ref{thrm2}}
\label{app:thrm2}

By the chain rule, we have
\begin{align}
\left\| \frac{\dd \bm{a}_{t|t}}{\dd \bm{a}_{0|0}'}\right\|_{\bm{I}} &= \left\| \frac{\dd \bm{a}_{t|t}}{\dd \bm{a}_{t|t-1}'} \frac{\dd \bm{a}_{t|t-1}}{\dd \bm{a}_{t-1|t-1}'}   \times  \ldots \times  \frac{\dd \bm{a}_{1|1}}{\dd \bm{a}_{1|0}'} \frac{\dd \bm{a}_{1|0}}{\dd \bm{a}_{0|0}'}   \right\|_{\bm{I}}
\; \leq
\;
 \left \| \frac{\dd \bm{a}_{t|t}}{\dd \bm{a}_{t|t-1}'} \right\|_{\bm{I}} \left\|\bm{T}\right\|_{\bm{I}}  \times \ldots \times  \left\| \frac{\dd \bm{a}_{1|1}}{\dd \bm{a}_{1|0}'}  \right\|_{\bm{I}} \left\|\bm{T}\right\|_{\bm{I}},\notag
\\
&\leq \left(\left\|\bm{T}\right\|_{\bm{I}}\right)^t \prod_{\tau=1}^t \left(1 -\frac{ \lambda_{\min}(\bm{H}_\tau)}{\lambda_{\max}(\bm{I})+\lambda_{\max}(\bm{H}_\tau)}\right)\; \leq\; \left(\left\|\bm{T}\right\|_{\bm{I}}\right)^t  \left(1 -\frac{ \mu_{\min}}{\nu_{\max}+\mu_{\max}}\right)^t. \label{submultiplicative inequality}
\end{align}
The inequality in the first line holds by the sub-multiplicative property of the induced matrix norm in combination with the linear prediction step. The second line holds by equation~\eqref{useful result for thrm2}, where $\bm{H}_t:=-\nabla^2 \ell(\bm{y}_t|\bm{a}_{t|t})$. The last inequality holds because $\lambda_{\max}(\bm{I})=\nu_{\max}$ and $0\leq \mu_{\min} \leq \lambda_{\min}(\bm{H}_t)\leq \lambda_{\max}(\bm{H}_t)\leq  \mu_{\max}$ by assumption.

To prove equation~\eqref{stability}, we must still bound the term $\|\bm{T}\|_{\bm{I}}$. To this end, we define $\delta:=\lambda_{\min}(\bm{I}-\bm{T}' \bm{I}\bm{T})'\in \mathbb{R}$, which could be positive or negative. Since $\bm{I}$ is positive definite, we must have
\begin{equation}
\label{starting point0}
\delta=\lambda_{\min}(\bm{I}-\bm{T}' \bm{I}\bm{T})\leq \lambda_{\min}(\bm{I})=\nu_{\min},
\end{equation}
so $\delta\leq \nu_{\min}$. Next, we have the inequality
\begin{equation}
\label{starting point}
\bm{0} \leq \bm{I}-\delta \,\mathds{1}_{m \times m}-\bm{T}' \bm{I}\bm{T} ,
\end{equation}
as we will use below. As $\bm{I}$ is positive definite with smallest and largest eigenvalues $\nu_{\min}$ and $\nu_{\max}$ respectively, we have 
$$
\frac{1}{\nu_{\max}} \bm{I}\;\leq \; \mathds{1}_{m\times m}\;\leq \; 
\frac{1}{\nu_{\min}}  \bm{I}.
$$
When $\delta>0$, multiplying this sequence of inequalities by $-\delta$ yields
$$
\frac{-\delta }{\nu_{\max}}\,\bm{I}\; \geq \; -\delta\, \mathds{1}_{m\times m}\; \geq\;\frac{-\delta}{\nu_{\min}} \bm{I}, \qquad \delta>0.
$$
When $\delta<0$, we obtain instead
$$
\frac{-\delta }{\nu_{\max}}\,\bm{I}\; \leq \; -\delta\, \mathds{1}_{m\times m}\; \leq\; \frac{-\delta}{\nu_{\min}} \bm{I}, \qquad \delta<0.
$$
Combining the last two results, we see that $-\delta \mathds{1}_{m\times m}$ is bounded above by $-\delta/\nu_{\max} \bm{I}$ when $\delta>0$ and $-\delta /\nu_{\min} \bm{I}$ when $\delta<0$. This means that for all $\delta \in \mathbb{R}$, we can write
\begin{equation}
\label{useful step}
-\delta \, \mathds{1}_{m\times m} \;\leq\; - \min\left\{\frac{\delta}{\nu_{\min}},\frac{\delta}{\nu_{\max}}\right\} \bm{I}, \qquad  \delta \in \mathbb{R}.
\end{equation}
Using inequality~\eqref{useful step}, inequality~\eqref{starting point} can be further extended as
\begin{equation}
\label{xxx}
\bm{0} \leq \bm{I}-\delta\, \mathds{1}_{m \times m}-\bm{T}' \bm{I}\bm{T}  \leq \left(1 - \min\left\{\frac{\delta}{\nu_{\min}},\frac{\delta}{\nu_{\max}}\right\}  \right) \bm{I}-\bm{T}' \bm{I}\bm{T}.
\end{equation}
Equation~\eqref{xxx} shows that $ z^2 \bm{I}-\bm{T}'\bm{I}\bm{T}\geq \bm{0}$ for a particular value of $z$.
This is useful because from \citet[p.\ 39]{jungers2009joint} we have
\begin{equation}
\label{inf formulation of weighted norm}
\|\bm{T}\|_{\bm{I}} = \text{inf} \big\{ z\geq 0 : z^2 \bm{I} - \bm{T}' \bm{I}\bm{T}\geq \bm{0}\big\},
\end{equation}
which says that $\|\bm{T}\|_{\bm{I}}$ is the infimum of such values. Hence equations~\eqref{xxx} and \eqref{inf formulation of weighted norm} together imply
\begin{equation}
\label{upper bound on norm}
\|\bm{T}\|_{\bm{I}}  \leq \sqrt{1 - \min\left\{\frac{\delta}{\nu_{\min}},\frac{\delta}{\nu_{\max}}\right\}}.
\end{equation}
As a sanity check, we may verify that the right-hand side is nonnegative, as when $\delta>0$ we have 
$\delta\leq \nu_{\min}$ by equation~\eqref{starting point0} above. Substituting equation~\eqref{upper bound on norm} in equation~\eqref{submultiplicative inequality} yields equation~\eqref{stability} in the main text. 

To prove equation~\eqref{log condition} in the main text, compute the derivative of the logarithm of the right-hand side of equation~\eqref{stability} as follows: 
\begin{align}
\frac{\dd}{\dd t} \log\left[ \left(1-\frac{\delta}{\nu_{\min}}
\right)^{t/2} \left(1-\frac{\mu_{\min} }{\nu_{\max}+\mu_{\max} } \right)^t\right] =\frac{1}{2} \log  \left(1-\frac{\delta}{\nu_{\min}}
\right) + \log \left(1-\frac{\mu_{\min} }{\nu_{\max}+\mu_{\max} } \right).
\label{log growth rates}
\end{align}
When this quantity is strictly negative, exponential almost sure convergence to zero follows. 

\section{Lemma involving quadratic functions}
\label{app:lemma}

\begin{lemma} 
\label{lemma1}
Let $\bm{x},\bm{y}\in \mathbb{R}^m$. Let $\bm{A},\bm{B}\in \mathbb{R}^{m\times m}$ be symmetric positive definite matrices. Define $f:\mathbb{R}^m\to \mathbb{R}$ as
\begin{align}
f(\bm{x}) &:= \max_{\bm{y}} \left\{-\frac{1}{2}\bm{x}'\bm{A}\bm{x}-\frac{1}{2}\bm{y}'\bm{B}\bm{y} +\bm{x}'\bm{C}\bm{y} +\bm{a}' \bm{x}+\bm{b}' \bm{y}\right\},
\\
&=  \max_{\bm{y}} \left\{ -\frac{1}{2} \left[\begin{array}{c} \bm{x} \\ \bm{y} \end{array} \right]' \left[\begin{array}{cc} \bm{A} & -\bm{C} \\ -\bm{C}' & \bm{B} \end{array} \right] \left[\begin{array}{c} \bm{x} \\ \bm{y} \end{array} \right] +\left[\begin{array}{c} \bm{a} \\ \bm{b} \end{array} \right]' \left[\begin{array}{c} \bm{x} \\ \bm{y} \end{array} \right]  , \right\}
\end{align}
for $\bm{C},\bm{a},\bm{b}$ of appropriate size. Then $f(\bm{x})$ is multivariate quadratic with negative Hessian matrix $\bm{A}-\bm{C}\bm{B}^{-1}\bm{C}'$. When this negative Hessian is positive definite, the argmax of $f(\bm{x})$ over $\bm{x}$ equals $(\bm{A}-\bm{C}\bm{B}^{-1}\bm{C}')^{-1}(\bm{a}+\bm{C}\bm{B}^{-1}\bm{b})$.
\end{lemma}

\begin{proof}
Take $\bm{x}$ as fixed. The first-order condition for the maximisation over $\bm{y}$ reads $\bm{0}=-\bm{B}\bm{y}+\bm{b}+\bm{C'}\bm{x}$, which leads to $\bm{y} = \bm{B}^{-1}(\bm{b}+\bm{C}'\bm{x})$. Substituting the optimised value of $\bm{y}$ into the expression for $f(\bm{x})$ gives
$$
f(\bm{x}) =-\frac{1}{2}\bm{x}'\bm{A}\bm{x}-\frac{1}{2}(\bm{b}+\bm{C}'\bm{x})'\bm{B}^{-1}(\bm{b}+\bm{C}'\bm{x})+ \bm{x}'\bm{C}\bm{B}^{-1}(\bm{b}+\bm{C}'\bm{x})+\bm{a}' \bm{x} + \bm{b}'\bm{B}^{-1} (\bm{b}+\bm{C}'\bm{x}).
$$
Several terms cancel and remaining terms can be grouped as
$$
f(\bm{x}) =-\frac{1}{2}\bm{x}'(\bm{A}-\bm{C}\bm{B}^{-1}\bm{C}')\bm{x}+(\bm{a}+\bm{C}\bm{B}^{-1}\bm{b})'\bm{x}+\text{constants},
$$
where constants independent of $\bm{x}$ are ignored. When $\bm{A}-\bm{C}\bm{B}^{-1}\bm{C}$ is positive definite, this quadratic function of $\bm{x}$ is maximised at $(\bm{A}-\bm{C}\bm{B}^{-1}\bm{C}')^{-1}(\bm{a}+\bm{C}\bm{B}^{-1}\bm{b})$, completing the proof.
\end{proof}

\section{Proof of Proposition~\ref{prop2}}
\label{app:RTS}

To derive a relation between $\bm{a}_{t|n}$ and $\bm{a}_{t+1|n}$ in the context of approximately quadratic value functions, it is useful to define a new value function $U_{t,t+1}(\cdot,\cdot): \mathbb{R}^m\times \mathbb{R}^m \to \mathbb{R}$, which takes two state variables as input. This value function is defined using the partial sum~\eqref{partial sum}, and can be rewritten using the value functions $V_t(\cdot)$ and $W_{t+1}(\cdot)$  defined in equations~\eqref{V} and~\eqref{W}, respectively, as follows:\begin{align}
 U_{t,t+1}(\bm{a}_t,\bm{a}_{t+1}) &:=\underset{\bm{a}_1,\ldots,\bm{a}_{t-1},\bm{a}_{t+2},\ldots,\bm{a}_{n}}{\max} \; L_{1:n}(\bm{a}_{1},\ldots, \bm{a}_{n}),\label{U def}
 \\
 &=\underset{\bm{a}_1,\ldots,\bm{a}_{t-1},\bm{a}_{t+2},\ldots,\bm{a}_{n}}{\max} \; \big[ L_{1:t}(\bm{a}_{1},\ldots, \bm{a}_{t})
+\ell(\bm{a}_{t+1}|\bm{a}_t)+ L_{t+1:n}(\bm{a}_{t+1},\ldots, \bm{a}_{n}) \big],
\\
 &=\left[\underset{\bm{a}_1,\ldots,\bm{a}_{t-1}}{\max} L_{1:t}(\bm{a}_{1},\ldots, \bm{a}_{t}) \right]+\ell(\bm{a}_{t+1}|\bm{a}_t) +\left[\underset{\bm{a}_{t+2},\ldots,\bm{a}_{n}}{\max} L_{t+1:n}(\bm{a}_{t+1},\ldots, \bm{a}_{n}) \right],
 \\
&=V_t(\bm{a}_t) +\ell(\bm{a}_{t+1}|\bm{a}_t)+ W_{t+1}(\bm{a}_{t+1}),\\
&=-\frac{1}{2} \| \bm{a}_t-\bm{a}_{t|t}\|^2_{\bm{I}_{t|t} }
-\frac{1}{2} \| \bm{a}_{t+1}-\bm{c}-\bm{T}\bm{a}_t\|^2_{\bm{Q}^{-1}}-\frac{1}{2}\|\bm{a}_{t+1}-\widehat{\bm{a}}_{t+1|t+1}\|^2_{\widehat{\bm{I}}_{t+1|t+1} }.
\label{U with norm}
\end{align}
In the last line, we take a linear Gaussian state equation as in Corollary~\ref{corol1}, and use the assumption that $V_t(\bm{a}_t)$ is multivariate quadratic with argmax $\bm{a}_{t|t}$ and negative Hessian matrix $\bm{I}_{t|t}$, while $W_{t+1}(\bm{a}_{t+1})$ is similarly multivariate quadratic with argmax $\widehat{\bm{a}}_{t+1|t+1}$ and negative Hessian matrix $\widehat{\bm{I}}_{t+1|t+1}$. Here, hats denote `backward filtered' quantities. It follows that $U_{t,t+1}(\cdot,\cdot)$ is a multivariate quadratic function in two state variables, $\bm{a}_t$ and $\bm{a}_{t+1}$. 

From definition~\eqref{U def}, it is clear that $Z_t(\cdot)$ and $Z_{t+1}(\cdot)$ defined in equation~\eqref{Z} can be recovered from $U_{t,t+1}(\cdot,\cdot)$ as follows:
\begin{align}
Z_t(\bm{a}_t) &=  \underset{\bm{a}_{t+1}}{\max}\quad U_{t,t+1}(\bm{a}_t,\bm{a}_{t+1}), \label{Z1}
\\
Z_{t+1}(\bm{a}_{t+1}) &=  \underset{\bm{a}_{t}}{\max} \quad U_{t,t+1}(\bm{a}_t,\bm{a}_{t+1}). \label{Z2}
\end{align}
Since $\bm{a}_{t|n}:=\arg \max_{\bm{a}} Z_t(\bm{a})$ while $\bm{a}_{t+1|n}:=\arg \max_{\bm{a}} Z_{t+1}(\bm{a})$, it is clear that $U_{t,t+1}(\cdot,\cdot)$ is maximised when $\bm{a}_t=\bm{a}_{t|n}$ and $\bm{a}_{t+1}=\bm{a}_{t+1|n}$. We evaluate $U_{t,t+1}(\cdot,\cdot)$ at $\bm{a}_{t+1}=\bm{a}_{t+1|n}$. Subsequently, the first-order condition with respect to $\bm{a}_t$ reads
$$
\bm{0}=\bm{I}_{t|t} (\bm{a}_t - \bm{a}_{t|t}) -\bm{T}'\bm{Q}^{-1}(\bm{a}_{t+1|n} -\bm{c} -\bm{T} \bm{a}_t) .
$$
Solving for $\bm{a}_t$ yields $\bm{a}_{t|n}$, which can be usefully rewritten as
\begin{align}
\bm{a}_{t|n} &=\big(\bm{I}_{t|t}\,+\,\bm{T}' \bm{Q}^{-1}\bm{T}\big)^{-1}\,\big(\bm{I}_{t|t}\,\bm{a}_{t|t}+ \bm{T}' \bm{Q}^{-1}(\bm{a}_{t+1|n}-\bm{c})\big),
\\
&= \bm{a}_{t|t} + (\bm{I}_{t|t}\,+\,\bm{T}' \bm{Q}^{-1}\bm{T}\big)^{-1} \bm{T}'\bm{Q}^{-1}\,\big(\bm{a}_{t+1|n}-\bm{c}-\bm{T} \bm{a}_{t|t}\big),
\\
&=\bm{a}_{t|t} + \bm{I}_{t|t}^{-1}\, \bm{T}'
\, \big(\bm{T}\bm{I}_{t|t}^{-1} \bm{T}'+\bm{Q}\big)^{-1}
\,\big(\bm{a}_{t+1|n}-\bm{c}-\bm{T} \bm{a}_{t|t}\big),
\\
&= \bm{a}_{t|t} +  \bm{I}_{t|t}^{-1}\, \bm{T}'
\, \bm{I}_{t+1|t} \,  \big(\bm{a}_{t+1|n}-\bm{a}_{t+1|t}\big).
\label{alphastar3}
\end{align}
This second line expresses $\bm{a}_{t|n}$ as the sum of $\bm{a}_{t|t}$ and a correction that is linear in $\bm{a}_{t+1|n} -\bm{c}- \bm{T}\bm{a}_{t|t}$. The third line uses matrix-inversion formulas by \citet[eqns. 9--11]{henderson1981deriving} to ensure that $\bm{Q}^{-1}$ no longer appears, such that by a limiting argument the result remains valid even when $\bm{Q}$ is singular. The last line employs the prediction step $\bm{a}_{t+1|t}:=\bm{c}+\bm{T} \bm{a}_{t|t}$  and $\bm{I}_{t+1|t}:=(\bm{T}\bm{I}_{t|t}^{-1} \bm{T}'+\bm{Q})^{-1}$. Equation~\eqref{alphastar3} is the Rauch-Tung-Striebel smoother expression, given in the main article in equation~\eqref{RTS1}.

To derive the backward recursion for the precision matrix, we note that $U_{t,t+1}(\cdot,\cdot)$ in equation~\eqref{U with norm} can be written using matrix notation as
\begin{align}
U_{t,t+1}(\bm{a}_t,\bm{a}_{t+1})&=-\frac{1}{2} \left[\begin{array}{c} \bm{a}_{t} \\ \bm{a}_{t+1} \end{array} \right]' \left[\begin{array}{cc} \bm{I}_{t|t}+\bm{T}' \bm{Q}^{-1}\bm{T} & -\bm{T}'\bm{Q}^{-1} \\ -\bm{Q}^{-1}\bm{T} & \widehat{\bm{I}}_{t+1|t+1}+\bm{Q}^{-1} \end{array} \right] \left[\begin{array}{c} \bm{a}_t \\ \bm{a}_{t+1} \end{array} \right] 
\\
&\hspace{1cm}+\left[\begin{array}{c} \bm{I}_{t|t} \bm{a}_{t|t}-\bm{T}'\bm{Q}^{-1}\bm{c} \\ \bm{Q}^{-1}\bm{c}+ \widehat{\bm{I}}_{t+1|t+1}\bm{a}_{t+1|t+1:n} \end{array} \right]' \left[\begin{array}{c} \bm{a}_t \\ \bm{a}_{t+1} \end{array} \right]+\text{constants},\notag
\end{align}
where any constants that do not depend on $\bm{a}_t$ and $\bm{a}_{t+1}$ are ignored. This representation together with Lemma~\ref{lemma1} implies that $Z_t(\cdot):=\max_{\bm{a}}U_{t,t+1}(\cdot,\bm{a})$ is multivariate quadratic functions with negative Hessian matrix given by the following Schur complement:
\begin{align}
\bm{I}_{t|n}&=\bm{I}_{t|t}+\bm{T}' \bm{Q}^{-1} \bm{T} - \bm{T}'\bm{Q}^{-1} (\widehat{\bm{I}}_{t+1|t+1}+\bm{Q}^{-1})^{-1}\bm{Q}^{-1}\bm{T},
\\
&=\bm{I}_{t|t} + \bm{T}' (\widehat{\bm{I}}^{-1}_{t+1|t+1}+\bm{Q})^{-1} \bm{T},
\end{align}
where the second line employs the Woodbury matrix equality (e.g.\ \citealp[eq. 1]{henderson1981deriving}). Similarly, $Z_{t+1}(\cdot):=\max_{\bm{a}}U_{t,t+1}(\bm{a},\cdot)$ is multivariate quadratic with a negative Hessian given by the other Schur complement as follows:
\begin{align}
\bm{I}_{t+1|n}&=\widehat{\bm{I}}_{t+1|t+1}+\bm{Q}^{-1}-\bm{Q}^{-1} \bm{T}(\bm{I}_{t|t}+\bm{T}' \bm{Q}^{-1} \bm{T})^{-1} \bm{T}' \bm{Q}^{-1},
\\
&= \widehat{\bm{I}}_{t+1|t+1} + (\bm{T}\bm{I}_{t|t}^{-1}\bm{T}' +\bm{Q})^{-1},
\\
&= \widehat{\bm{I}}_{t+1|t+1} + \bm{I}_{t+1|t},
\label{to use}
\end{align}
where the second line again follows by the Woodbury matrix identity, while the last line employs the definition $\bm{I}_{t+1|t}:=(\bm{T}\bm{I}_{t|t}^{-1}\bm{T}' +\bm{Q})^{-1}$. To derive equation~\eqref{RTS2}, we note that
\allowdisplaybreaks
\begin{align}
\bm{I}_{t|n}^{-1} &= \big[ \bm{I}_{t|t} + \bm{T}' \,(\widehat{\bm{I}}^{-1}_{t+1|t+1}+\bm{Q})^{-1}\, \bm{T}\big]^{-1},
\\
&= \bm{I}_{t|t}^{-1} - \bm{I}_{t|t}^{-1}\bm{T}' \,\big[\widehat{\bm{I}}_{t+1|t+1}^{-1}+\bm{T}\bm{I}_{t|t}^{-1}\bm{T}'+\bm{Q}\big]^{-1}\,\bm{T}\bm{I}_{t|t}^{-1}, \quad \text{by Woodbury},
\\
&= \bm{I}_{t|t}^{-1} - \bm{I}_{t|t}^{-1}\bm{T}' \,\big[\widehat{\bm{I}}_{t+1|t+1}^{-1}+\bm{I}_{t+1|t}^{-1}\big]^{-1}\,\bm{T}\bm{I}_{t|t}^{-1}, \quad \text{by Woodbury},
\\
&=\bm{I}_{t|t}^{-1} - \bm{I}_{t|t}^{-1}\bm{T}' \,\big[\bm{I}_{t+1|t} - \bm{I}_{t+1|t} (\widehat{\bm{I}}_{t+1|t+1}+\bm{I}_{t+1|t} )^{-1} \bm{I}_{t+1|t}\big]\,\bm{T}\bm{I}_{t|t}^{-1}, \quad \text{Woodbury again},
\\
&=\bm{I}_{t|t}^{-1} - \bm{I}_{t|t}^{-1}\bm{T}' \,\big[\bm{I}_{t+1|t} - \bm{I}_{t+1|t} \bm{I}_{t+1|n}^{-1} \bm{I}_{t+1|t}\big]\,\bm{T}\bm{I}_{t|t}^{-1}, \quad \text{by equation~\eqref{to use}},
\\
&=\bm{I}_{t|t}^{-1} - \bm{I}_{t|t}^{-1}\bm{T}' \bm{I}_{t+1|t}\big[\bm{I}_{t+1|t}^{-1} - \bm{I}_{t+1|n}^{-1} \big]\, \bm{I}_{t+1|t} \bm{T}\bm{I}_{t|t}^{-1}, 
\end{align}
confirming equation~\eqref{RTS2} in the main text.

\begin{landscape}
\section{Simulation study: Observation densities}
\label{sec:simulation}

\begin{table}[H]
\caption{\label{table2b} Overview of data-generating processes in simulation studies.}
\begin{footnotesize}
\begin{threeparttable}
\begin{tabular}{l@{\hspace{0.15cm}}l@{\hspace{0.15cm}}c@{\hspace{0.15cm}}c@{\hspace{0.15cm}}c@{\hspace{0.15cm}}c@{\hspace{0.15cm}}c}
  \toprule
 \multicolumn{2}{l}{\bf{DGP}} & \bf{Link function}  & \bf{Density} & \bf{Score} & \bf{Realised information}&  \bf{Information} 
 \\
Type & Distribution &   & $p(\bm{y}_t|\alpha_t)$ & $\displaystyle \frac{\dd\ell(\bm{y_t}|\alpha_t)}{\dd\alpha_t}$ & $\displaystyle  -\frac{\dd^2\ell(\bm{y_t}|\alpha_t)}{\dd\alpha_t^2}$ &
$\displaystyle \mathbb{E}\left[-\frac{\dd^2\ell(\bm{y_t}|\alpha_t)}{\dd\alpha_t^2}
\Big|\alpha_t\right]$
 \\
  \cmidrule(r{10pt}){1-2}  \cmidrule(lr){3-3} \cmidrule(lr){4-4} \cmidrule(lr){5-5} \cmidrule(lr){6-6} \cmidrule(lr){7-7}
  
  Count & Poisson & $\lambda_t=\exp(\alpha_t)$ &$\displaystyle\lambda_t^{y_t}\, \exp(-\lambda_t)/y_t! $ & $y_t-\lambda_t$ & $\lambda_t$ & $\lambda_t$  
\\
Count  & Negative bin.& $\lambda_t=\exp(\alpha_t)$ & $\displaystyle \frac{\Gamma(\kappa+y_t)\left(\frac{\kappa}{\kappa+\lambda_t}\right)^\kappa\left(\frac{\lambda_t}{\kappa+\lambda_t}\right)^{y_t}}{\Gamma(\kappa)\Gamma(y_t+1)}$ &$\displaystyle y_t-\frac{\lambda_t(\kappa+y_t)}{\kappa+\lambda_t}$ & $\displaystyle \frac{\kappa \lambda_t(\kappa+y_t)}{(\kappa+\lambda_t)^2}$ & $\displaystyle \frac{\kappa\,\lambda_t}{\kappa+\lambda_t}$ 
\\
Intensity & Exponential & $\lambda_t=\exp(\alpha_t)$ & $\displaystyle\lambda_t\, \exp(-\lambda_t y_t)$ &  $1-\lambda_t\,y_t$ & $y_t \lambda_t$ & $1$ 
\\
Duration & Gamma & $\beta_t=\exp(\alpha_t)$ & $\displaystyle
\frac{y_t^{\kappa-1}\exp(-y_t/\beta_t)}{\Gamma(\kappa)\beta_t^\kappa }
$ & $\displaystyle \frac{y_t}{\beta_t}-\kappa$& $\displaystyle \frac{y_t}{\beta_t}$ & $\kappa$ 
\\
Duration & Weibull  & $\beta_t=\exp(\alpha_t)$ & $\displaystyle\frac{\kappa\, \left(y_t/\beta_t \right)^{\kappa-1} }{\beta_t \exp\{(y_t/\beta_t)^\kappa\} }
$& $\displaystyle \kappa\left(\frac{y_t}{\beta_t}\right)^\kappa-\kappa$& $\displaystyle \kappa^2 \left(\frac{y_t}{\beta_t}\right)^\kappa$&$\kappa^2$  
\\
Volatility & Gaussian & $\sigma^2_t=\exp(\alpha_t)$ &$ \displaystyle
\frac{\exp\{-y_t^2/(2\sigma_t^2)\}}{ \{2\pi \sigma_t^2\}^{1/2} }$ & $\displaystyle \frac{y_t^2}{2\sigma_t^2}-\frac{1}{2}$
&$\displaystyle \frac{y_t^2}{2\sigma_t^2}$ &$\displaystyle\frac{1}{2}$  
\\
Volatility  & Student's \emph{t}& $\sigma^2_t=\exp(\alpha_t)$ &  $\displaystyle\frac{\Gamma\left(\frac{\nu+1}{2}\right)\left(1+\frac{y_t^2}{(\nu-2)\sigma_t^2}\right)^{-\frac{\nu+1}{2}}}{\sqrt{(\nu-2)\pi}\Gamma\left(\nu/2\right)\sigma_t}$ & 
$\displaystyle \frac{\omega_t \, y_t^2}{2\sigma_t^2}-\frac{1}{2}$
& $\displaystyle \frac{\nu-2}{\nu+1}\,\frac{\omega_t^2\,y_t^2}{2\sigma_t^2}$
& $\displaystyle \frac{\nu}{2\nu+6}$ \\ 
&&&& $\displaystyle \omega_t:=\frac{\nu+1}{\nu-2+y_t^2/\sigma_t^2}$ & &
\\
Dependence & Gaussian & $\displaystyle\rho_t=\frac{1-\exp(-\alpha_t)}{1+\exp(-\alpha_t)}$ & $\displaystyle\frac{\exp\left\{-\frac{y_{1t}^2+y_{2t}^2-2\rho_t y_{1t}y_{2t} }{2(1-\rho_t^2)} \right\}}{2\pi \sqrt{1-\rho_t^2}}$ & $
\displaystyle \frac{\rho_t}{2}+
\frac{1}{2}\frac{ z_{1t}\,z_{2t}}{1-\rho_t^2} $
& $\displaystyle 0\nleq \frac{1}{4}\frac{z_{1t}^2+z_{2t}^2}{1-\rho_t^2}- \frac{1-\rho_t^2}{4}$
& $\displaystyle \frac{1+\rho_t^2}{4}$
\\ 
&& $\displaystyle$ && $z_{1t}:=y_{1t}-\rho_t y_{2t}$
\\
&&&& $z_{2t}:=y_{2t}-\rho_t y_{1t}$
\\
Dependence & Student's \emph{t} & $\displaystyle\rho_t=\frac{1-\exp(-\alpha_t)}{1+\exp(-\alpha_t)}$ & $\displaystyle\frac{\nu \left(1+\frac{y_{1t}^2+y_{2t}^2-2\rho_t y_{1t}y_{2t} }{(\nu-2)(1-\rho_t^2)}\right)^{-\frac{\nu+2}{2}}}{2\pi(\nu-2) \sqrt{1-\rho_t^2}}$ & $\displaystyle \frac{\rho_t}{2}+
\frac{\omega_t}{2}\frac{ z_{1t}\,z_{2t}}{1-\rho_t^2} $
& $\displaystyle 0\nleq \frac{\omega_t}{4}\frac{z_{1t}^2+z_{2t}^2}{1-\rho_t^2}-\frac{1-\rho_t^2}{4}-\frac{1}{2}\frac{\omega_t^2}{\nu+2}\frac{z_{1t}^2 \, z_{2t}^2}{(1-\rho_t^2)^2}$
& $\displaystyle \frac{2+\nu(1+\rho_t^2)}{4(\nu+4)}$
\\ 
&& $\displaystyle$ && $z_{1t}:=y_{1t}-\rho_t y_{2t}$ &  \multirow{2}{*}{$\displaystyle \omega_t:=\frac{\nu+2}{\nu-2+\frac{y_{1t}^2+y_{2t}^2-2\rho_t y_{1t}y_{2t} }{1-\rho_t^2}}$}
\\
&&&& $z_{2t}:=y_{2t}-\rho_t y_{1t}$ 
\\
Local level & Student's \emph{t} & $\mu_t=\alpha_t$ & $\displaystyle\frac{\Gamma\left(\frac{\nu+1}{2}\right)\left(1+\frac{(y_t-\mu_t)^2}{(\nu-2)\sigma^2}\right)^{-\frac{\nu+1}{2}}}{\sqrt{(\nu-2)\pi}\Gamma\left(\frac{\nu}{2}\right)\sigma}$ & $\displaystyle \frac{1}{\sigma}
 \frac{(\nu+1)e_t}{\nu-2+e_t^2}$ & $\displaystyle 0\nleq  \frac{\nu+1}{\sigma^2}\frac{\nu-2-e_t^2}{(\nu-2+e_t^2)^2}$ & $\displaystyle \frac{\nu(\nu+1)}{\sigma^2(\nu-2)(\nu+3)}$
\\
& & & & $\displaystyle e_t:=\frac{y_t-\mu_t}{\sigma}$

  \\
  \bottomrule
\end{tabular}
\begin{tablenotes}
\item Note: The table contains ten data-generating processes (DGPs) and link functions, the first nine of which are adapted from \citet{koopman2016predicting}. For each model, the DGP is given by the linear Gaussian state equation~\eqref{DGP0.3} in combination with the observation density and link functions indicated in the table. The table further displays scores, realised information quantities and expected information quantities. The realised information quantities are nonnegative except for the bottom three models.
\end{tablenotes}
\end{threeparttable}
\end{footnotesize}
\end{table}

\end{landscape}

\section{Simulation study: Parameter-estimation results}
\label{S:parameter}

\begin{table}[H]
\center
\caption{Short-window parameter estimates}
\begin{footnotesize}
\renewcommand{\arraystretch}{1}
\begin{threeparttable}
\begin{tabular}{
l@{\hspace{0.3cm}} 
l@{\hspace{0.8cm}} 
c@{\hspace{0.2cm}} 
c@{\hspace{0.6cm}} 
r@{\hspace{0.2cm}} 
c@{\hspace{0.6cm}} 
r@{\hspace{0.2cm}} 
c@{\hspace{0.6cm}} 
r@{\hspace{0.2cm}} 
c@{\hspace{0.2cm}} 
}
\toprule
\multicolumn{2}{l}{\bf{DGP}} 	&	\multicolumn{2}{c}{ }	& 	\multicolumn{2}{c}{\bf{BF}$\qquad$} 	&	\multicolumn{2}{c}{\bf{PF}$\qquad$} &	\multicolumn{2}{c}{\bf{NAIS}$\qquad$}		\\ 
Type & Distribution & \multicolumn{2}{c}{Truth $\quad$} & Average & RMSE & Average & RMSE & Average & RMSE  
\\
  \cmidrule(l{0pt}r{5pt}){1-2}   \cmidrule(l{5pt}r{5pt}){3-4}   \cmidrule(l{5pt}r{5pt}){5-6} \cmidrule(l{5pt}r{5pt}){7-8}  \cmidrule(l{5pt}r{0pt}){9-10}
Count&	Poisson&	\multicolumn{1}{c}{$c$}&	$0.000$&	$-0.016$&	﻿ $[0.088]$&	$-0.003$&	﻿ $[0.042]$&	$-0.002$&	﻿ $[0.040]$	\\ 	
&	&	\multicolumn{1}{c}{$\phi$}&	$0.980$&	$0.932$&	﻿ $[0.132]$&	$0.941$&	﻿ $[0.099]$&	$0.945$&	﻿ $[0.084]$	\\ 	
&	&	\multicolumn{1}{c}{$\sigma_\eta$}&	$0.150$&	$0.182$&	﻿ $[0.083]$&	$0.170$&	﻿ $[0.070]$&	$0.168$&	﻿ $[0.060]$	\\ 	\midrule
Count&	Negative Bin.&	\multicolumn{1}{c}{$c$}&	$0.000$&	$-0.019$&	﻿ $[0.095]$&	$-0.008$&	﻿ $[0.080]$&	$-0.001$&	﻿ $[0.036]$	\\ 	
&	&	\multicolumn{1}{c}{$\phi$}&	$0.980$&	$0.925$&	﻿ $[0.147]$&	$0.929$&	﻿ $[0.153]$&	$0.946$&	﻿ $[0.099]$	\\ 	
&	&	\multicolumn{1}{c}{$\sigma_\eta$}&	$0.150$&	$0.194$&	﻿ $[0.123]$&	$0.176$&	﻿ $[0.098]$&	$0.158$&	﻿ $[0.055]$	\\ 	
&	&	\multicolumn{1}{c}{$1/\kappa$}&	$0.250$&	$0.205$&	﻿ $[0.138]$&	$0.227$&	﻿ $[0.122]$&	$0.298$&	﻿ $[0.141]$	\\ 	\midrule
Intensity&	Exponential&	\multicolumn{1}{c}{$c$}&	$0.000$&	$-0.006$&	﻿ $[0.033]$&	$0.000$&	﻿ $[0.030]$&	$0.000$&	﻿ $[0.030]$	\\ 	
&	&	\multicolumn{1}{c}{$\phi$}&	$0.980$&	$0.943$&	﻿ $[0.070]$&	$0.946$&	﻿ $[0.079]$&	$0.948$&	﻿ $[0.064]$	\\ 	
&	&	\multicolumn{1}{c}{$\sigma_\eta$}&	$0.150$&	$0.180$&	﻿ $[0.070]$&	$0.168$&	﻿ $[0.063]$&	$0.169$&	﻿ $[0.059]$	\\ 	\midrule
Duration&	Gamma&	\multicolumn{1}{c}{$c$}&	$0.000$&	$0.002$&	﻿ $[0.041]$&	$-0.003$&	﻿ $[0.036]$&	$-0.003$&	﻿ $[0.037]$	\\ 	
&	&	\multicolumn{1}{c}{$\phi$}&	$0.980$&	$0.944$&	﻿ $[0.072]$&	$0.948$&	﻿ $[0.072]$&	$0.949$&	﻿ $[0.062]$	\\ 	
&	&	\multicolumn{1}{c}{$\sigma_\eta$}&	$0.150$&	$0.175$&	﻿ $[0.062]$&	$0.166$&	﻿ $[0.054]$&	$0.166$&	﻿ $[0.054]$	\\ 	
&	&	\multicolumn{1}{c}{$\kappa$}&	$1.500$&	$1.541$&	﻿ $[0.160]$&	$1.531$&	﻿ $[0.156]$&	$1.532$&	﻿ $[0.155]$	\\ 	\midrule
Duration&	Weibull&	\multicolumn{1}{c}{$c$}&	$0.000$&	$0.005$&	﻿ $[0.041]$&	$-0.003$&	﻿ $[0.034]$&	$-0.003$&	﻿ $[0.033]$	\\ 	
&	&	\multicolumn{1}{c}{$\phi$}&	$0.980$&	$0.939$&	﻿ $[0.079]$&	$0.946$&	﻿ $[0.069]$&	$0.947$&	﻿ $[0.064]$	\\ 	
&	&	\multicolumn{1}{c}{$\sigma_\eta$}&	$0.150$&	$0.188$&	﻿ $[0.075]$&	$0.172$&	﻿ $[0.064]$&	$0.173$&	﻿ $[0.060]$	\\ 	
&	&	\multicolumn{1}{c}{$\kappa$}&	$1.200$&	$1.225$&	﻿ $[0.080]$&	$1.215$&	﻿ $[0.075]$&	$1.215$&	﻿ $[0.075]$	\\ 	\midrule
Volatility&	Gaussian&	\multicolumn{1}{c}{$c$}&	$0.000$&	$0.000$&	﻿ $[0.068]$&	$-0.004$&	﻿ $[0.063]$&	$-0.003$&	﻿ $[0.073]$	\\ 	
&	&	\multicolumn{1}{c}{$\phi$}&	$0.980$&	$0.905$&	﻿ $[0.200]$&	$0.906$&	﻿ $[0.218]$&	$0.914$&	﻿ $[0.184]$	\\ 	
&	&	\multicolumn{1}{c}{$\sigma_\eta$}&	$0.150$&	$0.202$&	﻿ $[0.119]$&	$0.174$&	﻿ $[0.112]$&	$0.183$&	﻿ $[0.099]$	\\ 	\midrule
Volatility&	Student's \emph{t}&	\multicolumn{1}{c}{$c$}&	$0.000$&	$-0.010$&	﻿ $[0.113]$&	$-0.008$&	﻿ $[0.106]$&	$-0.005$&	﻿ $[0.070]$	\\ 	
&	&	\multicolumn{1}{c}{$\phi$}&	$0.980$&	$0.870$&	﻿ $[0.261]$&	$0.872$&	﻿ $[0.311]$&	$0.914$&	﻿ $[0.162]$	\\ 	
&	&	\multicolumn{1}{c}{$\sigma_\eta$}&	$0.150$&	$0.249$&	﻿ $[0.198]$&	$0.190$&	﻿ $[0.151]$&	$0.192$&	﻿ $[0.116]$	\\ 	
&	&	\multicolumn{1}{c}{$1/\nu$}&	$0.100$&	$0.063$&	﻿ $[0.069]$&	$0.088$&	﻿ $[0.041]$&	$0.082$&	﻿ $[0.057]$	\\ 	\midrule
Dependence&	Gaussian&	\multicolumn{1}{c}{$c$}&	$0.020$&	$0.082$&	﻿ $[0.103]$&	$0.142$&	﻿ $[0.292]$&	$0.165$&	﻿ $[0.350]$	\\ 	
&	&	\multicolumn{1}{c}{$\phi$}&	$0.980$&	$0.916$&	﻿ $[0.102]$&	$0.859$&	﻿ $[0.278]$&	$0.834$&	﻿ $[0.339]$	\\ 	
&	&	\multicolumn{1}{c}{$\sigma_\eta$}&	$0.100$&	$0.124$&	﻿ $[0.090]$&	$0.155$&	﻿ $[0.185]$&	$0.144$&	﻿ $[0.132]$	\\ 	\midrule
Dependence&	Student's \emph{t}&	\multicolumn{1}{c}{$c$}&	$0.020$&	$0.148$&	﻿ $[0.321]$&	$0.263$&	﻿ $[0.540]$&	$0.189$&	﻿ $[0.349]$	\\ 	
&	&	\multicolumn{1}{c}{$\phi$}&	$0.980$&	$0.854$&	﻿ $[0.303]$&	$0.744$&	﻿ $[0.501]$&	$0.810$&	﻿ $[0.344]$	\\ 	
&	&	\multicolumn{1}{c}{$\sigma_\eta$}&	$0.100$&	$0.136$&	﻿ $[0.128]$&	$0.201$&	﻿ $[0.225]$&	$0.146$&	﻿ $[0.139]$	\\ 	
&	&	\multicolumn{1}{c}{$1/\nu$}&	$0.100$&	$0.100$&	﻿ $[0.031]$&	$0.096$&	﻿ $[0.033]$&	$0.091$&	﻿ $[0.066]$	\\ 	\midrule
Level&	Student's \emph{t}&	\multicolumn{1}{c}{$c$}&	$0.000$&	$0.000$&	﻿ $[0.016]$&	$0.000$&	﻿ $[0.019]$&	&		\\ 	
&	&	\multicolumn{1}{c}{$\phi$}&	$0.980$&	$0.965$&	﻿ $[0.027]$&	$0.959$&	﻿ $[0.034]$&	&		\\ 	
&	&	\multicolumn{1}{c}{$\sigma_\eta$}&	$0.150$&	$0.131$&	﻿ $[0.028]$&	$0.155$&	﻿ $[0.027]$&	&		\\ 	
&	&	\multicolumn{1}{c}{$\sigma$}&	$0.450$&	$0.433$&	﻿ $[0.061]$&	$0.484$&	﻿ $[0.147]$&	&		\\ 	
&	&	\multicolumn{1}{c}{$1/\nu$}&	$0.333$&	$0.237$&	﻿ $[0.121]$&	$0.324$&	﻿ $[0.083]$&	&		\\ 	
\bottomrule
 \end{tabular}
\begin{tablenotes}
\item \emph{Note}: BF = Bellman  filter. PF = Particle filter. NAIS = Numerically accelerated importance sampler. RMSE = root mean squared error. For the simulation setting, see the note to Table~\ref{table5} in the main text. 
\end{tablenotes}
\end{threeparttable}
\end{footnotesize}
\end{table}

\begin{table}[H]
\center
\caption{Medium-window parameter estimates}
\begin{footnotesize}
\renewcommand{\arraystretch}{1}
\begin{threeparttable}
\begin{tabular}{
l@{\hspace{0.3cm}} 
l@{\hspace{0.8cm}} 
c@{\hspace{0.2cm}} 
c@{\hspace{0.6cm}} 
r@{\hspace{0.2cm}} 
c@{\hspace{0.6cm}} 
r@{\hspace{0.2cm}} 
c@{\hspace{0.6cm}} 
r@{\hspace{0.2cm}} 
c@{\hspace{0.2cm}} 
}
\toprule
\multicolumn{2}{l}{\bf{DGP}} 	&	\multicolumn{2}{c}{ }	& 	\multicolumn{2}{c}{\bf{BF}$\qquad$} 	&	\multicolumn{2}{c}{\bf{PF}$\qquad$} &	\multicolumn{2}{c}{\bf{NAIS}$\qquad$}		\\ 
Type & Distribution & \multicolumn{2}{c}{Truth $\quad$} & Average & RMSE & Average & RMSE & Average & RMSE  
\\
  \cmidrule(l{0pt}r{5pt}){1-2}   \cmidrule(l{5pt}r{5pt}){3-4}   \cmidrule(l{5pt}r{5pt}){5-6} \cmidrule(l{5pt}r{5pt}){7-8}  \cmidrule(l{5pt}r{0pt}){9-10}
Count&	Poisson&	\multicolumn{1}{c}{$c$}&	$0.000$&	$-0.007$&	﻿ $[0.010]$&	$0.000$&	﻿ $[0.006]$&	$0.000$&	﻿ $[0.006]$	\\ 	
&	&	\multicolumn{1}{c}{$\phi$}&	$0.980$&	$0.974$&	﻿ $[0.013]$&	$0.975$&	﻿ $[0.011]$&	$0.975$&	﻿ $[0.011]$	\\ 	
&	&	\multicolumn{1}{c}{$\sigma_\eta$}&	$0.150$&	$0.155$&	﻿ $[0.023]$&	$0.154$&	﻿ $[0.022]$&	$0.151$&	﻿ $[0.021]$		
\\ 
\midrule
Count&	Negative Bin.&	\multicolumn{1}{c}{$c$}&	$0.000$&	$-0.004$&	﻿ $[0.008]$&	$0.000$&	﻿ $[0.007]$&	$0.001$&	﻿ $[0.006]$	\\ 	
&	&	\multicolumn{1}{c}{$\phi$}&	$0.980$&	$0.976$&	﻿ $[0.012]$&	$0.974$&	﻿ $[0.013]$&	$0.976$&	﻿ $[0.011]$	\\ 	
&	&	\multicolumn{1}{c}{$\sigma_\eta$}&	$0.150$&	$0.152$&	﻿ $[0.027]$&	$0.155$&	﻿ $[0.027]$&	$0.147$&	﻿ $[0.025]$	\\ 	
&	&	\multicolumn{1}{c}{$1/\kappa$}&	$0.250$&	$0.236$&	﻿ $[0.058]$&	$0.245$&	﻿ $[0.051]$&	$0.288$&	﻿ $[0.066]$	
\\ 
\midrule
Intensity&	Exponential&	\multicolumn{1}{c}{$c$}&	$0.000$&	$-0.007$&	﻿ $[0.010]$&	$0.000$&	﻿ $[0.007]$&	$0.000$&	﻿ $[0.007]$	\\ 	
&	&	\multicolumn{1}{c}{$\phi$}&	$0.980$&	$0.972$&	﻿ $[0.014]$&	$0.974$&	﻿ $[0.013]$&	$0.974$&	﻿ $[0.013]$	\\ 	
&	&	\multicolumn{1}{c}{$\sigma_\eta$}&	$0.150$&	$0.162$&	﻿ $[0.027]$&	$0.154$&	﻿ $[0.023]$&	$0.154$&	﻿ $[0.023]$	
\\ 
\midrule
  Duration&	Gamma&	\multicolumn{1}{c}{$c$}&	$0.000$&	$0.007$&	﻿ $[0.010]$&	$0.000$&	﻿ $[0.007]$&	$0.000$&	﻿ $[0.007]$	\\ 	
&	&	\multicolumn{1}{c}{$\phi$}&	$0.980$&	$0.973$&	﻿ $[0.013]$&	$0.974$&	﻿ $[0.012]$&	$0.974$&	﻿ $[0.012]$	\\ 	
&	&	\multicolumn{1}{c}{$\sigma_\eta$}&	$0.150$&	$0.159$&	﻿ $[0.023]$&	$0.154$&	﻿ $[0.021]$&	$0.153$&	﻿ $[0.020]$	\\ 	
&	&	\multicolumn{1}{c}{$\kappa$}&	$1.500$&	$1.510$&	﻿ $[0.070]$&	$1.503$&	﻿ $[0.069]$&	$1.503$&	﻿ $[0.069]$	
\\ 
\midrule
  Duration&	Weibull&	\multicolumn{1}{c}{$c$}&	$0.000$&	$0.009$&	﻿ $[0.012]$&	$0.000$&	﻿ $[0.007]$&	$0.000$&	﻿ $[0.007]$	\\ 	
&	&	\multicolumn{1}{c}{$\phi$}&	$0.980$&	$0.971$&	﻿ $[0.015]$&	$0.974$&	﻿ $[0.012]$&	$0.974$&	﻿ $[0.012]$	\\ 	
&	&	\multicolumn{1}{c}{$\sigma_\eta$}&	$0.150$&	$0.163$&	﻿ $[0.027]$&	$0.154$&	﻿ $[0.021]$&	$0.154$&	﻿ $[0.021]$	\\ 	
&	&	\multicolumn{1}{c}{$\kappa$}&	$1.200$&	$1.209$&	﻿ $[0.037]$&	$1.201$&	﻿ $[0.035]$&	$1.202$&	﻿ $[0.035]$	
\\ 
\midrule
  Volatility&	Gaussian&	\multicolumn{1}{c}{$c$}&	$0.000$&	$0.007$&	﻿ $[0.010]$&	$0.000$&	﻿ $[0.007]$&	$0.000$&	﻿ $[0.007]$	\\ 	
&	&	\multicolumn{1}{c}{$\phi$}&	$0.980$&	$0.970$&	﻿ $[0.019]$&	$0.973$&	﻿ $[0.016]$&	$0.973$&	﻿ $[0.016]$	\\ 	
&	&	\multicolumn{1}{c}{$\sigma_\eta$}&	$0.150$&	$0.169$&	﻿ $[0.040]$&	$0.156$&	﻿ $[0.032]$&	$0.156$&	﻿ $[0.031]$	
\\ 
\midrule
  Volatility&	Student's \emph{t}&	\multicolumn{1}{c}{$c$}&	$0.000$&	$0.004$&	﻿ $[0.010]$&	$0.000$&	﻿ $[0.007]$&	$0.000$&	﻿ $[0.007]$	\\ 	
&	&	\multicolumn{1}{c}{$\phi$}&	$0.980$&	$0.969$&	﻿ $[0.023]$&	$0.974$&	﻿ $[0.015]$&	$0.973$&	﻿ $[0.015]$	\\ 	
&	&	\multicolumn{1}{c}{$\sigma_\eta$}&	$0.150$&	$0.173$&	﻿ $[0.059]$&	$0.157$&	﻿ $[0.037]$&	$0.158$&	﻿ $[0.038]$	\\ 	
&	&	\multicolumn{1}{c}{$1/\nu$}&	$0.100$&	$0.083$&	﻿ $[0.045]$&	$0.098$&	﻿ $[0.021]$&	$0.094$&	﻿ $[0.034]$	
\\ 
\midrule
Dependence&	Gaussian&	\multicolumn{1}{c}{$c$}&	$0.020$&	$0.028$&	﻿ $[0.024]$&	$0.035$&	﻿ $[0.055]$&	$0.034$&	﻿ $[0.039]$	\\ 	
&	&	\multicolumn{1}{c}{$\phi$}&	$0.980$&	$0.972$&	﻿ $[0.023]$&	$0.965$&	﻿ $[0.056]$&	$0.966$&	﻿ $[0.038]$	\\ 	
&	&	\multicolumn{1}{c}{$\sigma_\eta$}&	$0.100$&	$0.101$&	﻿ $[0.033]$&	$0.113$&	﻿ $[0.054]$&	$0.113$&	﻿ $[0.049]$		
\\ 
\midrule
Dependence&	Student's \emph{t}&	\multicolumn{1}{c}{$c$}&	$0.020$&	$0.034$&	﻿ $[0.059]$&	$0.042$&	﻿ $[0.088]$&	$0.039$&	﻿ $[0.052]$	\\ 	
&	&	\multicolumn{1}{c}{$\phi$}&	$0.980$&	$0.966$&	﻿ $[0.063]$&	$0.958$&	﻿ $[0.082]$&	$0.961$&	﻿ $[0.053]$	\\ 	
&	&	\multicolumn{1}{c}{$\sigma_\eta$}&	$0.100$&	$0.107$&	﻿ $[0.044]$&	$0.121$&	﻿ $[0.072]$&	$0.122$&	﻿ $[0.074]$	\\ 	
&	&	\multicolumn{1}{c}{$1/\nu$}&	$0.100$&	$0.102$&	﻿ $[0.017]$&	$0.099$&	﻿ $[0.013]$&	$0.095$&	﻿ $[0.039]$	
\\ 	\midrule
Level&	Student's \emph{t}&	\multicolumn{1}{c}{$c$}&	$0.000$&	$0.000$&	﻿ $[0.005]$&	$0.000$&	﻿ $[0.006]$&	&		
\\
&	&	\multicolumn{1}{c}{$\phi$}&	$0.980$&	$0.979$&	﻿ $[0.007]$&	$0.975$&	﻿ $[0.010]$&	&		\\ 	
&	&	\multicolumn{1}{c}{$\sigma_\eta$}&	$0.150$&	$0.129$&	﻿ $[0.023]$&	$0.152$&	﻿ $[0.012]$&	&		\\ 	
&	&	\multicolumn{1}{c}{$\sigma$}&	$0.450$&	$0.431$&	﻿ $[0.033]$&	$0.455$&	﻿ $[0.053]$&	&		\\ 	
&	&	\multicolumn{1}{c}{$1/\nu$}&	$0.333$&	$0.246$&	﻿ $[0.094]$&	$0.330$&	﻿ $[0.043]$&	&		\\ 	
\bottomrule
 \end{tabular}
\begin{tablenotes}
\item \emph{Note}: BF = Bellman  filter. PF = Particle filter. NAIS = Numerically accelerated importance sampler. RMSE = root mean squared error. For the simulation setting, see the note to Table~\ref{table5} in the main text. 
\end{tablenotes}
\end{threeparttable}
\end{footnotesize}
\end{table}

\section{Simulation study: Root mean squared errors}
\label{app:F}

\begin{table}[H]
\caption{\label{table6_RMSE}Root mean squared errors (RMSEs)  of filtered states in the out-of-sample period.}
\begin{footnotesize}
\begin{threeparttable}
\begin{tabular}{l@{\hspace{0.2cm}}l@{\hspace{0.2cm}}c@{\hspace{0.2cm}}c@{\hspace{0.2cm}}c@{\hspace{0.2cm}}c@{\hspace{0.2cm}}c@{\hspace{0.2cm}}c@{\hspace{0.2cm}}c@{\hspace{0.2cm}}c@{\hspace{0.2cm}}c@{\hspace{0.2cm}}c@{\hspace{0.2cm}}c@{\hspace{0.2cm}}c@{\hspace{0.2cm}}c}
\toprule
& & & 	\multicolumn{4}{c}{\bf Short estimation} & 	\multicolumn{4}{c}{\bf  Medium estimation }& 	\multicolumn{4}{c}{\bf  Long estimation}
\\
 & &	 {\bf Infeasible} & 	\multicolumn{4}{c}{\bf  window (250 obs.)} & 	\multicolumn{4}{c}{\bf  window (1{,}000 obs.) }& 	\multicolumn{4}{c}{\bf  window (2{,}500 obs.)}
\\
{\bf DGP} & & {\bf estimator}&BF &	PF &	NAIS &	KF &	BF &	PF &	NAIS &	KF &	BF &	PF &	NAIS &	KF
\\
\cmidrule(r{5pt}l{5pt}){3-3}
\cmidrule(r{5pt}l{5pt}){4-7}\cmidrule(r{5pt}l{5pt}){8-11}\cmidrule(r{5pt}l{5pt}){12-15}
Type &	Distribution & Absolute RMSE & \multicolumn{4}{c}{Relative RMSE}	& \multicolumn{4}{c}{Relative RMSE}	 & \multicolumn{4}{c}{Relative RMSE}		\\
\cmidrule(r{5pt}l{5pt}){1-2} \cmidrule(r{5pt}l{5pt}){3-3}\cmidrule(r{5pt}l{5pt}){4-7}\cmidrule(r{5pt}l{5pt}){8-11}\cmidrule(r{5pt}l{5pt}){12-15}
Count & 	Poisson &	$0.360$&	$1.163$&	$1.157$&	$1.155$&	&	$1.015$&	$1.015$&	$1.015$&	&	$1.000$&	$1.000$ &	$1.001$ &		\\
Count &	Neg. Bin.  &	$0.379$&	$1.177$&	$1.171$&	$1.173$&	&	$1.019$&	$1.019$&	$1.020$&	&	$1.005$&	$1.005$ &	$1.006$ &		\\
Intensity  &	Exponential  &	$0.361$&	$1.139$&	$1.141$&	$1.137$&	&	$1.013$&	$1.012$&	$1.012$&	&	$1.001$&	$1.001$ &	$1.000$ &		\\
Duration  &	Gamma  &	$0.326$&	$1.169$&	$1.165$&	$1.163$&	&	$1.023$&	$1.022$&	$1.022$&	&	$1.006$&	$1.005$ &	$1.005$ &		\\
Duration  &	Weibull  &	$0.332$&	$1.126$&	$1.123$&	$1.120$&	&	$1.010$&	$1.009$&	$1.009$&	&	$0.999$&	$0.998$ &	$0.998$ &		\\
Volatility  &	Gaussian  &	$0.425$&	$1.218$&	$1.221$&	$1.220$&	$1.497$&	$1.022$&	$1.022$&	$1.022$&	$1.229$&	$1.003$&	$1.003$ &	$1.002$ &	$1.229$	\\
Volatility  &	Student's \emph{t}  &	$0.442$&	$1.250$&	$1.231$&	$1.235$&	$1.593$&	$1.039$&	$1.028$&	$1.029$&	$1.338$&	$1.012$&	$1.028$ &	$1.009$ &	$1.275$	\\
Dependence  &	Gaussian  &	$0.362$&	$1.307$&	$1.313$&	$1.321$&	&	$1.057$&	$1.056$&	$1.054$&	&	$1.017$&	$1.014$ &	$1.014$ &		\\
Dependence  &	Student's \emph{t}  &	$0.371$&	$1.314$&	$1.327$&	$1.303$&	&	$1.065$&	$1.066$&	$1.068$&	&	$1.022$&	$1.021$ &	$1.021$ &		\\
Level  &	Student's \emph{t}  &	$0.204$&	$1.058$&	$1.045$&	n/a &	$1.233$&	$1.007$&	$1.000$&	n/a &	$1.156$&	$0.998$&	$0.996$ &	n/a &	$1.148$	\\

\bottomrule
 \end{tabular}
\begin{tablenotes}
\item \emph{Note}: MAE = mean absolute error. BF = Bellman  filter. PF = particle filter. NAIS = numerically accelerated importance sampler. KF = Kalman filter. See the note to Table~\ref{table3} in the main text. The only difference is that here we report root mean squared errors (RMSEs), not mean absolute errors (MAEs).
\end{tablenotes}
\end{threeparttable}
\end{footnotesize}
\end{table}

\section{\citeauthor{catania2022stochastic}'s (\citeyear{catania2022stochastic}) model: State-space representation}
\label{app:Catania}

Fix $t>k+1$. Conditional on the information set at time $t-k-1$, denoted $\mathcal{F}_{t-k-1}$, \citeauthor{catania2022stochastic}'s (\citeyear{catania2022stochastic}) model~\eqref{catania eqn 1}--\eqref{catania eqn 3} implies that the volatility shock $\eta_t$ and the return shocks $\varepsilon_{t},\ldots,\varepsilon_{t-k}$ are jointly normally distributed as
\begin{equation}
\left[
\begin{array}{c}
\eta_t\\
\varepsilon_t\\
\varepsilon_{t-1}\\
\vdots\\
\varepsilon_{t-k}
\end{array}
\right] \; \Big| \mathcal{F}_{t-k-1}\; \sim\; \mathrm{N}\left(\left[
\begin{array}{c}
0\\
0\\
0\\
\vdots\\
0
\end{array}
\right] , 
\left[
\begin{array}{ccccc}
1 & \rho_0 & \rho_1 & \hdots & \rho_k
\\
\rho_0 & 1 & 0 & \hdots & 0
\\
\rho_1 & 0 & 1 & \hdots & 0
\\
\vdots & \vdots & \vdots & \ddots & \vdots  
\\
\rho_k & 0 & 0  & \hdots & 1
\end{array}
\right]
\right).
\end{equation}
Next, we compute the distribution of both current shocks, i.e.\ $\eta_t$ and $\varepsilon_t$, conditional on the past shocks, $\varepsilon_{t-1},\hdots,\varepsilon_{t-k}$. 
From a well-known lemma regarding conditional Gaussian distributions (e.g.\ \citealp[p.\ 165]{harvey1990forecasting}), it follows that $\eta_t,\varepsilon_t$ conditional on $\varepsilon_{t-1},\hdots,\varepsilon_{t-k}$, or, equivalently, $\mathcal{F}_{t-1}$ and $\bm{a}_{t-1}$, are jointly normally distributed as
\begin{equation}
\label{bivariate normal}
\left[
\begin{array}{c}
\eta_t\\
\varepsilon_t
\end{array}
\right] \; \Big| \mathcal{F}_{t-1},\bm{a}_{t-1}\; \sim\; \mathrm{N}\left(\left[
\begin{array}{c}
\sum_{j=1}^{k} \rho_{j} \varepsilon_{t-j}
\\
0
\end{array}
\right] , 
\left[
\begin{array}{cc}
1-\sum_{j=1}^{k} \rho_{j}^2 & \rho_0 
\\
\rho_0 & 1 
\end{array}
\right]
\right).
\end{equation}
The marginal distribution of $\eta_t$ is again Gaussian, with a mean and variance that can be read off.
Next, the state-transition equation implies that $h_t=c+\varphi h_{t-1}+\sigma_\eta\eta_t$, being a linear transformation of $\eta_t$,   is 
distributed as
\begin{align}
&h_t | \mathcal{F}_{t-1}, \bm{a}_{t-1}\; \sim\; \mathrm{N}(\mu_{h,t},\sigma^2_{h,t}),\quad \text{where}
\\
&\mu_{h,t}=c+\varphi h_{t-1}+\sigma_\eta\,
 \sum_{j=1}^k \rho_j \frac{y_{t-j}-\mu}{\exp(h_{t-j}/2)}
,\qquad \sigma_{h,t}=\sigma_\eta\,\sqrt{1-\sum_{j=1}^k \rho_j^2},
\end{align}
where we have used $\varepsilon_{t-j}=(y_{t-j}-\mu)\exp(-h_{t-j}/2)$ for $j=1,\hdots,k$ in the expression for $\mu_{h,t}$. This confirms the non-degenerate part of the state-transition density~\eqref{degenerate state dynamics}. To derive the observation density, we note that the bivariate distribution~\eqref{bivariate normal} with another application of the conditional-Gaussian lemma (\citealp[p.\ 165]{harvey1990forecasting}) gives 
\begin{align}
&\varepsilon_t |\mathcal{F}_{t-1},\bm{a}_{t-1},\eta_t\; \sim\; \mathrm{N}(\mu_{\varepsilon,t},\sigma_{\varepsilon,t}^2),\quad \text{where} \label{joint1}
\\
&\mu_{\varepsilon,t}=\frac{\rho_0}{1-\sum_{j=1}^k\rho_j^2}\left(\eta_t - \sum_{j=1}^{k}  \rho_j \varepsilon_{t-j}\right),\quad  \sigma_{\varepsilon,t}=\sqrt{1-\frac{\rho_0^2}{1-\sum_{j=1}^k\rho_j^2}}.\label{joint2}
\end{align}
Noting that neither $\mu_{\varepsilon,t}$ nor $\sigma_{\varepsilon,t}$ depend on $h_{t-k-1}$, while $\bm{a}_{t-1}$ and $\eta_t$ together imply $\bm{a}_t$, the conditioning set $(\mathcal{F}_{t-1},\bm{a}_{t-1},\eta_t)$ can be simplified to $(\mathcal{F}_{t-1},\bm{a}_t)$. Further, by substituting $\eta_t = (h_t-c-\varphi h_{t-1})/\sigma_\eta$ and $\varepsilon_{t-j}=(y_{t-j}-\mu)\exp(-h_{t-j}/2)$ for $j=1,\hdots,k$,
 equations~\eqref{joint1}--\eqref{joint2} become
\begin{align}
&\varepsilon_t |\mathcal{F}_{t-1},\bm{a}_{t},\; \sim\; \mathrm{N}(\mu_{\varepsilon,t},\sigma_{\varepsilon,t}^2) ,\quad \text{where}
\\
&\mu_{\varepsilon,t}=\frac{\rho_0}{1-\sum_{j=1}^k\rho_j^2}\left(\frac{h_t-c-\varphi h_{t-1}}{\sigma_\eta} - \sum_{j=1}^k \rho_j \frac{y_{t-j}-\mu}{\exp(h_{t-j}/2)}\right),\quad  \sigma_{\varepsilon,t}=\sqrt{1-\frac{\rho_0^2}{1-\sum_{j=1}^k\rho_j^2}}.\label{mu_eps}
\end{align}
Finally, the distribution of the observation $y_t=\mu+\exp(h_t/2)\varepsilon_t$ conditional on $\mathcal{F}_{t-1}$ and $\bm{a}_t$ is Gaussian with mean $\mu_{y,t}=\mu+\exp(h_t/2)\mu_{\varepsilon,t}$ and variance $\sigma_{y,t}^2=\exp(h_t)\sigma_{\varepsilon,t}^2$, where $\mu_{\varepsilon,t}$ and $\sigma_{\varepsilon,t}$ are given in equation~\eqref{mu_eps}. This confirms observation density~\eqref{Catania observation density}.

\section{\citeauthor{catania2022stochastic}'s (\citeyear{catania2022stochastic}) model: Bellman-filter implementation}
\label{app:Catania2}

Bellman's equation~\eqref{bivariate optimisiation} at time $t$ involves the maximisation over two state variables, i.e.\ $\bm{a}_t$ and $\bm{a}_{t-1}$, which in general contain independent components.  For the specific case of \citeauthor{catania2022stochastic}'s (\citeyear{catania2022stochastic}) model, as described in section~\ref{section8}, the state vector is $\bm{a}_{t}=(h_{t},h_{t-1},\ldots,h_{t-k})'\in \mathbb{R}^{k+1}$, which contains the log-volatility $h_t$ as well as $k$ lags. This implies that the state variables $\bm{a}_t$ and $\bm{a}_{t-1}$ have $k$ elements in common, namely $h_{t-1}$ through $h_{t-k}$. Further, $h_t$ appears only in $\bm{a}_t$, while $h_{t-k-1}$ appears only in $\bm{a}_{t-1}$. 
Taking into account these restrictions, optimisation~\eqref{bivariate optimisiation} specialised to  \citeauthor{catania2022stochastic}'s (\citeyear{catania2022stochastic}) model reads
\begin{equation}
\label{Catania optimisation}
\resizebox{.9\hsize}{!}{$
\left[ \begin{array}{c} \bm{a}_{t|t} \\ h_{t-k-1|t}
\end{array}
\right]
= \left[\begin{array}{c} h_{t|t} \\ h_{t-1|t} \\ \vdots \\ h_{t-k|t} \\ h_{t-k-1|t} \end{array} \right]
= \underset{h_t,h_{t-1},\ldots,h_{t-k-1}}{\arg \max} \Big\{ \ell(y_{t}|\bm{a}_{t},\mathcal{F}_{t-1})+\ell(h_t|\bm{a}_{t-1},\mathcal{F}_{t-1}) + V_{t-1}(\bm{a}_{t-1}) \Big\},
$
}
\end{equation}
where $\ell(\cdot|\cdot):=\log p(\cdot|\cdot)$ and the observation and state-transition densities are given in equations~\eqref{Catania observation density} and \eqref{degenerate state dynamics}, respectively. In equation~\eqref{Catania optimisation}, we have dropped the degenerate part of the state-transition density, which is permitted given that the optimisation variables are taken to be $h_t,\ldots,h_{t-k-1}$, such that the restrictions on the components of $\bm{a}_t$ and $\bm{a}_{t-1}$ are automatically satisfied. Value function $V_{t-1}:\mathbb{R}^{k+1}\to \mathbb{R}$ on the right-hand side is approximated by the quadratic form~\eqref{normalapproximation00}. 

To simplify the analysis of optimisation~\eqref{Catania optimisation}, we introduce three notational conventions. First, 
the $k+2$ optimisation variables in optimisation~\eqref{Catania optimisation} are collected in a single vector:
\begin{equation}
\label{notation1}
\bm{x}_t \;:=\; (h_t,h_{t-1},\ldots,h_{t-k-1})' =  (h_t,\bm{a}_{t-1}')' \;=\; (\bm{a}_t',h_{t-k-1})'\;\in\; \mathbb{R}^{k+2}.
\end{equation}
Second, we write the observation log density as $f:=\ell(y_t|\bm{a}_t,\mathcal{F}_{t-1})$, such that by equation~\eqref{Catania observation density} we have
\begin{align}
\allowdisplaybreaks
&f(\bm{a}_t):= -\frac{1}{2}\log(2\pi)-\log( \sigma_{y,t}) -\frac{(y_t-\mu_{y,t})^2}{2\sigma_{y,t}^2},\quad \sigma_{y,t}  = \exp(h_t/2) \sqrt{1-\frac{\rho_0^2}{1-\sum_{j=1}^k \rho_j^2}},\label{notation2}
\\
\allowdisplaybreaks
&\mu_{y,t} = \mu + \frac{\rho_0\,\exp(h_t/2) }{1-\sum_{j=1}^k \rho_j^2}\left[ \frac{h_t-c-\varphi\, h_{t-1}}{\sigma_\eta}- \sum_{j=1}^k \rho_j\, \frac{y_{t-j}-\mu}{\exp(h_{t-j}/2)}  \right].
\notag
\end{align}
Third, for the state-transition log density we use the short-hand $g:=\ell(h_t|\bm{a}_{t-1},\mathcal{F}_{t-1})$ and note from equation~\eqref{degenerate state dynamics} that it does not depend on $h_{t-k-1}$, such that we may write $g=g(\bm{a}_t)$ as follows:
\begin{align}
&g(\bm{a}_t):=-\frac{1}{2}\log(2\pi)-\log(\sigma_{h,t})-\frac{(h_t-\mu_{h,t})^2}{2\sigma_{h,t}^2} ,
\label{notation3}
 \\
&\mu_{h,t} = c +\varphi\, h_{t-1} + \sigma_\eta\,\sum_{j=1}^k \rho_j\, \frac{y_{t-j}-\mu}{ \exp(h_{t-j}/2)} ,\qquad
\sigma_{h,t} = \sigma_\eta \sqrt{1-\sum_{j=1}^k \rho_j^2}.
\notag
\end{align}
Notation~\eqref{notation1} through \eqref{notation3} allows us to write optimisation~\eqref{Catania optimisation} as
\begin{equation}
\label{Catania optimisation v2 }
\hat{\bm{x}}_{t|t}
= \underset{\bm{x}_t}{\arg \max} \Big\{ f(\bm{a}_t) + g(\bm{a}_t) - \frac{1}{2}(\bm{a}_{t-1}-\bm{a}_{t-1|t-1})' \bm{I}_{t-1|t-1} (\bm{a}_{t-1}-\bm{a}_{t-1|t-1})\Big\}.
\end{equation}
The Newton scoring algorithm for optimisation~\eqref{Catania optimisation v2 } reads
\begin{align}
\bm{x}_t  &\leftarrow  \bm{x}_t +\left[\left(\begin{array}{cc}-\frac{\dd^2 f}{\dd \bm{a}_t \dd \bm{a}_t'}-\frac{\dd^2 g}{\dd \bm{a}_t \dd \bm{a}_t'} & \bm{0}_{k+1} \\
\bm{0}_{k+1}' & 0 \end{array} \right)+ \left(\begin{array}{cc}0 & \bm{0}_{k+1}' \\
\bm{0}_{k+1} & \bm{I}_{t-1|t-1} \end{array} \right)
\right]^{-1} \notag
\\
& \hspace{4cm} \left[ \left(\begin{array}{c} \frac{\dd (f+g)}{\dd \bm{a}_t} \\0  \end{array}\right)-\left(\begin{array}{c} 0 \\ \bm{I}_{t-1|t-1}(\bm{a}_{t-1}-\bm{a}_{t-1|t-1})\end{array}\right)  \right],
\label{Newton iteration for Catania}
\end{align}
where  $\bm{0}_{k+1}$ is a column vector consisting of $k+1$ zeroes. Fisher scoring steps are obtained by replacing $\dd^2 f/(\dd \bm{a}_t \dd \bm{a}_t')$ by $ \mathbb{E}[\dd^2 f/(\dd \bm{a}_t \dd \bm{a}_t')| \bm{a}_t,\mathcal{F}_{t-1}]$. Iterating Newton step~\eqref{Newton iteration for Catania} or its Fisher equivalent requires (expectations of) first and second derivatives of $f,g$, as derived next.

\textbf{Derivatives of $f$:} By the chain rule, first and second derivatives of the function $f$ defined in equation~\eqref{notation2} with respect to $\bm{a}_t=(h_t,\ldots,h_{t-k})'$ read
\begin{align}
\frac{\dd f }{\dd\bm{a}_t} 
&= \frac{\dd f}{\dd\mu_{y,t}}  \frac{\dd\mu_{y,t}}{\dd\bm{a}_t} \,+\, \frac{\dd f}{\dd\sigma_{y,t}}   \frac{\dd\sigma_{y,t}}{\dd\bm{a}_t},
\label{f1}
\\
 \frac{\dd^2 f}{\dd\bm{a}_t\dd\bm{a}_t'}
 &=
\frac{\dd^2f}{(\dd\mu_{y,t})^2 }
\frac{\dd\mu_{y,t}}{\dd\bm{a}_t}\frac{\dd\mu_{y,t}}{\dd\bm{a}_t'}
+
\frac{\dd^2f}{(\dd\sigma_{y,t})^2 }
\frac{\dd\sigma_{y,t}}{\dd\bm{a}_t}\frac{\dd\sigma_{y,t}}{\dd\bm{a}_t'}
+
\frac{\dd^2f}{\dd\mu_{y,t}\dd\sigma_{y,t}} 
\frac{\dd\mu_{y,t}}{\dd\bm{a}_t}\frac{\dd\sigma_{y,t}}{\dd\bm{a}_t'}
\label{f2}
\\
&\hspace{2cm}
+
\frac{\dd^2f}{\dd\mu_{y,t}\dd\sigma_{y,t}} 
\frac{\dd\sigma_{y,t}}{\dd\bm{a}_t}\frac{\dd\mu_{y,t}}{\dd\bm{a}_t'} 
+ 
\frac{\dd f}{\dd \mu_{y,t}} \frac{\dd^2 \mu_{y,t}}{\dd \bm{a}_t \dd \bm{a}_t'}+\frac{\dd f}{\dd \sigma_{y,t}} \frac{\dd^2 \sigma_{y,t}}{\dd \bm{a}_t \dd \bm{a}_t'}.\notag
\\
\label{f2_v2}
\mathbb{E}\left[ \frac{\dd^2 f}{\dd\bm{a}_t\dd\bm{a}_t'}\Big| \bm{a}_{t},\mathcal{F}_{t-1}\right]
 &=
\frac{\dd^2f}{(\dd\mu_{y,t})^2 }
\frac{\dd\mu_{y,t}}{\dd\bm{a}_t}\frac{\dd\mu_{y,t}}{\dd\bm{a}_t'}
+
\mathbb{E}\left[ \frac{\dd^2f}{(\dd\sigma_{y,t})^2 }\Big| \bm{a}_{t},\mathcal{F}_{t-1}\right]
\frac{\dd\sigma_{y,t}}{\dd\bm{a}_t}\frac{\dd\sigma_{y,t}}{\dd\bm{a}_t'}
\\
&\hspace{0cm}
+
\mathbb{E}\left[\frac{\dd^2f}{\dd\mu_{y,t}\dd\sigma_{y,t}}\Big| \bm{a}_{t},\mathcal{F}_{t-1}\right] 
\frac{\dd\mu_{y,t}}{\dd\bm{a}_t}\frac{\dd\sigma_{y,t}}{\dd\bm{a}_t'}
+
\mathbb{E}\left[ \frac{\dd^2f}{\dd\mu_{y,t}\dd\sigma_{y,t}} \Big| \bm{a}_{t},\mathcal{F}_{t-1}\right] 
\frac{\dd\sigma_{y,t}}{\dd\bm{a}_t}\frac{\dd\mu_{y,t}}{\dd\bm{a}_t'}. \notag
\end{align}
Equation~\eqref{f2_v2} contains two fewer terms than equation~\eqref{f2}, because the expectation of the last two terms in equation~\eqref{f2} is zero. In equations~\eqref{f1} through \eqref{f2_v2}, derivatives of $f$ with respect $\mu_{y,t}$ and $\sigma_{y,t}$ are given by
\begin{align}
&\hspace{1cm} \frac{\dd f}{\dd \mu_{y,t}}=\frac{y_t-\mu_{y,t}}{\sigma_{y,t}^2}, \hspace{1.5cm}  \frac{\dd f}{\dd \sigma_{y,t}}=\frac{(y_t-\mu_{y,t})^2}{\sigma_{y,t}^3}-\frac{1}{\sigma_{y,t}},
\\
& \frac{\dd^2 f}{(\dd \mu_{y,t})^2}=\frac{-1}{\sigma_{y,t}^2}, \hspace{1cm}  \frac{\dd^2 f}{\dd \mu_{y,t}\dd \sigma_{y,t}}=-2\frac{y_t-\mu_{y,t}}{\sigma_{y,t}^3}
,\hspace{1cm} \frac{\dd^2 f}{(\dd \sigma_{y,t})^2}=\frac{1}{\sigma_{y,t}^2}-\frac{3(y_t-\mu_{y,t})^2}{\sigma_{y,t}^4},
\\
& \hspace{2cm} \mathbb{E}\left[ \frac{\dd^2 f}{\dd \mu_{y,t}\dd \sigma_{y,t}}\Big| \mathcal{F}_{t-1},\bm{a}_t\right]=0
,\hspace{1cm} \mathbb{E}\left[\frac{\dd^2 f}{(\dd \sigma_{y,t})^2}\Big| \mathcal{F}_{t-1},\bm{a}_t\right]=\frac{-2}{\sigma_{y,t}^2},
\end{align}
where we also give expectations when relevant for Fisher scoring steps. 
In equations~\eqref{f1} and \eqref{f2}, first  derivatives of $\mu_{y,t}$ with respect to the elements of $\bm{a}_t$ read
\begin{equation}
\resizebox{.9\hsize}{!}{$
\displaystyle \frac{\dd \mu_{y,t}}{\dd\bm{a}_t}
=
\left[
\begin{array}{c} 
 (\mu_{y,t}-\mu)/2\\ 0 \\ \vdots \\ 0 
 \end{array}
 \right]
+ \frac{\rho_0\,\exp(h_t/2)}{1-\sum_{j=1}^k \rho_j^2}
 \left[
 \begin{array}{r}
1/\sigma_\eta \hphantom{xxxxxxx} 
\\ 
-\varphi/\sigma_\eta+  \rho_1/2 \, \frac{y_{t-1}-\mu}{\exp(h_{t-1}/2)} 
\\
  \rho_2/2 \,  \frac{y_{t-2}-\mu}{\exp(h_{t-2}/2)} 
 \\ 
 \vdots\hphantom{xxxxxxxx} 
  \\
    \rho_k/2 \,  \frac{y_{t-k}-\mu}{\exp(h_{t-k}/2)} 
\end{array}
\right] =: \left[
\begin{array}{c}
 \displaystyle 
 (\mu_{y,t}-\mu)/2\\ 0 \\ \vdots \\ 0 
 \end{array}
 \right] +\bm{b}_t,
 $}
 \end{equation}
where the second equality entails a definition of $\bm{b}_t$. For second derivatives of $\mu_{y,t}$, we have
\begin{equation}
\resizebox{.9\hsize}{!}{$
\displaystyle
\frac{\dd^2 \mu_{y,t}}{\dd\bm{a}_t \dd \bm{a}_t'}=
\text{diag} 
\left[
\begin{array}{c} 
\displaystyle (\mu_{y,t}-\mu)/4\\ 0 \\ \vdots \\ 0 \end{array}
\right]
-\frac{1}{4}
 \frac{\rho_0\,\exp(h_t/2)}{1-\sum_{j=1}^k \rho_j^2}
\text{diag} 
\left[
\begin{array}{r}
0 \hphantom{xxx} 
\\ 
\rho_1 \, \frac{y_{t-1}-\mu}{\exp(h_{t-1}/2)} 
\\
 \rho_2 \,  \frac{y_{t-2}-\mu}{\exp(h_{t-2}/2)} 
 \\ 
 \vdots\hphantom{xxxxxx} 
  \\
 \rho_k\,  \frac{y_{t-k}-\mu}{\exp(h_{t-k}/2)} 
\end{array}
\right] +\left[\begin{array}{c} 1/2\\0 \\ \vdots \\0 \end{array} \right] \bm{b}_t' + \bm{b}_t \left[ \frac{1}{2}\; 0\; \hdots \; 0\right],
 $}
 \end{equation}
where the $\text{diag}$ operator creates a diagonal matrix from a given vector. The derivatives of $\sigma_{y,t}$ read
\begin{equation}
\frac{\dd \sigma_{y,t}}{\dd\bm{a}_t}
= \left[\begin{array}{c}\sigma_{y,t}/2 \\ 0 \\ \vdots \\0 \end{array} \right],
 \hspace{2cm}
\frac{\dd^2 \sigma_{y,t}}{\dd\bm{a}_t\dd\bm{a}_t'}
= \text{diag}\left[\begin{array}{c}\sigma_{y,t}/4 \\ 0 \\ \vdots \\0 \end{array} \right] .
\end{equation}
All components of equations~\eqref{f1} and~\eqref{f2} have now been specified.

\textbf{Derivatives of $g$:} By the chain rule, first and second derivatives of the function $g$ given in equation~\eqref{notation3} with respect to $\bm{a}_t=(h_t,\ldots,h_{t-k})'$ are
\begin{align}
\frac{\dd g }{\dd\bm{a}_t} 
&=  \frac{h_t-\mu_{h,t}}{\sigma_{h,t}^2}  
 \left[\begin{array}{r} -1 \hphantom{xxxxx} \\  \varphi-\frac{\sigma_\eta}{2}\, \rho_1 \frac{y_{t-1}-\mu}{\exp(h_{t-1}/2)} \\ -\frac{\sigma_\eta}{2}\, \rho_2 \frac{y_{t-2}-\mu}{\exp(h_{t-2}/2)}\\
  \vdots  \hphantom{xxxxx} \\ -\frac{\sigma_\eta}{2}\, \rho_k \frac{y_{t-k}-\mu}{\exp(h_{t-k}/2)} \end{array} \right]=: \frac{h_t-\mu_{h,t}}{\sigma_{h,t}^2}  \bm{c}_t,
\label{g1}
\\
\frac{\dd^2 g }{\dd\bm{a}_t \dd \bm{a}_t'} 
&=\frac{-1}{\sigma_{h,t}^2}  \bm{c}_t \bm{c}_t' +\frac{h_t-\mu_{h,t}}{\sigma_{h,t}^2} \,\frac{\sigma_\eta}{4}\, \text{diag}\left[\begin{array}{r} 0 \hphantom{xxxxx} \\   \rho_1\, \frac{y_{t-1}-\mu}{\exp(h_{t-1}/2)} \\  \rho_2\, \frac{y_{t-2}-\mu}{\exp(h_{t-2}/2)}\\
  \vdots  \hphantom{xxxxx} \\   \rho_k\, \frac{y_{t-k}-\mu}{\exp(h_{t-k}/2)} \end{array} \right].
\label{g2}
\end{align}
Jointly, equations~\eqref{f1} through~\eqref{g2} specify all components of the Fisher scoring step~\eqref{Newton iteration for Catania}.

Finally, the updated information matrix $\bm{I}_{t|t}$ is determined by the Schur complement of the bottom-right element of the negative Hessian matrix used in Newton's scoring step, which is given by
$$
\left(\begin{array}{cc}-\frac{\dd^2 f}{\dd \bm{a}_t \dd \bm{a}_t'}-\frac{\dd^2 g}{\dd \bm{a}_t \dd \bm{a}_t'} & \bm{0}_{k+1} \\
\bm{0}_{k+1}' & 0 \end{array} \right)+ \left(\begin{array}{cc}0 & \bm{0}_{k+1}' \\
\bm{0}_{k+1} & \bm{I}_{t-1|t-1} \end{array} \right),
$$
Taking Schur complement of the bottom-right element and evaluating the result at the peak, i.e.\ at $\bm{a}_{t|t}$, gives the updated information matrix $\bm{I}_{t|t}$. The Fisher version of the updating steps is obtained by replacing $\dd^2 f/(\dd \bm{a}_t \dd \bm{a}_t')$ by $ \mathbb{E}[\dd^2 f/(\dd \bm{a}_t \dd \bm{a}_t')| \bm{a}_t,\mathcal{F}_{t-1}]$.

\clearpage
\section{Full estimation results for the S$\&$P500}
\label{additional estimation results}

\begin{table}[h!]
\center
\caption{\label{tableR} Full estimation results for the Bellman filter (top panel) and particle filter (bottom panel).}
\begin{footnotesize}
\begin{threeparttable}
\begin{tabular}{l@{\hspace{0.2cm}}r@{\hspace{0.2cm}}r@{\hspace{0.2cm}}r@{\hspace{0.2cm}}r@{\hspace{0.2cm}}r@{\hspace{0.2cm}}r@{\hspace{0.2cm}}r@{\hspace{0.2cm}}r@{\hspace{0.2cm}}r@{\hspace{0.2cm}}r@{\hspace{0.2cm}}r@{\hspace{0.2cm}}r@{\hspace{0.2cm}}r@{\hspace{0.2cm}}r@{\hspace{0.2cm}}r@{\hspace{0.2cm}}r}
  \toprule
    \multicolumn{1}{c}{$\mu$} 
   &   \multicolumn{1}{c}{$c$} &  \multicolumn{1}{c}{$\varphi$} &  \multicolumn{1}{c}{$\sigma_\eta$} &  \multicolumn{1}{c}{$\rho_0$} &  \multicolumn{1}{c}{$\rho_1$} &  \multicolumn{1}{c}{$\rho_2$} &  \multicolumn{1}{c}{$\rho_3$} &  \multicolumn{1}{c}{$\rho_4$} &  \multicolumn{1}{c}{$\rho_5$} &  \multicolumn{1}{c}{$\rho_6$} &  \multicolumn{1}{c}{$\rho_7$} &  \multicolumn{1}{c}{$\rho_8$} &  \multicolumn{1}{c}{$\rho_9$} &  \multicolumn{1}{c}{$\rho_{10}$} & LogL & BIC\\
     \cmidrule(r{5pt}){1-15}    \cmidrule(r{5pt}){16-17}      
\tiny $﻿.0696  $ & \tiny $.0004 $ & \tiny $.9839 $ & \tiny $.2006  $ & \tiny $-.7189		 $ & \tiny $  $ & \tiny $  $ & \tiny $  $ & \tiny $  $ & \tiny $  $ & \tiny $  $ & \tiny $  $ & \tiny $  $ & \tiny $  $ & \tiny $  $ & \tiny $									-9555.1	 $ & \tiny $2.5344$
\\
﻿\tiny $.0519 $ & \tiny $	-.0017	 $ & \tiny $.9759	 $ & \tiny $.2058	 $ & \tiny $-.4830 $ & \tiny $	-.4028										  $ & \tiny $ $ & \tiny $ $ & \tiny $ $ & \tiny $ $ & \tiny $ $ & \tiny $ $ & \tiny $ $ & \tiny $ $ & \tiny $ $ & \tiny $-9531.7 $ & \tiny $2.5294$
\\
﻿\tiny $.0518 $ & \tiny $	-.0013	 $ & \tiny $.9776 $ & \tiny $	.2447	 $ & \tiny $-.4020 $ & \tiny $	-.5945 $ & \tiny $	.2910  $ & \tiny $ $ & \tiny $ $ & \tiny $ $ & \tiny $ $ & \tiny $ $ & \tiny $ $ & \tiny $ $ & \tiny $ $ & \tiny $									-9524.3	 $ & \tiny $2.5286$
\\
﻿\tiny $.0513	 $ & \tiny $-.0006 $ & \tiny $	.9815	 $ & \tiny $.2582	 $ & \tiny $-.3770 $ & \tiny $	-.5828 $ & \tiny $	-.0913 $ & \tiny $	.4633		 $ & \tiny $ $ & \tiny $ $ & \tiny $ $ & \tiny $ $ & \tiny $ $ & \tiny $ $ & \tiny $ $ & \tiny $						-9503.2 $ & \tiny $	\mathbf{2.5242}$
\\
﻿\tiny $.0509	 $ & \tiny $-.0003 $ & \tiny $	.9826 $ & \tiny $	.2456	 $ & \tiny $-.3989 $ & \tiny $	-.6108 $ & \tiny $	-.0926 $ & \tiny $	.3612	 $ & \tiny $.1463  $ & \tiny $ $ & \tiny $ $ & \tiny $ $ & \tiny $ $ & \tiny $ $ & \tiny $ $ & \tiny $					-9500.3 $ & \tiny $	2.5246$
\\
﻿\tiny $.0509	 $ & \tiny $ -.0001  $ & \tiny $	.9842  $ & \tiny $	.2456  $ & \tiny $	-.4016  $ & \tiny $ 	-.6037	 $ & \tiny $ -.0962 $ & \tiny $	.3665 $ & \tiny $	-.0382	 $ & \tiny $.2132		 $ & \tiny $ $ & \tiny $ $ & \tiny $ $ & \tiny $ $ & \tiny $ $ & \tiny $				-9494.5	 $ & \tiny $ 2.5243$
\\
﻿\tiny $.0503 $ & \tiny $	.0002	 $ & \tiny $.9852 $ & \tiny $	.2412 $ & \tiny $	-.4136  $ & \tiny $	-.6107 $ & \tiny $	-.0921 $ & \tiny $	.3715 $ & \tiny $	-.0424 $ & \tiny $	.0808 $ & \tiny $	.1616	 $ & \tiny $ $ & \tiny $ $ & \tiny $ $ & \tiny $ $ & \tiny $				-9490.9 $ & \tiny $	2.5245$
\\
﻿\tiny $.0499 $ & \tiny $	.0005 $ & \tiny $	.9862	 $ & \tiny $.2397	 $ & \tiny $-.4193  $ & \tiny $	-.6115  $ & \tiny $	-.0936 $ & \tiny $	.3750 $ & \tiny $	-.0478 $ & \tiny $	 .0916 $ & \tiny $	.0186	 $ & \tiny $.1644  $ & \tiny $ $ & \tiny $ $ & \tiny $ $ & \tiny $				-9487.6 $ & \tiny $	2.5248$
\\
﻿\tiny $.0508 $ & \tiny $	.0002 $ & \tiny $	.9867 $ & \tiny $	.2376 $ & \tiny $	-.4204 $ & \tiny $	-.6163 $ & \tiny $	-.0955 $ & \tiny $	.3817	 $ & \tiny $-.0511	 $ & \tiny $.0968 $ & \tiny $	.0159 $ & \tiny $	.0540	 $ & \tiny $.1242		 $ & \tiny $ $ & \tiny $ $ & \tiny $	-9482.0 $ & \tiny $	2.5245$
\\
﻿\tiny $.0502 $ & \tiny $	.0006	 $ & \tiny $.9875 $ & \tiny $	.2384	 $ & \tiny $-.4223	 $ & \tiny $-.6096	 $ & \tiny $-.0897 $ & \tiny $	.3791 $ & \tiny $	-.0572 $ & \tiny $	.0986	 $ & \tiny $.0188	 $ & \tiny $.0553	 $ & \tiny $-.0462 $ & \tiny $	.1901	 $ & \tiny $ $ & \tiny $	-9477.4	 $ & \tiny $ 2.5245$
\\
﻿\tiny $.0500	 $ & \tiny $ .0007	 $ & \tiny $.9881	 $ & \tiny $.2353	 $ & \tiny $-.4309 $ & \tiny $	-.6126 $ & \tiny $	-.0912	 $ & \tiny $.3828	 $ & \tiny $-.0616	 $ & \tiny $.1031	 $ & \tiny $.0175 $ & \tiny $	.0597	 $ & \tiny $-.0471 $ & \tiny $	.0804	 $ & \tiny $.1277	 $ & \tiny $-9474.5 $ & \tiny $	2.5249$
\\
    \cmidrule(r{5pt}){1-15}    \cmidrule(r{5pt}){16-17}     
    \tiny $﻿.0680 $ & \tiny $  -.0042 $ & \tiny $ 	.9850	$ & \tiny $  .1926 $ & \tiny $ 	-.7319 $ & \tiny $ 	 $ & \tiny $ 	 $ & \tiny $ 	 $ & \tiny $ 	 $ & \tiny $ 		 $ & \tiny $ $ & \tiny $ 		$ & \tiny $ $ & \tiny $ 		$ & \tiny $ $ & \tiny $ 	-9562.1$ & \tiny $ 	2.5362$
\\
\tiny $ .0517$ & \tiny $ 	-.0071$ & \tiny $ 	.9784$ & \tiny $ 	.1932	$ & \tiny $ -.5071$ & \tiny $ 	-.4149	$ & \tiny $ $ & \tiny $ 	$ & \tiny $ 		$ & \tiny $ $ & \tiny $ 	$ & \tiny $ 	$ & \tiny $ 	$ & \tiny $ 	$ & \tiny $ 	$ & \tiny $ 	-9539.3$ & \tiny $ 	2.5314$
\\
\tiny $.0511	$ & \tiny $ -.0065$ & \tiny $ 	.9796	$ & \tiny $ .2262	$ & \tiny $ -.4278$ & \tiny $ 	-.5935$ & \tiny $ 	.2732$ & \tiny $ 	$ & \tiny $ 	$ & \tiny $ 	$ & \tiny $ 	$ & \tiny $ 	$ & \tiny $ 	$ & \tiny $ 		$ & \tiny $ $ & \tiny $ 	-9534.2$ & \tiny $ 	2.5312$
\\
\tiny $.0519$ & \tiny $ 	-.0056$ & \tiny $ 	.9828$ & \tiny $ 	.2395$ & \tiny $ 	-.3979$ & \tiny $ 	-.5707$ & \tiny $ 	-.1141$ & \tiny $ 	.4593$ & \tiny $ 		$ & \tiny $ $ & \tiny $ 	$ & \tiny $ 	$ & \tiny $ 		$ & \tiny $ $ & \tiny $ 	$ & \tiny $ 	-9516.9$ & \tiny $ 	\mathbf{2.5278}$
\\
\tiny $.0513$ & \tiny $ 	-.0065$ & \tiny $ 	.9826	$ & \tiny $ .2420	$ & \tiny $ -.3743$ & \tiny $ 	-.6300$ & \tiny $ 	-.0624 $ & \tiny $ 	.4107 $ & \tiny $ 	.0501 $ & \tiny $ 	 $ & \tiny $ 	 $ & \tiny $ 	 $ & \tiny $ 	 $ & \tiny $ 	 $ & \tiny $ 	 $ & \tiny $ 	-9516.2 $ & \tiny $ 	2.5288$
\\
\tiny $.0502 $ & \tiny $ 	-.0051 $ & \tiny $ 	.9837 $ & \tiny $ 	.2284 $ & \tiny $ 	-.4059 $ & \tiny $ 	-.6137 $ & \tiny $ 	-.1062 $ & \tiny $ 	.3489 $ & \tiny $ 	.1464 $ & \tiny $ 	.0044	$ & \tiny $   $ & \tiny $ 	 $ & \tiny $ 	 $ & \tiny $ 	 $ & \tiny $ 	 $ & \tiny $ 	-9515.1	$ & \tiny $  2.5297$
\\
\tiny $.0491 $ & \tiny $  -.0041	$ & \tiny $  .9853 $ & \tiny $ 	.2267 $ & \tiny $ 	-.4217 $ & \tiny $ 	-.5909 $ & \tiny $ 	-.1206 $ & \tiny $ 	.3700	$ & \tiny $  -.0808 $ & \tiny $ 	.1629 $ & \tiny $ 	.1019	 $ & \tiny $  	 $ & \tiny $   $ & \tiny $ 	 $ & \tiny $ 		$ & \tiny $  -9509.1 $ & \tiny $ 	2.5293$
\\
\tiny $.0489 $ & \tiny $ 	-.0038	$ & \tiny $  .9860 $ & \tiny $  	.2301	$ & \tiny $  -.4171 $ & \tiny $ 	-.600 1 $ & \tiny $ 	-.1134 $ & \tiny $ 	.3845 $ & \tiny $ 	-.0756 $ & \tiny $ 	.1106 $ & \tiny $ 	-.0147 $ & \tiny $ 	.1842 $ & \tiny $ 		$ & \tiny $   $ & \tiny $ 	 $ & \tiny $ 	-9505.9 $ & \tiny $ 	2.5296$
\\
\tiny $.0495 $ & \tiny $ 	-.0039 $ & \tiny $ 	.9864 $ & \tiny $ 	.2294 $ & \tiny $ 	-.4165 $ & \tiny $ 	-.5988	$ & \tiny $  -.1126 $ & \tiny $ 	.3838 $ & \tiny $ 	-.0760 $ & \tiny $ 	.1102 $ & \tiny $ 	-.0146 $ & \tiny $ 	.1846 $ & \tiny $ 	.0001	$ & \tiny $   $ & \tiny $ 		$ & \tiny $  -9505.9 $ & \tiny $ 	2.5308$
\\
\tiny $.0495 $ & \tiny $ 	-.0039	$ & \tiny $  .9863 $ & \tiny $  	.2294$ & \tiny $ 	-.4163	$ & \tiny $ -.5991	$ & \tiny $  -.1128 $ & \tiny $ 	.3831 $ & \tiny $ 	-.0761 $ & \tiny $ 	.1104	$ & \tiny $  -.0144 $ & \tiny $ 	.1848 $ & \tiny $ 	.0001	$ & \tiny $  .0003 $ & \tiny $ 		$ & \tiny $  -9505.9$ & \tiny $ 	2.5320$
\\
\tiny $.0471 $ & \tiny $ 	-.0037 $ & \tiny $ 	.9874 $ & \tiny $ 	.2204 $ & \tiny $ 	-.4107 $ & \tiny $ 	-.6221 $ & \tiny $ 	-.1563 $ & \tiny $ 	.3621 $ & \tiny $ 	.0545 $ & \tiny $ 	.0495 $ & \tiny $ 	.0157$ & \tiny $	.0727 $ & \tiny $ 	-.0021 $ & \tiny $ 	.0014	 $ & \tiny $  .1236 $ & \tiny $ 	-9501.9 $ & \tiny $ 	2.5321$
    \\
  \bottomrule
\end{tabular}
\begin{tablenotes}
\item \emph{Note}: LogL = log likelihood. BIC = Bayesian information criterion. For each panel, the best BIC is indicated in bold. The data are $100\times$ the log returns of the S$\&$P500 from $3$ Jan $1990$ to $31$ Dec $2019$ ($7{,}558$ observations). The Bellman filter is implemented as described in Appendix~\ref{app:Catania2} and estimated using estimator~\eqref{approximate estimator}. The particle filter is estimated as in \cite{catania2022stochastic}, who uses the continuous sampling importance resampling (CSIR) method of \cite{malik2011particle}.
\end{tablenotes}
\end{threeparttable}
\end{footnotesize}
\end{table}

\end{document}